\documentclass[10pt]{article}
\usepackage{geometry}
\usepackage{amssymb} 
\usepackage{amsmath} 
\usepackage{stmaryrd} 
\usepackage{graphicx}
\usepackage{textcomp}
\usepackage{comment}
\usepackage{float}
\usepackage{mathrsfs}
\usepackage{cancel}
\usepackage{varioref}
\usepackage{subfigure}
\usepackage{algorithm}
\usepackage{algorithmic}
\usepackage{stmaryrd} 
\usepackage[numbers]{natbib}
\usepackage{tikz}

\newcommand{\vect}[1]
{\underline{\bold{#1}}}
\newcommand{\mat}[1]
{\underline{\underline{\bold{#1}}}}

\newcommand{\V}[1]
{\underline{#1}}
\newcommand{\M}[1]
{\underline{\underline{#1}}}

\newcommand{\Def}{\overset{\text{def}}{=}}

\title{A partitioned model order reduction approach to rationalise computational expenses in nonlinear fracture mechanics}

\author{P. Kerfriden$^{1}$\footnote{
email: \textit{kerfridenp@cardiff.ac.uk}, tel: +44 (0)29 20874071, fax: +44 (0)29 20874716
} , O. Goury$^{1}$, T. Rabczuk$^{2}$, S.P.A. Bordas$^{1}$
\\ \\
$\begin{array}{cl}
^{1} & \textrm{Cardiff University, School of Engineering} \\
& \textrm{Queen's Buildings, The Parade, Cardiff CF24 3AA, Wales, UK} \\
^{2} & \textrm{Institute of Structural Mechanics, Bauhaus-University Weimar} \\
& \textrm{Marienstra\ss e 15, 99423 Weimar, Germany} 
\end{array}$
}

\begin{document}
\maketitle

\begin{abstract}
We propose in this paper a reduced order modelling technique based on domain partitioning for parametric problems of fracture. We show that coupling domain decomposition and projection-based model order reduction permits to focus the numerical effort where it is most needed: around the zones where damage propagates. No \textit{a priori} knowledge of the damage pattern is required, the extraction of the corresponding spatial regions being  based solely on  algebra. The efficiency of the proposed approach is demonstrated numerically with an example relevant to engineering fracture. \\

\end{abstract}

\section{Introduction}

Engineering problems are very often characterised by a large ratio between the scale of the structure and the scale at which the phenomena of interest need to be described. In fracture mechanics, the initiation and propagation of cracks is the result of localised microscopic phenomena. These phenomena are usually represented in a homogenised manner at a scale which is suitable for the simulation: the scale of the coarser material heterogeneities (meso-scale), or the engineering scale when such a coarse representation allows for predictive results. In any case, the local nature of fracture leads to large numerical models  because sharp local gradients need to be correctly represented or because the meso-structure needs to be described in an explicit manner. To some extent, the availability of super-computing facilities alleviate this difficulty. However, in engineering design processes, a prohibitively high number of solutions might be of interest, for a range of values of design parameters, or to take into account the effect of randomness in the model for instance. Therefore, one needs to devise efficient strategies for the solution to parametric multiscale problems. In doing so, the availability of a range of efficient numerical methods for the solution to one particular realisation of the parametric problem (homogenisation techniques, advanced discretisation tools, domain decomposition and multiscale-based preconditioners for parallel computing) should not be ignored.


Model order reduction techniques that are based on the projection of fine scale problems in reduced spaces are a potential solution to this issue. Such strategies rely on the fact that the solutions to the fine-scale problem obtained for different values of the input parameters can be often represented accurately in low-dimensional subspaces spanned by well-chosen basis functions at the fine scale. Applying this idea, the numerous unknowns that arise from the discretisation of the fine-scale problem are reduced to a few state variables (i.e. the amplitude associated to each of the basis functions). Of course, obtaining the aforementioned global basis functions still requires heavy computations at the fine scale. Therefore, this class of methods is of interest if (i) the goal is to interact with a model (one can afford expensive ``offline'' computations in order to allow the user to interact with the reduced model in real or quasi-real time) or (ii)  the cost of computing the global basis remains small when compared to the cost of solving the fine-scale problem for a large range of  input parameters. This paper addresses the latter case, with a restriction to the design of structural components under extreme loading conditions.

Projection-based reduction methods have been extensively studied in system engineering (see the review proposed in \cite{antoulassorensen2001}), fluid mechanics \cite{sirovich1987,beattieborggaard2006,ansallemfarhat2008,nguyenpatera2008,buffonitelib2009} and structural dynamics \cite{craigbampton1968,dickensnakagawa1997,meyermatthies2003,barbonegivoli2003,rixen2004,markovicpark2007}. The theory and applicability of various projection-based model order reduction methods such as component mode synthesis \cite{hurty1960,craigbampton1968}, the reduced basis method \cite{prudhommerovas2002,barraultmaday2004,constantinewang2012}, the proper orthogonal decomposition \cite{pearson1901,hotelling1933,sirovich1987} which will be used in this work, the \textit{a priori} hyperreduction method \cite{ryckelynck2008,ryckelynckbenziane2010} or the proper generalised decomposition \cite{ladevezepassieux2009,chinestaammar2010,nouy2010} are now well-established in the linear to mildly nonlinear cases. Some attempts have been proposed to extend this concept to strong nonlinearities, in particular in structural mechanics \cite{yvonnethe2007,ryckelynck2008,kerfridengosselet2010,gallandgravouil2010}. This background makes it conceivable to use such methods in complex engineering problems such as fracture mechanics.

Fracture mechanics is characterised by an intrinsic lack of separation of scales between the engineering scale and the scale at which damage initiation is described. Consequently, these problems are not directly reducible by the aforementioned methods (this fact will be illustrated in the core of the paper). More precisely, the level of reducibility of such multiscale problems depends on the region of the domain which is considered. Typically, the solution in the zones where damage initiates and propagates will not be correctly approximated in low-dimensional subspaces. To circumvent this difficulty, the idea followed in this work is to use a partition of the structural components into substructures and perform a reduction of the resulting subproblems only if such a reduction can be done without sacrificing accuracy.

\begin{figure}[p]
\centering
\includegraphics[width=0.85 \textwidth]{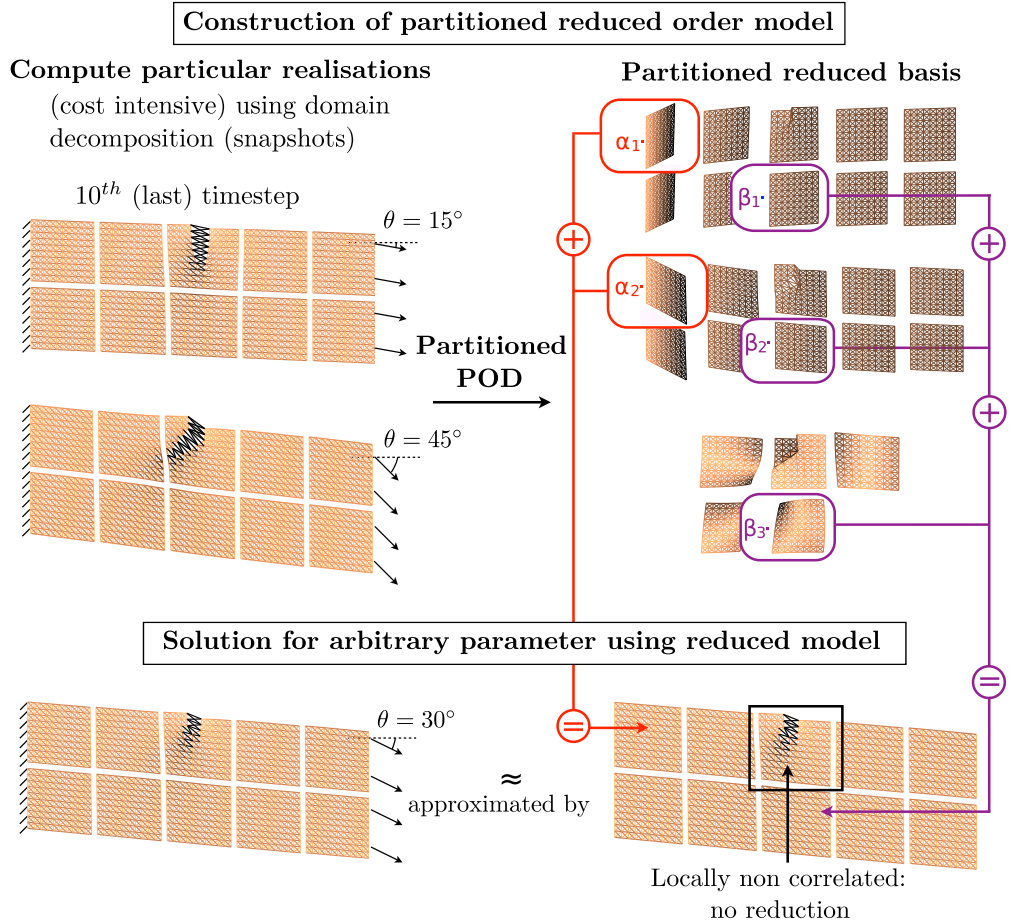}
\caption{Schematic representation of the partitioned POD-based model order reduction strategy. A Snapshot POD is performed locally for each subdomain in an "offline" phase, which requires the "truth" solution corresponding to a set of particular parameter values. 
In the``online" phase, the solution corresponding to any value of the parameter is approximated by making use of a Galerkin projection of the governing equations in the local POD subspaces. If the convergence of the local POD transforms is not satisfying in the``offline" phase, the corresponding subproblems are systematically solved without reduction in the ``online" phase (Galerkin projection of the governing equations in the local ``truth" space). The darkest bars correspond to a completely damaged state of the material, while the lightest bars are undamaged}
\label{PODwithDDM}
\end{figure}

The concept of local reduced basis itself is not new. It probably originates from the work of Craig and Bampton \cite{craigbampton1968}, who proposed a reduction by projection on a modal basis defined over predefined subdomains. This idea has been explored and improved in \cite{parkpark2004,rixen2004,markovicpark2007}, or coupled with other reduction methods, as in the case of the proper generalised decomposition \cite{ladevezepassieux2009}. A closely related family of solvers uses this concept within local/global approaches: only part of the domain is reduced (sufficiently far away from the sources of nonlinearity) \cite{barbonegivoli2003,rickeltreese2006,kerfridenpassieux2011, buffonitelib2009}, or the global reduced model is locally enriched by a fine-scale description \cite{haryadikapania1998,legresleyalonso2003,ammar2011} (these two approaches are equivalent when the reduced model is used as a preconditioner for the local fine-scale problem in the former group of methods \cite{kerfridenpassieux2011}). The work presented here is novel in the sense that (i) it is the first formal coupling between Schur-based domain decomposition approaches and model order reduction by the Proper Orthogonal Decomposition  and (ii) it is, to the authors' knowledge, the first application of systematic partitioned model order reduction for multiscale fracture.

Reduced order models obtained by the proper orthogonal decomposition (see for instance \cite{kunishvolkwein2003, legresleyalonso2003,astridweiland2008,xiaobreitkopf2010,carlbergbou-mosleh2011}) are powerful tools to reduce the computational burden associated with the repetitive analysis of parametrised nonlinear problems. The principle is to build the projection basis from the knowledge of a set of fine-scale solutions corresponding to a certain number of chosen values of the input parameters (the so-called ``snapshots''). The proper orthogonal decomposition (POD) is used to extract attractive reduced spaces from these fine scale solutions in an ``offline'' phase (we use here the terminology developed for interactivity). Classical Galerkin-based reduction is finally deployed to compute a reliable approximation of the solution to the boundary value problem for arbitrary values of the input parameters at reduced cost (``online'' phase). Let us emphasize the fact that, by construction, this family of reduction techniques rely on the ``offline'' computation of fine-scale solutions (like the reduced-basis method, and as opposed to the proper generalised decomposition and \textit{a priori} hyperreduction methods, which only require cheap fine-scale predictors).

These ``offline'' computations are potentially expensive in the case of multiscale problems, and our conception of the design process is that domain decomposition methods \cite{farhatroux1991,mandel1993,le-tallec1994a,ladevezedureisseix2000}, which are, to date, probably the most efficient family of parallel solvers, could be used to make them tractable. Examples of parallel computations using domain decomposition methods in the case of fracture can be found in \cite{germainbesson2007,allixkerfriden2010}. The purpose of this work is to reuse the substructured nature of the information generated during the ``offline'' stage to accelerate the solution process of the ``online'' stage. The choice of the domain decomposition method itself is not of prime interest here. Conceptually, we believe that the work presented in this paper can be extended to Schwartz-based methods, as done for the proper generalised decomposition in the LaTin framework \cite{ladevezepassieux2009}, or to other Schur-dual based domain decomposition methods, as presented in \cite{rixen2004} for component mode synthesis. We will focus in this work on the primal Schur-based domain decomposition method proposed in \cite{farhatroux1991,mandel1993}. This method relies on a static condensation of the subproblems on the interface degrees of freedom, and a solution of the resulting problem by a projected, preconditioned conjugate gradient in order to ensure a certain level of scalability. We propose to use the snapshot POD method to construct reduced models of the sub-problems corresponding to the interior degrees of freedom of each subdomain.

The proposed substructured approach to model order reduction (see a schematic representation in figure \ref{PODwithDDM}) is adapted to the multiscale nature of fracture problems and provides benefits in terms of applicability of POD-based reduction techniques, along the following lines. Firstly, the POD transform, even when using the snapshot technique proposed in \cite{sirovich1987} can be prohibitively expensive to compute. This issue was treated in \cite{beattieborggaard2006} by preserving the distributed nature of the snapshot data and reconstructing an approximation of the first modes of the global POD transform from local transforms computed independently for each subdomain. In our case, the POD bases will be used locally, and therefore, their parallel construction is natural. Secondly, using local reduced bases means that the dimension of the reduced spaces, can be adapted to the level of nonlinearity of the subproblems (seen as a statistic correlation of the snapshot data by the POD transform). As mentioned previously, the domain decomposition framework makes it natural to switch from a model order reduction solver to a full scale solver for the solution of subproblems for which no relevant low-dimensional reduced space can be constructed. Notice that similar ideas have been used in the context of domain decomposition methods without reduction for the treatment of localised nonlinearities arising in fracture mechanics. In \cite{lloberas-vallsrixen2011}, subproblems corresponding to domains far away from the zones of interest are treated as linear, and the local finite element discretisation is coarsened to allow for computational savings. In \cite{guidaultallix2008} and \cite{allixkerfriden2010b}, the preconditioner of the domain decomposition method is used for the coarse solution of subproblems that are far away from the process zones. At last, we believe that the systematic decomposition of the domain makes the solution of propagating nonlinearities by reduced order techniques more amenable than local refinements around evolving zones of interest.

The paper is organised as follows. In section \ref{sec:problStatementSection}, we give the general assumptions regarding the class of nonlinear problems which are addressed in this paper. Section \ref{sec:MORAndPOD} introduces classical model order reduction by projection. We focus on the snapshot POD methodology and establish the state-of-the-art of system approximations for nonlinear problems. An example of application of POD-based model order reduction in the case of fracture mechanics is presented to highlight the difficulties due to the local lack of correlation in the data. In section \ref{sec:DDM-MOR}, we introduce the primal domain decomposition method, and formally develop a POD-based model order reduction of the sub-problems in a Galerkin context. An inductive method is proposed to determine the set of fine-scale solutions that should be used to obtain a certain level of accuracy in the partitioned snapshot POD.
A system approximation strategy for the partitioned POD approach is developed in section \ref{sec:SA-PPOD}.
Finally, we propose results in terms of running time in section \ref{sec:results} (as a first step, the partitioned POD is used in a serial computing approach), and discuss further improvements for the proposed strategy.

\section{General problem statement}
\label{sec:problStatementSection}

We consider the evolution of a structure described by the partial differential equations of continuum mechanics (mechanical equilibrium and constitutive law with appropriate boundary conditions) on a bounded spatial domain $\Omega$, over time interval $\mathcal{T}=[ 0 , T ]$. The evolution in time is supposed to be quasi-static. We focus on nonlinear constitutive material models representing the progressive failure of structures, such as plasticity or damage. We assume that the damage processes are rate-independent. The mechanical problem is parametrised by a set of real variables $\vect{\boldsymbol \mu}$ that evolves in parameter domain $\mathcal{P} \subset \mathbb{R}^{n_\mu}$. 

Performing a space discretisation (finite element in our examples) of such a problem leads to a system of coupled nonlinear (ordinary differential of viscous effects were  described) equations. We look for the parametric evolution of the state variables $\vect{U}(t;\vect{\boldsymbol \mu}) \in \mathbb{R}^{n_\textrm{u}}$ satisfying the following semi-discrete problem
\begin{equation}
\forall \, (t,\vect{\boldsymbol \mu}) \in \mathcal{T} \times \mathcal{P},  \qquad \vect{F}_{\textbf{int}} \left( \left( \vect{U}(\tau; \vect{\boldsymbol \mu}) \right)_{\tau \in [0,t]} ; \vect{\boldsymbol \mu} \right) + \vect{F}_{\textbf{ext}}(t;\vect{\boldsymbol \mu}) = \vect{0} \, .
 \label{eq:semidiscrete}
\end{equation}
The vector of internal forces, $\vect{F}_{\textbf{int}} \in \mathbb{R}^{n_\textrm{u}} $, is a non-linear function of the current state variables $\vect{U}(t;\vect{\boldsymbol \mu})$ (e.g. vector of nodal values of the displacement field in finite element; we will therefore refer to it as ``displacement''). $n_\textrm{u}$ is the number of spatial unknowns in system \eqref{eq:semidiscrete}. As we model structural damage, the vector of internal forces at time $t$ also depends on the history of the state variables $\left( \vect{U}(\tau; \vect{\boldsymbol \mu}) \right)_{\tau \in [0,t[}$ over the past time interval $[ 0 , t [$. 
Typically, the dependence of $\vect{F}_{\textbf{int}}$ to the history of the displacement is due to non-reversible material processes. In the context of parametric problems, $\vect{F}_{\textbf{int}}$ may additionally depend on the design variables (design-dependent elastic constants for instance). $\vect{F}_{\textbf{ext}} \in \mathbb{R}^{n_\textrm{u}}$ is the vector of external forces, which may depend on time and on the design variables (design-dependant external load for instance). 

A classical time discretisation of semi-discrete system \eqref{eq:semidiscrete} is performed. We search for a sequence of solutions $\left( \vect{U}(t;\vect{\boldsymbol \mu}) \right)_{t \in \mathcal{T}^\textrm{h}}$, where we introduce the discrete time space $\mathcal{T}^\textrm{h} = \{  t_0, \, t_1 , \, ... , \, t_{n_\textrm{t}} \} $ such that $t_0=0$ and $t_{n_\textrm{t}}=T$, which satisfies the fully discrete set of equations
\begin{equation}
\label{eq:discrete}
\forall \, (t,\vect{\boldsymbol \mu}) \in \mathcal{T}^\textrm{h} \times \mathcal{P}, \qquad \vect{F}_{\textbf{int}} \left( \vect{U}(t;\vect{\boldsymbol \mu}) , 
\left( \vect{U}(\tau;\vect{\boldsymbol \mu}) \right)_{\tau \in \mathcal{T}^h , \,  \tau < t} ; 
\vect{\boldsymbol \mu} \right) + \vect{F}_{\textbf{ext}} (t ; \vect{\boldsymbol \mu}) = \vect{0}  \, 
\end{equation}
System \eqref{eq:discrete} is solved sequentially in time, and we assume that the structure is undamaged and at rest at $t_0$. At an arbitrary time $t \in \mathcal{T}^\text{h}$, the discrete history of the displacement $ \left( \vect{U}(\tau ; \vect{\boldsymbol \mu})  \right)_{\tau \in \mathcal{T}^h , \,  <  t}$ is known, which allows to compute vector $\vect{U}(t ;  \vect{\boldsymbol \mu})$. For readability, the dependence of the system of equations and of the solution vector to the discrete history of the variables, to the time and to the parameter will be explicitly written only if necessary.

The space and time discretisation are assumed to be sufficiently fine for our purpose (e.g.: extraction of an engineering quantity of interest). In this context, $\left( \vect{U}(t ; \vect{\boldsymbol \mu}) \right)_{t \in \mathcal{T}^\textrm{h}}$ is referred to as the ``truth" solution as it is the one that will be approximated in the reduced order modelling approach.

Discrete system \eqref{eq:discrete} at current time $t \in \mathcal{T}^\text{h}$ is \textit{a priori} nonlinear. It is solved by a usual Newton-Raphson algorithm. At iteration $i+1$ of the nonlinear solver, a tangent linear system is solved:
\begin{equation}
\label{eqNewton}
\text{Find} \ \vect{   \boldsymbol \Delta  U }^{i+1} \in \mathbb{R}^{n_\text{u}} \ \text{such that} \quad \mat{K}^i \, \vect{   \boldsymbol \Delta  U }^{i+1} = -\vect{R}^i \, ,
\end{equation}
where $\vect{   \boldsymbol \Delta  U}^{i+1} = \vect{U}^{i+1} - \vect{U}^{i}$ is an increment in the displacement vector ($\vect{U}^{i+1}$ is the actual solution of  linear prediction $i+1$), $\mat{K}^i = \left.\frac{\partial \vect{F}_{\textbf{int}}(\vect{U})}{\partial\vect{U}}\right|_{\vect{U} = \vect{U}^i}$ is the tangent operator and $\vect{R}^i =  \vect{F}_{\textbf{int}}(\vect{U}^i) + \vect{F}_{\textbf{ext}}$ is the residual of the fully discrete system of equations. The Newton algorithm is stopped if the relative euclidean norm of the residual at iteration $i+1$, $\frac{ \| \vect{R}^{i+1} \|_2 }{ \| \vect{F}_{\textbf{ext}} \|_2}$, is lower than a chosen tolerance $\epsilon_{\textrm{new}}$. 

\section{Model Order Reduction and Proper Orthogonal Decomposition}
\label{sec:MORAndPOD}

Let us recall that our goal is to solve problem \eqref{eq:discrete} for a range of admissible values of the design parameter. In this context, the property underlying the applicability of projection-based MOR is that variations in the design variables generate variations in the solution which can be represented in an attractive low-dimensional subspace of $\mathbb{R}^{n_\textrm{u}}$. Supposing that we can obtain a basis for this subspace, called "reduced space'', for instance by a particular application of the Proper Orthogonal Decomposition (``offline phase" consisting of ``truth" computations), then the evolution problem \eqref{eq:discrete} can be solved approximately for any value of the parameter by looking for the solution in the reduced space (``online phase", whose complexity must not depend on $n_\text{u}$).


\subsection{Projection-based model order reduction}
\label{sec:Proj-based_MOR}

Let us write that the solution of \eqref{eq:discrete} can be approximated, at any time $t \in \mathcal{T}^\text{h}$ and for any value of the parameter $\mu \in \mathcal{P}$, in a subspace of $\mathbb{R}^{n_\textrm{u}}$ spanned by (a few) identified basis vectors $(\vect{C}_i(t;\mu))_{i \in \llbracket 1 , n_\textrm{c} \rrbracket }$ belonging to $\mathbb{R}^{n_\textrm{u}}$:
\begin{equation}
\forall \, (t,\vect{\boldsymbol \mu}) \in \mathcal{T}^\textrm{h} \times \mathcal{P}, \qquad  \vect{U}(t;\vect{\boldsymbol \mu}) \approx \sum_{i=1}^{n_\textrm{c}} \vect{C}_i(t;\vect{\boldsymbol \mu}) \, \alpha_i(t,\vect{\boldsymbol \mu}) = \mat{C}(t;\vect{\boldsymbol \mu}) \, \vect{ \boldsymbol \alpha}(t,\vect{\boldsymbol \mu}) \, .
\end{equation}
where $\mat{C}(t;\vect{\boldsymbol \mu}) \in \mathbb{R}^{n_\textrm{u}} \times \mathbb{R}^{n_\textrm{c}}$ is a matrix whose columns are the basis vectors $(\vect{C}_i(t;\vect{\boldsymbol \mu}))_{i \in \llbracket 1 , n_\textrm{c} \rrbracket }$ and $\vect{\boldsymbol \alpha}(t,\vect{\boldsymbol \mu})$ is a vector of reduced state variables $( \alpha_i(t,\vect{\boldsymbol \mu}))_{i \in \llbracket 1 , n_\textrm{c} \rrbracket }$ that needs to be computed ``online". We emphasize that the reduced space $\textrm{Im}(\mat{C}(t;\vect{\boldsymbol \mu}))$ might depend on time and parameter, depending on the method of extraction performed ``offline''.


Injecting this approximation into \eqref{eq:discrete} at a particular point $(t,\vect{\boldsymbol \mu})$ of the time-parameter domain $\mathcal{T}^\textrm{h} \times \mathcal{P}$, one obtains an over-constrained set of equations in the $n_\textrm{c}$  reduced state variables $\vect{ \boldsymbol \alpha}$ ($n_\textrm{c} \ll n_\textrm{u}$). Let us define the residual of \eqref{eq:discrete} by
\begin{equation}
\label{eq:Discretereduced}
\forall \, \vect{ \boldsymbol \alpha }^\star \in \mathbb{R}^{n_\text{c}}, \quad \widetilde{\vect{R}} (\vect {\boldsymbol \alpha}^\star ) \overset{\text{def}}{=}  {\vect{R}} ( \mat{C} \, \vect {\boldsymbol \alpha} ) = \vect{F}_{\textbf{int}} \left( \mat{C} \, \vect{ \boldsymbol \alpha }^\star   \right) + \vect{F}_{\textbf{ext}}  \, 
\end{equation}

Determining optimal values for the reduced variables can be done in different ways in the ``online phase", depending on the physical quantities of interest and on computational tractability and stability issues. The most widely used methods are the Galerkin projection of the residual \eqref{eq:Discretereduced} and its least-square minimisation. The latter reads:
\begin{equation}
\vect {\boldsymbol \alpha} = \underset{\vect{ \boldsymbol \alpha}^{*} \in \mathbb{R}^{n_\textrm{c}}}{\operatorname{argmin}} \left( \left\| \widetilde{\vect{R}}(\vect{ \boldsymbol \alpha}^{*}) \right\|_{\mat{\Theta}} \right) \, ,
\end{equation}
where $\| \, \widetilde{\vect{R}} \, \|_{\mat{\Theta}} = \sqrt{ \widetilde{\vect{R}}^T \, \mat{\Theta} \, \widetilde{\vect{R}} } $ denotes a $\Theta$-norm of the residual vector $\widetilde{\vect{R}}$ (${\mat{\Theta}}$ is a symmetric, positive definite operator). Alternatively, in a Galerkin projection framework, $\vect {\boldsymbol \alpha}$ is defined as the solution of
\begin{equation}
\label{eq:DiscreteReducedGalerkin}
\mat{C}^{T} \, \widetilde{\vect{R}}(\vect {\boldsymbol \alpha}) = \vect{0}  \, .
\end{equation}
We use the Galerkin approach. Nonlinear problem \eqref{eq:DiscreteReducedGalerkin} can be solved by a classical Newton algorithm. The linearisation of reduced problem \eqref{eq:DiscreteReducedGalerkin} at iteration $i+1$ of a Newton solver (see for instance \cite{kerfridengosselet2010} for more details) leads to the following problem:
\begin{equation}
\begin{array}{l}
\label{eq:LeastSquareReduc}
\mat{C}^T \left( \widetilde{\vect{R}}^i + \mat{K}^i \, \mat{C} \, \vect{ \boldsymbol {\Delta \alpha} }^{i+1} \right) = \vect{0} 
\quad \Longleftrightarrow \quad
 \vect{ \boldsymbol {\Delta \alpha}}^{i+1} = \underset{\vect{ \boldsymbol {\Delta \alpha}}^{*} \in \mathbb{R}^{n_\textrm{c}}  }{\operatorname{argmin}} \left( \left\| \widetilde{\vect{R}}^i + \mat{K}^i \, \mat{C} \, \vect{\boldsymbol {\Delta \alpha}}^{*} \right\|_{{ \left( \mat{K}^i \right)}^{-1}} \right) \, ,
\end{array}
\end{equation}
where $\vect{ \boldsymbol {\Delta \alpha} }^{i+1} = \vect{ \boldsymbol{ \alpha} }^{i+1} - \vect{ \boldsymbol{ \alpha} }^{i}$ is the unknown quantity of the linear prediction and $\widetilde{\vect{R}}^i \overset{\text{def}}{=} \widetilde{\vect{R}}(\vect{ \boldsymbol{ \alpha} }^{i})$. Linearised system \eqref{eq:LeastSquareReduc} is a Galerkin reduction (or a least-square reduction as these two approaches are equivalent for the linearised problem when using a ${\mat{K}^{-1}}$-norm) of linearised equation \eqref{eqNewton} with the kinematic constraint
$\vect{\boldsymbol \Delta  U}^{i+1} = \mat{C} \, \vect{ \boldsymbol {\Delta \alpha} }^{i+1} $. The solution to \eqref{eq:LeastSquareReduc} reads
\begin{equation}
\vect{ \boldsymbol {\Delta \alpha}}^{i+1} = -\left( \mat{C}^T \, \mat{K}^i \, \mat{C} \right)^{-1} \mat{C}^T \widetilde{\vect{R}}^i \, ,
\end{equation}
providing the reduced linearised operator $\mat{C}^T \, \mat{K}^i \, \mat{C}$ (of very small size $n_\textrm{c}$) is invertible. 

At this point, we can notice the two following classical issues in projection-based model order reduction:
\begin{itemize}
\item The well-posedness of tangent problems \eqref{eq:LeastSquareReduc} and the accuracy of the solution strongly depends on the choice of the reduced space.
\item The Galerkin projection framework presented previously is inefficient. The tangent and residual of the initial problem of evolution must be evaluated at each iteration of the Newton solver. The evaluation of nonlinear function $\vect{F}_\textbf{int}$ requires a global integration over domain $\Omega$. As a result, the numerical complexity of the reduction technique does not only depend on the dimension of the reduced space but also on the size of the initial problem, which results in insignificant speed-up.
\end{itemize}

Therefore, a reduction method should provide a ``good'' reduced space (in the sense of accuracy and stability of the solution), as well as an ``efficient'' strategy to obtain the ``online" solution (significant speed-up compared to the full model, without sacrificing the accuracy expected when using a good reduced space). These two points are discussed in the following sections.

\subsection{Proper Orthogonal Decomposition in projection-based model order reduction}

\subsubsection{Proper Orthogonal Decomposition}
\label{sec:POD}

The proper orthogonal decomposition (POD) is a popular transform which is classically used to generate relevant bases for projection-based reduced order models. Applied to our parametric evolution problem, the POD decomposes the solution  of the problem over the full time-parameter domain $\tilde{ \mathcal{P}} \overset{\text{def}}{=}\mathcal{T}^\textrm{h} \times \mathcal{P}$ as
\begin{equation}
\label{eq:SepVar}
\forall \, (t,\vect{\boldsymbol \mu}) \in \tilde{ \mathcal{P}}, \qquad \vect{U}(t;\vect{\boldsymbol \mu})  = \vect{\bar{U}}(t;\vect{\boldsymbol \mu}) +  \vect{\boldsymbol{\epsilon}}(t;\vect{\boldsymbol \mu}) 
\end{equation}
\begin{equation}
\nonumber
 \vect{\bar{U}}(t;\vect{\boldsymbol \mu}) = \sum_{i=1}^{n_\textrm{p}} \vect{\boldsymbol \phi}_i \, \gamma_i (t,\vect{\boldsymbol \mu}) = \mat{\boldsymbol \phi} \, \vect{\boldsymbol \gamma}(t,\vect{\boldsymbol \mu}) \, ,
\end{equation}
such that $\vect{\bar{U}}$ is the function of separable form \eqref{eq:SepVar} that is the closest to the exact solution,
\begin{equation}
 \vect{\bar{U}} = \underset{   
 \vect{\bar{U}}^*  \in \, \{  \vect{Z} \, |  \,  \vect{Z}(t;\vect{\boldsymbol \mu}) = \mat{\boldsymbol \phi} \, \vect{\boldsymbol \gamma}(t,\vect{\boldsymbol \mu}) , \, \forall \, (t,\vect{\boldsymbol \mu}) \in \tilde{ \mathcal{P}} \}
 }{\operatorname{argmin}} d(\vect{U},\vect{\bar{U}}^*) \, ,
\end{equation}
with the metric $d$ defined on the space $\mathcal{\bar{U}}$ of functions defined over $\tilde{ \mathcal{P}}$ with values in $\mathbb{R}^{n_\textrm{u}}$:
\begin{equation} \begin{array}{cccc}
d: & \mathcal{\bar{U}} \times \mathcal{\bar{U}} & \rightarrow & \mathbb{R}
\\ 
& (\vect{U}, \vect{\bar{U}}) & \mapsto & d(\vect{U},\vect{\bar{U}})
\end{array}  \end{equation}
\begin{equation}
 \label{eq:DistancePOD}
 d(\vect{U},\vect{\bar{U}}) =  \int_{\vect{\boldsymbol \mu} \in \mathcal{P}}  \sum_{t \in \mathcal{T}^\textrm{h}} \| \vect{U}(t;\vect{\boldsymbol \mu}) - \vect{\bar{U}}(t;\vect{\boldsymbol \mu}) \|_2^2 \ d\vect{\boldsymbol \mu} \, .
 \end{equation}

\noindent $(\vect{\boldsymbol \phi}_i)_{i \in \llbracket 1,n_\textrm{p} \rrbracket}$ are ``space'' vectors that belong to $\mathbb{R}^{n_\textrm{u}}$ and are further constrained to be orthonormal with respect to the usual euclidean scalar product of $\mathbb{R}^{n_\textrm{u}}$, while $(\gamma_i)_{i \in \llbracket 1,n_\textrm{p} \rrbracket}$ are scalar functions of time and parameter.
We emphasise here the fact that the spatial basis $\mat{\boldsymbol \phi}$ is not known \textit{a priori} but is assumed to be independent on time and parameter (i.e.: we perform a separation of variables).
The POD essentially delivers a decomposition of the exact solution $\vect{U}$ into bi-orthonormal modes  $\left( (\vect{\boldsymbol \phi}_i) , \gamma_i  \right)_{i \in \llbracket 1,n_\textrm{p} \rrbracket}$ of decreasing importance. The truncation of those modes at order $n_\textrm{p}$ provides the best representation of the solution with a basis of $n_\textrm{p}$ modes in the sense that the sum over the time-parameter domain of all distances between the exact solution and its $n_\textrm{p}$-order approximation is minimised. Distance $d(\vect{U},\vect{\bar{U}})$ is expected to decrease quickly with the order of the decomposition. 


\subsubsection{Snapshot POD}

The POD transform (\ref{eq:SepVar},\ref{eq:DistancePOD}) requires the knowledge of the exact solution over $\tilde{ \mathcal{P}}$, which is not compatible with our desired usage. However, one can derive a similar transform that computes an optimal decomposition of the solution $\vect{U}$ over a discrete subset $\tilde{ \mathcal{P}}^\textrm{s} = \mathcal{T}^\textrm{h} \times \mathcal{P}^\textrm{s}$ of $\tilde{ \mathcal{P}}$. 
\begin{equation}
\label{eq:SnapshotPOD}
\forall \, (t,\vect{\boldsymbol \mu}) \in \tilde{ \mathcal{P}}^\textrm{s}  , \qquad \vect{\bar{U}}(t;\vect{\boldsymbol \mu}) =  \vect{\bar{U}}^\textrm{s}(t;\vect{\boldsymbol \mu}) + \vect{\boldsymbol \epsilon}^\textrm{s}(t;\vect{\boldsymbol \mu})
\end{equation}
\begin{equation}
\nonumber
\vect{\bar{U}}^\textrm{s}(t;\vect{\boldsymbol \mu}) = \sum_{i=1}^{n_\textrm{p}} \vect{\boldsymbol \phi}_i \, \gamma_i (t;\vect{\boldsymbol \mu}) =  \mat{\boldsymbol \phi} \, \vect{\boldsymbol \gamma}(t;\vect{\boldsymbol \mu}) \, ,
\end{equation}
such that $\vect{\bar{U}}^\textrm{s}$ is solution to the optimisation problem:
\begin{equation}
 \label{eq:DistanceSnapshotPOD}
 \vect{\bar{U}}^\textrm{s} = \underset{\vect{\bar{U}}^* \in \,\{  \vect{Z} \, |  \, \vect{Z}(t;\vect{\boldsymbol \mu}) = \mat{\boldsymbol \phi} \, \vect{\boldsymbol \gamma}(t;\mu) , \, \forall \, (t;\vect{\boldsymbol \mu}) \in \tilde{ \mathcal{P}}^\textrm{s}\} }{\operatorname{argmin}} d^\textrm{s}(\vect{U},\vect{\bar{U}}^*)
\end{equation}
with $d^\textrm{s}$ the metric defined on the space $\mathcal{\bar{U}}^\textrm{s}$ of functions defined over $\tilde{ \mathcal{P}}^\textrm{s}$ with values in $\mathbb{R}^{n_\textrm{u}}$:
\begin{equation} \begin{array}{cccc}
d^\textrm{s}: & \mathcal{\bar{U}}^\textrm{s} \times \mathcal{\bar{U}}^\textrm{s}& \rightarrow & \mathbb{R}
\\ 
& ( \vect{U} ,  \vect{\bar{U}} ) & \mapsto & d(\vect{U},\vect{\bar{U}})
\end{array}  \end{equation}
with
\begin{equation}
\label{eq:SnapPODMetric}
 d^\textrm{s}(\vect{U},\vect{\bar{U}}^\textrm{s}) = \sum_{\vect{\boldsymbol \mu} \in \mathcal{P}^\textrm{s}} \sum_{t \in \mathcal{T}^\textrm{h}}  \| \vect{U}(t;\vect{\boldsymbol \mu}) - \vect{\bar{U}}^\textrm{s}(t;\vect{\boldsymbol \mu}) \|_2^2
\end{equation}


\noindent $\mathcal{P}^\textrm{s} = \{ \vect{\boldsymbol \mu}_1 , \, ... \, , \vect{\boldsymbol \mu}_{n_\mu} \}$ is a discrete subset of the parameter domain $\mathcal{P}$. $\left( \vect{U}(t;\vect{\boldsymbol \mu}) \right)_{(t;\vect{\boldsymbol \mu})  \in \tilde{ \mathcal{P}}^\textrm{s}}$ are particular ``truth" solutions of problem \eqref{eq:discrete} for some parameters $\vect{\boldsymbol \mu} \in \tilde{\mathcal{P}}$, called snapshot. The snapshot POD metric \eqref{eq:SnapPODMetric} can be viewed as a quadrature rule for its integral counterpart \eqref{eq:DistancePOD}.

Optimal reduced spatial space $\textrm{span} \left( (\vect{\boldsymbol \phi}_i)_{i \in \llbracket 1, n_\textrm{p}\rrbracket} \right)$, with the additional constraint of orthonormality of $(\vect{\boldsymbol \phi}_i)_{i \in \llbracket 1, n_\textrm{p}\rrbracket}$, and corresponding scalar weighting functions $( \gamma_i)_{i \in \llbracket 1, n_\textrm{p}\rrbracket}$ are given, at any order $n_\textrm{p}$, by
\begin{itemize}
\item $\vect{\boldsymbol \phi}_i$ is the eigenvector of the POD operator $\mat{H}$ (covariance operators if the snapshot vectors were centred) associated to its i$^\textrm{th}$ largest eigenvalue $\lambda_i$. $\mat{H}$ is defined by
\begin{equation}
\mat{H} = \sum_{\vect{\boldsymbol \mu} \in \mathcal{P}^\textrm{s}} \sum_{t \in \mathcal{T}^\textrm{h}}  \vect{U}(t;\vect{\boldsymbol \mu}) \, \vect{U}(t;\vect{\boldsymbol \mu})^T \, .
\end{equation}
\item $\forall \, (t,\vect{\boldsymbol \mu}) \in \tilde{ \mathcal{P}}^\textrm{s} , \quad \gamma_i (t;\vect{\boldsymbol \mu}) = \vect{\boldsymbol \phi}_i^T \vect{U}(t;\vect{\boldsymbol \mu}) $
\end{itemize}
\noindent The truncation error of a POD transform of order $n_\textrm{p}$ is given by
\begin{equation}\label{errSnapshot1}
d^\textrm{s}(\vect{U},\vect{\bar{U}}^\textrm{s}) =  \displaystyle \sum_{i=n_\textrm{p}+1}^{n_\textrm{s}} {\lambda_i} \, ,
\end{equation}
where $n_\textrm{s} = n_\textrm{t} \times n_\mu$ is the number of snapshot solutions, and therefore the maximum possible rank of operator $\mat{H}$.

The eigenvalue decomposition of $\mat{H}$ is obtained at relatively cheap costs when $n_\text{t} \times n_\mu < n_\text{u}$ by exploiting the discrete nature of the available information (which is essentially the idea proposed in \cite{sirovich1987}). One computes the singular value decomposition (SVD) of the snapshot operator $\mat{S} = \left( \vect{U}(t_1,\vect{\boldsymbol \mu}_1) \ \vect{U}(t_2,\vect{\boldsymbol \mu}_1) \  \, ... \, \ \vect{U}(t_{n_\text{t}},\vect{\boldsymbol \mu}_{n_\mu}) \right)$. The SVD reads $\mat{S} = \mat{Q} \, \mat{\Sigma} \, \mat{W}^T$ with $\mat{Q}$ and $\mat{W}$ unitary matrices and $\mat{\Sigma}$ a rectangular matrix with diagonal upper block. We then have $\mat{H} = \mat{S} \, \mat{S}^T =  \mat{Q} \,\mat{\Sigma} \, \mat{W}^T \, \mat{W} \, \mat{\Sigma}^T \, \mat{Q}^T = \mat{Q} \, \mat{\Sigma} \, \mat{\Sigma}^T \, \mat{Q}^T$, which is the eigenvalue decomposition of $\mat{H}$ and the  eigenvalues are the squares of the singular values of $\mat{S}$. The values of the weighting functions $( \gamma_i)_{i \in \llbracket 1, n_\textrm{p}\rrbracket}$ over $\tilde{\mathcal{P}}^\textrm{s}$ can be readily extracted from matrix $\mat{W} $ if necessary, but this information is not of particular interest the present context.

\subsubsection{Reduced spaces in POD-based model order reduction}

The snapshot POD essentially provides an optimal decomposition of the solution in the discrete space $\tilde{ \mathcal{P}}^\textrm{s}$. It can be truncated at an order $n_\textrm{p} \leq n_\textrm{s}$ for which the normalised truncation error
\begin{equation}\label{errSnapshot}
\nu_{\textrm{snap}}^2 = d^\textrm{s}(\vect{U},\vect{\bar{U}}^\textrm{s}) = \frac{ \displaystyle \sum_{i=n_\textrm{p}+1}^{n_\textrm{s}} {\lambda_i}}{\displaystyle \sum_{i=1}^{n_\textrm{s}} {\lambda_i}} \, ,
\end{equation}
is sufficiently low. 

POD-based reduced order modelling proposes to simply discard functions $(\gamma_i)_{i \in \llbracket 1,n_\textrm{p} \rrbracket}$ (which are only defined for a discrete set of parameter values anyway), and look for the solution of the evolution problem for any value of parameter $\vect{\boldsymbol \mu} \in \tilde{ \mathcal{P} }$, in the reduced space $\textrm{span}( (\vect{\boldsymbol \phi}_i)_{i \in \llbracket 1,n_\textrm{p} \rrbracket})$. The amplitude associated with the basis vectors are computed optimally by the ``online" projection technique given in section \ref{sec:Proj-based_MOR}. 
In this context, it is clear that the snapshot POD is used to define a reduced space for projection-based reduced order modelling (which is therefore independent on time and parameter):
 \begin{equation}
\forall \, (t,\vect{\boldsymbol \mu}) \in \mathcal{T}^\textrm{h} \times \mathcal{P}, \, \forall \, i \in \llbracket 1,n_\textrm{c} \rrbracket \qquad \vect{C}_i(t;\vect{\boldsymbol \mu}) = \vect{\boldsymbol \phi}_i \qquad (n_\textrm{c} = n_\textrm{p})
 \end{equation}
 
 \vspace{1 \baselineskip} \noindent \underline{Remark:} \textit{A solution over the initial time-parameter domain $\tilde{ \mathcal{P}}$ could be reconstructed by an explicit interpolation of the functions  $(\gamma_i)_{i \in \llbracket 1,n_\textrm{p} \rrbracket}$ (i.e.: interpolation by an arbitrary polynomial basis) or by other implicit interpolation techniques such as Kriging or Moving Least-Squares for instance, as proposed in \cite{xiaobreitkopf2010,braconnierferrier2001}), which would lead to a decomposition of type \eqref{eq:SepVar}. However, such an explicit interpolation approach in $\tilde{ \mathcal{P}}$ is suboptimal as the behaviour of the governing equations between the pre-computed snapshot solutions is unknown. In addition, the Galerkin projection framework defined in section \ref{sec:Proj-based_MOR} permits to reuse the error estimates available in finite element schemes for the certification of the implicitly interpolated solution (see \cite{prudhommerovas2002,legresleyalonso2003,meyermatthies2003,quarteronirozza2011} for instance), at least in the linear case. The extension of this idea to nonlinear problems is currently an active area of research and will not be addressed in this contribution. \vspace{1 \baselineskip}}


An important point to emphasise is the requirement to perform cost-intensive simulations to compute the snapshot in the ``offline" phase. We assume in this work that the initial problem of evolution involves a large number of degrees of freedom in space and time and requires high-performance computing for the ``truth" solutions to be at reach. In particular, these solutions can be obtained efficiently on parallel architecture by using domain decomposition methods, which are, to date, probably the best parallel solvers for structural mechanics. This requirement will actually serve our needs in the case of fracture, as shown later.


\subsection{System approximation}

As stated in section \ref{sec:Proj-based_MOR}, an approximation of the fully discrete system of equations \eqref{eq:discrete} must be associated with the choice of the reduced space. In order to limit the computational expense due to the evaluation of the nonlinear functions  $\vect{F}_\textbf{int}$, two families of strategies have been intensively studied in the literature.

\subsubsection{Linearisation}

The first family proposes to linearise \cite{chenwhite2000,meyermatthies2003}, or perform a higher-order Taylor expansion \cite{rewienskiwhite2003,yvonnethe2007,niroomandialfaro2008} of the nonlinear terms in the system of equations governing the ``truth" solutions. The reduced linearised operators can be computed once and for all ``offline'' and reused ``online'' in the Newton solver. Obviously, the validity of Taylor expansions is only local along the trajectory of the reduced state variables. The authors of \cite{rewienskiwhite2003} proposed an elegant ``offline'' linearisation of the nonlinear terms of the discrete set of equations that depends on the value of the reduced state variables. In the ``online'' phase, the nonlinear terms of the discrete set of equations are approximated as a weighted combination of the ``offline'' trajectory-dependent linearisations.

\subsubsection{Evaluation of nonlinear terms on reduced spatial domains}
\label{sec:collocation}

The second family of system approximations proposes to only evaluate the nonlinear function at particular points of the domain. In a first subset of these strategies, the nonlinear function is reconstructed by interpolation over an other POD basis (``gappy'' technique) \cite{nguyenpatera2008,astridweiland2008,chaturantabutsorensen2010,carlbergbou-mosleh2011}. The expansion of the nonlinear term reads:

\begin{equation}
\begin{array}{lll}
\displaystyle
\forall \, t \in \mathcal{T}^h , \, \forall \, \vect{ \boldsymbol \alpha}^\star \in \mathbb{R}^{n_\textrm{c}}, \, 
& & \\ \displaystyle 
\qquad  \vect{F}_\textbf{int} \left(  \mat{C} \, \vect{ \boldsymbol \alpha }^\star  , \left( \mat{C} \, \vect{ \boldsymbol \alpha }(\tau,\mu) \right)_{\tau \in \mathcal{T}^h, \, \tau < t}   ; \vect{\boldsymbol \mu} \right) 
& \displaystyle \approx &  \displaystyle
 \sum_{i=1}^{n_\textrm{d}} \vect{D}_i \, \beta_i( \vect{ \boldsymbol \alpha}^\star, ( \vect{ \boldsymbol \alpha }(\tau,\mu) )_{\tau \in \mathcal{T}^h, \, \tau < t} ; \vect{\boldsymbol \mu} ) 
\\
& \displaystyle \approx & \displaystyle \mat{D} \, \vect{ \boldsymbol \beta}\left( \vect{ \boldsymbol \alpha}^\star, \left( \vect{ \boldsymbol \alpha }(\tau,\mu) \right)_{\tau \in \mathcal{T}^h, \, \tau < t} ; \vect{\boldsymbol \mu} \right) \,.
\end{array}
\end{equation}

The columns of $\mat{D} \in {\mathbb{R}}^{n_{\textrm{u}}} \times {\mathbb{R}}^{n_{\textrm{d}}} $
are spatial functions corresponding to a truncated snapshot POD expansion of the image of the reduced space by $\vect{F}_\textbf{int}$, which is performed ``offline". In practice, Newton iterates  obtained while solving the reduced model without system approximation are used to define the ``static" snapshot space $\{\vect{F}_\textbf{int} \left(  \mat{C} \, \vect{ \boldsymbol \alpha }^\star  , ( \mat{C} \, \vect{ \boldsymbol \alpha }(\tau,\mu) )_{\tau \in \mathcal{T}^h, \, \tau < t}   ; \vect{\boldsymbol \mu} \right) \, | \, t \in \mathcal{T}^h, \, \vect{ \boldsymbol \alpha}^\star \in \mathbb{R}^{n_\textrm{c}}\}$.
Interpolation coefficients $\vect{ \boldsymbol \beta}$ are found by enforcing that at any point $(t,\vect{\boldsymbol \mu})$ of $\tilde{\mathcal{P}}$, the interpolation must be optimal with respect to a limited number $n_\textrm{sa}$ of spatial degrees of freedom: 
\begin{equation}
\vect{ \boldsymbol \beta} \left( \vect{ \boldsymbol \alpha}^\star, \left( \vect{ \boldsymbol \alpha }(\tau,\mu) \right)_{\tau \in \mathcal{T}^h, \, \tau < t} ; \vect{\boldsymbol \mu} \right) 
= \underset{ \vect{\boldsymbol \beta}^\star \in \mathbb{R}^{n_\text{d}}}{\textrm{argmin}} \left( \| \mat{D} \, \vect{ \boldsymbol \beta }^\star - \vect{F}_\textbf{int} \left(  \mat{C} \, \vect{ \boldsymbol \alpha }^\star  , \left( \mat{C} \, \vect{ \boldsymbol \alpha }(\tau,\mu) \right)_{\tau \in \mathcal{T}^h, \, \tau < t}   ; \vect{\boldsymbol \mu} \right) \|_{\mat{P}} \right)
\label{eq:gappyLeastSquare}
\end{equation}
$\mat{P}$ is a boolean diagonal operator with $n_\textrm{sa}$ non-zero entries ($n_\textrm{sa} \geq n_\textrm{d}$ and $n_\textrm{sa} \ll n_\text{u}$) corresponding to the evaluation degrees of freedom of the spatial interpolation of the nonlinear term. $\| \vect{X} \|_{\mat{P}} = \sqrt{\vect{X}^T \mat{P} \, \mat{X} }$ is the semi-norm associated with $\mat{P}$ for an arbitrary vector $\vect{X} \in \mathbb{R}^{n_\text{u}}$. Substituting this approximation into the full system of equation \eqref{eq:discrete}, together with the reduced basis approximation for the displacement vector, the following reduced expression is obtained for the approximation of the ``truth" residual \eqref{eq:Discretereduced} at a particular point of the time-parameter domain:
\begin{equation}
\label{eq:DEIM}
\forall \, \vect{\boldsymbol \alpha}^\star \in \mathbb{R}^{n_\text{c}}, \qquad \vect{R}_\textbf{gap}(\vect{\boldsymbol \alpha}^\star)   \overset{\text{def}}{=}  \mat{D} \left( \mat{D}^T \, \mat{P} \, \mat{D} \right)^{-1} \mat{D}^T \mat{P} \,  \vect{F}_\textbf{int} \left(\mat{C} \, \vect{ \boldsymbol \alpha }^\star \right) + \vect{F}_\textbf{ext} \, ,
\end{equation}
where operator $\mat{D}^T \, \mat{P} \, \mat{D}$ is assumed to be invertible. The reduced variables can then be obtained in the ``offline" phase by minimising the norm of the modified residual, or by solving the Galerkin projection of the governing equations $\mat{C}^T \vect{R}_\textbf{gap}(\vect{\boldsymbol \alpha}) = \vect{0}$. Only a restriction to the evaluation degrees of freedom of the nonlinear function is calculated to evaluate the residual of the system, which allows the ``online" phase of the interpolation scheme to have a numerical complexity that does not depend on the ``truth" discretisations.  \\


The second subset of these strategies, comprising  the method proposed in \cite{legresleyalonso2001}, the Hyperreduction method \cite{ryckelynck2005} and an early version of the Missing Point Estimation technique \cite{astrid2004} can be qualified as collocations-based strategies. These methods do not reconstruct the nonlinear function over the domain. They propose instead to look for a solution that is optimal with respect to a few of the equations of the initial system \eqref{eq:discrete}. This can be expressed in a least-square approach:
\begin{equation}
 \vect{\boldsymbol \alpha} = \underset{\vect{\boldsymbol \alpha}^\star \in \mathbb{R}^{n_\text{c}}}{\textrm{argmin}} \left( \| \vect{R}_\textbf{gap}(\vect{\boldsymbol \alpha}^\star )  \|_{\mat{P}} \right) \, ,
\end{equation}
or in the (Petrov-) Galerkin framework
\begin{equation}
\label{eq:MP}
\text{Find} \ \vect{\boldsymbol \alpha} \in \mathbb{R}^{n_\text{c}} \ \text{such that} \quad \mat{C}^T \mat{P} \, \vect{R}_\textbf{gap}(\vect{\boldsymbol \alpha}) = \vect{0} \, .
\end{equation}

The strategies proposed in the literature for this two subset of techniques differ in the way of building operator $\mat{P}$, which requires a critical trade-off between optimality, stability and tractability. In \cite{astridweiland2008}, $\mat{P}$ is constructed such that the condition number of operator $\mat{D}^T \, \mat{P} \, \mat{D}$ is minimised. In the hyperreduction method \cite{ryckelynck2005}, the non-zero entries of $\mat{P}$ correspond to the largest entries (in some sense) of the approximated nonlinear vector function. In \cite{chaturantabutsorensen2010}, the points are selected to limit the growth of the residual error between a solution and its snapshot reconstruction.

\subsubsection{Chosen strategy}

We will focus in this work on the ``gappy'' technique, as used in \cite{chaturantabutsorensen2010} and \cite{carlbergbou-mosleh2011}. Since the main objective of this paper is not the system approximation strategy but the introduction of the partitioned POD technique, this method is selected as the most widely used and studied. We note for the following developments that at a particular point of the time-parameter domain, Newton iteration $i+1$ applied to reduced system \eqref{eq:DEIM}, in the Galerkin framework, reads:
\begin{equation}
\vect{ \boldsymbol {\Delta \alpha} }^{i+1} = -\left( \mat{C}^T \mat{D} (\mat{D}^T \mat{P} \mat{D})^{-1}\mat{D}^T \mat{P} \, \mat{K} \, \mat{C} \right)^{-1} \mat{C}^T \vect{R}_\textbf{gap}^i \, ,
\end{equation}
where $\vect{R}_\textbf{gap}^i \overset{\text{def}}{=} \vect{R}_\textbf{gap} (\vect{\boldsymbol { \alpha}}^i)$.


The application of this technique will be further addressed in the last section of this paper. Meanwhile, we focus on the issue of computing and using relevant POD-based reduced spaces in the particular case of fracture mechanics, using a Partitioned POD approach.

\subsection{Example of application of the POD in fracture mechanics}
\label{sec:example}

\subsubsection{Lattice model}

We consider a lattice structure made of $n_\text{b}$ damageable bars in uniaxial tension or compression. A bar marked $b \in \mathcal{B} \overset{\text{def}}{=} \llbracket 1 , n_\text{b}\rrbracket$ occupies a 1D linear domain $\Omega^{(b)}$ embedded in $\mathbb{R}^2$, such that $ \Omega \overset{\text{def}}{=} \displaystyle \bigcup_{ b \in \llbracket 1,n_\textrm{b} \rrbracket }  \Omega^{(b)}$.
We will denote by $\mathcal{P} = \{ P_i \, | \, i \in \llbracket 1,n_\textrm{pt} \rrbracket  \}$ the set of nodes of the lattice structure. Let us define the unit vector $\V{n}^{(b)}$ attached to bar $b \in \mathcal{B}$ such that if $P_i$ and $P_j$ are the two extremities of $\Omega^{(b)}$ and $i<j$, then $\V{n}^{(b)}=\frac{\V{P_i P_j}}{ \| \V{P_iP_j}  \| }$. We denote the local coordinate of point $M \in \Omega^{(b)}$ by $s^{(b)} = \| \V{P_i M} \|$. The global reference frame associated to the physical space is denoted by $\mathcal{R}(0,\V{e}_x,\V{e}_y).$

We look for a two dimensional displacement field $\V{u}$, and a scalar stress field $N$ defined over $\Omega$ that satisfy the system of equations given below. The restriction of these fields to bar $b \in \mathcal{B}$ will be denoted by $\V{u}^{(b)}$ and $N^{(b)}$ respectively. 

\paragraph{Equilibrium.}
The local mechanical equilibrium of bar $b \in \mathcal{B}$ reads, at any point of domain $\Omega^{(b)}$:
\begin{equation}
\label{eq:EquBeam}
\frac{\partial N^{(b)}}{\partial s^{(b)}} + \V{f} \cdot \V{n}^{(b)} = 0 \, ,
\end{equation}
where the body force $\V{f}$ is a two-dimensional vector field. At a lattice node $P \in \mathcal{P}$ between a set of bars denoted by $\mathcal{B}^{(i)}_{\text{n}} \subset \mathcal{B}$, the stresses are required to satisfy the nodal equilibrium, which reads, if no pointwise external force is applied at point $P$,
\begin{equation}
\label{eq:eq_joint}
\sum_{b \in \mathcal{B}^{(i)}_{\text{n}}} N^{(b)}_{|P} \ \V{\bar{n}}_{|P}^{(b)} = 0 \, ,
\end{equation}
or if $P$ belong to the set of points $\mathcal{P^\text{F}} \subset{\mathcal{P}}$ that are subjected to Neumann boundary conditions,
\begin{equation}
\label{eq:eq_joint2}
\sum_{b \in \mathcal{B}^{(i)}_{\text{n}}} N^{(b)}_{|P} \ \V{\bar{n}}_{|P}^{(b)} + \V{N}_{\textrm{d} | P} = 0 \, .
\end{equation}
In the previous equation $\V{N}_{\textrm{d} | P} \in \mathbb{R}^2$ is a prescribed force. In equilibrium equation \eqref{eq:eq_joint} and \eqref{eq:eq_joint2}, $\V{\bar{n}}^{(b)}_{|P} =  \V{n}^{(b)}$ if $s^{(b)}_{|P} = 0$ (first extremity of the bar), and  $\V{\bar{n}}^{(b)}_{|P} = - \V{n}^{(b)}$ otherwise (second extremity of the bar). 

\paragraph{Displacement admissibility.}
We assume that the restriction $\V{u}^{(b)}$ to beam $b$ of the displacement $\V{u}$ is linear. Furthermore, at any node $P \in \mathcal{P}$, the continuity of the displacement field between connected beams must be satisfied:
\begin{equation}
\forall (b,b') \in \mathcal{B}^{(i)}_{\text{n}} , \qquad \V{u}^{(b)}_{|P} = \V{u}^{(b')}_{|P} = \V{u}_{|P}  \, .
\end{equation}
The displacement field also satisfies Dirichlet boundary conditions at any node $P \in \mathcal{P^\text{u}} \subset{\mathcal{P}}$ satisfying $\mathcal{P^\text{u}} \cap \mathcal{P^\text{F}} = \{ \}$, which reads
\begin{equation}
\label{eq:Diri}
\V{u}_{|P} = \V{u}_{\textrm{d} | P}  \, ,
\end{equation}
where $\V{u}_{\textrm{d} | P} \in \mathbb{R}^2$ is a prescribed displacement.


\paragraph{Constitutive law.}
The constitutive law relates the stress and displacement fields locally. At time $t \in \mathcal{T}$, and for any $b \in \mathcal{B}$, the constitutive law expressed at an arbitrary point of domain $\Omega^{(b)}$ reads formally
\begin{equation}
N^{(b)} = N^{(b)} \left( \left\{ \epsilon^{(b)} \left(\V{u}^{(b)}_{| \tau} \right) \right\}_{\tau \leq t} \right) \, ,
\end{equation}
where the deformation $\epsilon^{(b)}$ is defined by 
\begin{equation}
\epsilon^{(b)} \left(\V{u}^{(b)} \right) = \frac{\partial \V{u}^{(b)}}{\partial s^{(b)}} \cdot \V{n}^{(b)}
\end{equation}

\subsubsection{Damage model}
\label{sec:damageModel}
The fracture of the lattice structure is described by classical damage mechanics \cite{lemaitrechaboche1990}. We postulate the existence of a free Helmholtz energy at any time $t \in \mathcal{T}$:
\begin{equation}
\psi(\epsilon^{(b)},d) = \frac{1}{2} \, E (1-d) S  \left( {\epsilon^{(b)}} \right)^2
\end{equation}
$E$ is the Young's modulus of bar $b$, $S$ is its section (assumed constant), and $d$ is a damage variable that ranges from $0$ (safe material), to $1$ (completely damaged material point). The state equations are obtained by derivation of the free energy with respect to the state variables.
\begin{equation}
N = \frac{\partial \psi}{\partial \epsilon^{(b)}} =  E(1-d) S  {\epsilon^{(b)}}  \, ,
\end{equation}
\begin{equation}
Y= - \frac{\partial \psi}{\partial d} = \frac{1}{2} \, E S \, \left( {\epsilon^{(b)}} \right)^2 \, .
\end{equation}
$Y$ is a driving force associated with the damage variable $d$. To close the system, a simple evolution law is formulated as follows
\begin{equation}
d = \textrm{min} \left\{ \underset{\tau \leq t}{ \textrm{max}} \left\{ \alpha \left( \frac{Y_{| \tau}}{Y_\text{c}} \right)^{\beta} \right\} , 1 \right\} \, ,
\end{equation}
where $Y_\text{c}$ , $\alpha$ and $\beta$ are parameters of the damage model. Notice that the history dependency in the previous equation (non-reversibility of the damage process) is inherited by the discretised system of equations. Regarding classical localisation issues related to damage models, we note that the lattice model is naturally nonlocal, the length of the beams being a length scale used as a regularisation parameter. Using shorter beams or higher order will provide material models that dissipate less energy when cracks propagate. \\


\paragraph{Variational form and discrete system of equations.}

Let us weigh the residual of the local equilibrium \eqref{eq:EquBeam} by a kinematically admissible displacement field $\V{u}^\star$, integrate over $\Omega^{(b)}$ and sum over $\mathcal{B}$:
\begin{equation}
\sum_{b \in \mathcal{B}} \int_{\Omega^{(b)}} \frac{\partial N^{(b)}}{\partial s^{(b)}} \  \V{u}^{(b)^\star} \cdot \V{n}^{(b)} \, ds^{(b)} + 
\sum_{b \in \mathcal{B}} \int_{\Omega^{(b)}} \V{f} \cdot \V{n}^{(b)} \  \V{u}^{(b)^\star} . \V{n}^{(b)} \, ds^{(b)} = 0 \, .
\end{equation}
Integrating by part the summands of first term of the last equation, and taking into both the continuity of $\V{u}^{\star}$ at any node of the lattice structure and the nodal equilibrium, one gets the variational form of the lattice problem
\begin{equation}
- \sum_{b \in \mathcal{B}} \int_{\Omega^{(b)}} N^{(b)} \frac{\partial \V{u}^{(b)^\star}}{\partial s^{(b)}} \cdot \V{n}^{(b)} \, ds^{(b)} + \sum_{b \in \mathcal{B}} \int_{\Omega^{(b)}}  \V{f} \cdot \V{n}^{(b)} \  \V{u}^{(b)^\star} . \V{n}^{(b)} \, ds^{(b)} + 
\sum_{P \in \mathcal{P}^\text{F}} \V{N}_{\textrm{d}|P} \cdot \V{u}_{|P}^{\star} = 0 \, ,
\, 
\end{equation}
where we have additionally enforced the condition that test function $\V{u}^{\star}$ vanishes at every node belonging to $P^\text{u}$. Last, by writing the piecewise linearity of the displacement field of bar $b \in \mathcal{B}$ in the form:
\begin{equation}
\V{u}^{(b)}(s^{(b)}) = \M{\Lambda}^{(b)} \left( s^{(b)} \right) \, {\mat{\tilde{A}}^{(b)}}^T \vect{U} \quad  \text{with} \quad 
\M{\Lambda}^{(b)}\left(s^{(b)} \right) =
\begin{pmatrix}
1-\tilde{s}^{(b)} & 0 & \tilde{s}^{(b)} & 0 \\
0 & 1-\tilde{s}^{(b)} & 0 & \tilde{s}^{(b)} 
\end{pmatrix}
\end{equation}
where $\tilde{s}^{(b)} \overset{\text{def}}{=} \frac{s^{(b)}}{\| \V{P_i P_j} \|}$ and $\mat{\tilde{A}}^{(b)}$ the assembly operator such that $\vect{U}^{(b)} = {\mat{\tilde{A}}^{(b)}}^T \, \vect{U}$ with \\
$\vect{U}^{(b)} = \left( \V{u}_{|P_i} \cdot \V{e}_x \quad \V{u}_{|P_i} \cdot \V{e}_y  \quad \V{u}_{|P_j} \cdot \V{e}_x \quad  \V{u}_{|P_j} \cdot \V{e}_y \right)^T$ the vector of nodal values of the restriction of the displacement to bar $b$, $\vect{U}$ the global vector of nodal displacement values and $P_i$ and $P_j$ ($i<j$) the two extremities of bar $b$, we get the expression of the semi-discrete problem at time $t \in \mathcal{T}$:
\begin{equation}
\label{eq:VarPrinc}
\begin{array}{l}
\displaystyle
\forall \, \vect{U}^\star  \in \mathbb{R}^{\tilde{n}_\text{u}}  \  \text{such that} \  \left( {\mat{\hat{A}}^{(P)}}^T   \vect{U}^\star =0, \, \forall \, P \in \mathcal{P}^\text{u} \right) , \quad { \vect{U}^\star }^T \left( \vect{\tilde{F}}_{\textbf{int}} \left((\vect{U}(\tau))_{\tau \in [0,t]} \right)  +  \vect{\tilde{F}}_{\text{ext}}(t)  \right) = 0 
\\ \displaystyle \text{with} \quad 
\left\{ \begin{array}{l}
\displaystyle
\vect{\tilde{F}}_{\textbf{int}} \left((\vect{U}(\tau))_{\tau \in [0,t]} \right) = - \sum_{b \in \mathcal{B}} {\mat{\tilde{A}}^{(b)}} \int_{\Omega^{(b)}} \frac{\partial \M{\Lambda}^{(b)^T}}{\partial s^{(b)}}  \,\V{n}^{(b)} N^{(b)} \left( (\vect{U}(\tau))_{\tau \in [0,t]} \right)  \, ds^{(b)} 
\\  \displaystyle
\vect{\tilde{F}}_{\text{ext}}(t)  = 
\sum_{b \in \mathcal{B}}  {\mat{\tilde{A}}^{(b)}} \int_{\Omega^{(b)}} \M{\Lambda}^{(b)^T} {\V{n}^{(b)}} \, {\V{n}^{(b)}}^T \V{f}(t)  \, ds^{(b)} + 
\sum_{P \in \mathcal{P}^\text{F}} {\mat{\hat{A}}^{(P)}} \V{N}_{\textrm{d}|P}(t)  \, ,
\end{array} \right.
\end{array}  
\end{equation}
where ${\mat{\hat{A}}^{(P)}}$ an assembly defined such that at any node $P \in \mathcal{P}$, we have $ \left( \V{u}_{|P} \cdot \V{e}_x \quad \V{u}_{|P} \cdot \V{e}_y  \right)^T = {\mat{\hat{A}}^{(P)}}^T   \vect{U} $. Variational principle \eqref{eq:VarPrinc} needs to be complemented by the Dirichlet boundary conditions \eqref{eq:Diri}, and the resulting problem can be parametrised and discretised in time to obtain the ``truth" problem \eqref{eq:discrete}.
 

\subsubsection{Parametrised problem of fracture}

\label{subsubsec:example}
\begin{figure}[htb]
\centering
\includegraphics[width=0.7 \textwidth]{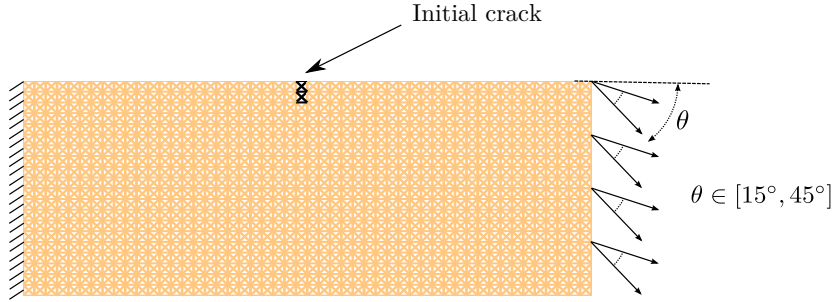}
\caption{Definition of the nonlinear lattice problem used for the numerical experiments of this paper. The loss of stiffness of each bar while increasing local strain is described by a damage model. The direction of the prescribed displacement on the right-hand edge of the rectangular lattice structure is a parameter of the model. The aim is to predict the propagation of the damage onset (initially damaged bars represented in black) for any angle of the prescribed load.}
\label{presentLattice}
\end{figure}

The leftmost part of the structure is fixed (null Dirichlet boundary conditions) while a prescribed displacement, which puts the structure in tension, is gradually applied on the rightmost part. The direction of the load is controlled by an input parameter $ \theta (\equiv \V{\boldsymbol \mu}) \in \mathbb{R}$ which ranges in $\mathcal{P} = [15^\circ, 45^\circ]$. An initial crack (notch) is defined at the top centre of the structure by initially setting the damage fields of the corresponding bars to 1, as illustrated in figure \ref{presentLattice}. As the load is progressively applied to the damageable structure, the crack propagates. The time evolution of the crack propagation problem is discretised using 10 homogeneous load steps. The lattice structure is built up using $1071$ nodes linked by $4070$ bars. The Young's moduli, bar sections and and lengths of the horizontal and vertical bars of the regular lattice are set to unity. The body force field is null.

Our goal is to predict the damage state in the lattice for any  arbitrary angle $\theta \in \mathcal{P}$ without solving the ``truth" model. The solution will be looked for in a space generated by a spectral analysis of precomputed solutions (Snapshot POD) corresponding to a number $n_\mu$ of particular parameters distributed homogeneously in the unidimensional parameter domain and including the two extrema values of $\theta$, $15^\circ$ and $45^\circ$.



\begin{figure}[p]
\centering
\includegraphics[width=0.75 \textwidth]{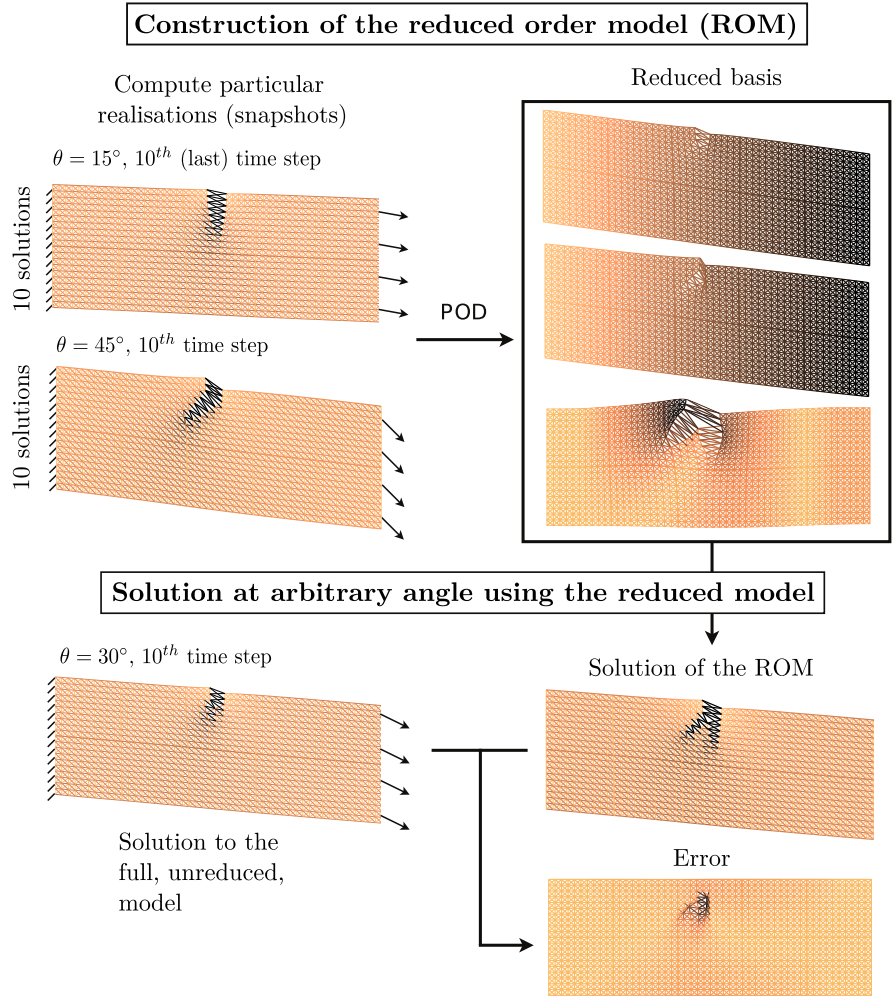}
\caption{Schematic representation of the Snapshot POD model order reduction technique for the proposed parametrised problem of fracture. The ``truth" time evolution of the problem is computed ``offline" for a certain number of values of the parameter. A reduced space is generated by performing a spectral analysis of this snapshot (POD). In the ``online" phase, the ``truth" problem is solved approximately by making use of a Galerkin projection of the governing equations in this reduced space, for any parameter value of interest. In the case of fracture mechanics, the projection error localises in the ``process zone'' surrounding the crack. Far away from it, a reduced space of small dimension associated to a relatively coarse exploration of the parameter domain is sufficient to capture the solution with a high level of accuracy. The darkest bars correspond to a completely damaged state of the material, while the lightest bars are undamaged.}
\label{PODnoDDM}
\end{figure}

\begin{figure}[htb]
\centering
\includegraphics[width=0.6 \textwidth]{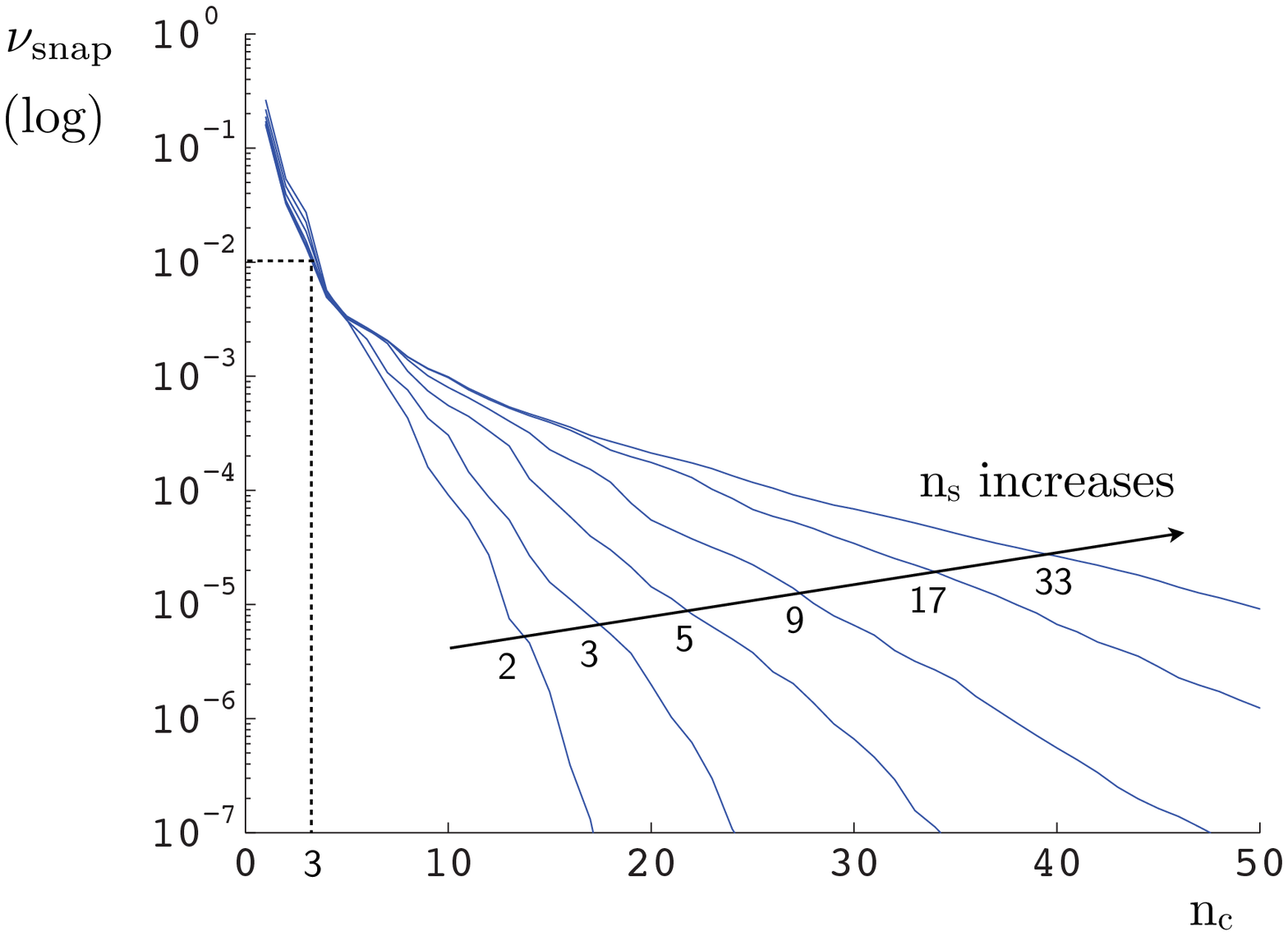}
\caption{Convergence of the normalised POD error indicator as a function of the order of truncation, for increasing size of the number of parameter values used to build the snapshot. The lack of correlation due to the crack propagation introduces a local error of projection, which appears here as a decrease in the convergence rate of the spectral decomposition below a certain value of the snapshot POD error indicator. This threshold is relatively low due to the global nature of the metric used to evaluate the accuracy of the projection. The numbers displayed on the graph are the number of load angles used to create the snapshot.}
\label{fig:PODConv}
\end{figure}

Results displayed in figure \ref{PODnoDDM} illustrate the behaviour of the reduced order modelling approach for $n_\mu = 2$. The normalised truncation error $\nu_\textrm{snap}$ of the snapshot POD as given in equation \eqref{errSnapshot} is arbitrarily set to $10^{-2}$ (see figure \ref{fig:PODConv}), which leads to the definition of a reduced space of dimension. 


It is noticed that each load angle $\theta$ leads to a crack/damage zone propagating approximately orthogonally to the load direction, as is commonly observed in fracture mechanics. Consequently, each and every load angle leads to a different damage pattern which cannot be well represented by a linear combination of the cracks obtained for a limited number of snapshot solutions (figure \ref{fig:PODConv}, bottom). In fact, the solution to parametric problems involving the evolution of topological changes cannot, in general, be obtained efficiently using a method based on the separation of variables (unless one manages to map the physical space to a reference space were correlation in the data can be retrieved \cite{gallandgravouil2010}). One systematic way to circumvent the problem would be to enrich the snapshot ``online" \cite{kerfridengosselet2010, ryckelynckbenziane2011}, but this leads to difficulties related to the cost of evaluating the projection error.

Despite these apparent difficulties, the topological changes are localised in space. In the regions that are far away from the crack, the solution is indeed well approximated by a linear combination of the pre-computed basis vectors. Consequently, a classical model reduction can still be performed but only over selected regions of the domain. The following section presents a possible strategy to implement this idea based on a domain decomposition method where the subdomains are selectively and independently reduced, based on a criterion described in section \ref{subsec:learning}. 

\vspace{1 \baselineskip} \noindent \underline{Remark:}
\textit{The initial crack is meant to provide a stress concentration zone from which fracture will initiates. We emphasise here that this is an idealisation of a general situation in realistic engineering components. Cracks initiates from joints, supports, free edges (large shear stresses due to a mismatch between elastic properties in composite laminates for instance), non-smooth parts of the boundary of the component (corner), or from interior regions which are subjected to extreme stress concentration under particular external loading conditions. Therefore, the regions of potential initiations are not arbitrary for a given parametric problem. In the particular example treated in this paper, fracture propagates from the notch which was introduced in the geometry. However, in all the following developments, we do not make use of the knowledge of the position of this initial defect, which emulates the existence of \textit{a priori} unknown zones of stress concentration in the structure. \\}



\section{Partitioned model order reduction approach}
\label{sec:DDM-MOR}

\subsection{Principle of the primal Schur-based domain decomposition method}
\label{sec:BDD}


Schur-based non-overlapping domain decomposition methods (see a review in \cite{gosseletrey2006}) are dedicated to the solution of large scale linear systems. In our case, we use the primal Schur-based domain decomposition (balancing domain decomposition (BDD) \cite{farhatroux1991,mandel1993,le-tallec1994a})) to calculate successive Newton iterates for the solution of the reference nonlinear time-dependant problem. Schur-based domain decomposition methods propose to condense the linearised balance equations on the interface degrees of freedom (degrees of freedom that are shared by at least two subdomains), by eliminating  the interior degrees of freedom. The resulting interface problem is solved by an iterative solver, usually by a preconditioned Krylov subspace algorithm, which is particularly well-suited to parallel computing. The condensation realises a first step of preconditioning, but the derivation of a preconditioner for the condensed interface problem is a key point to obtain an efficient and scalable domain decomposition method. 

\begin{figure}[htb]
\centering
\includegraphics[width=0.8 \textwidth]{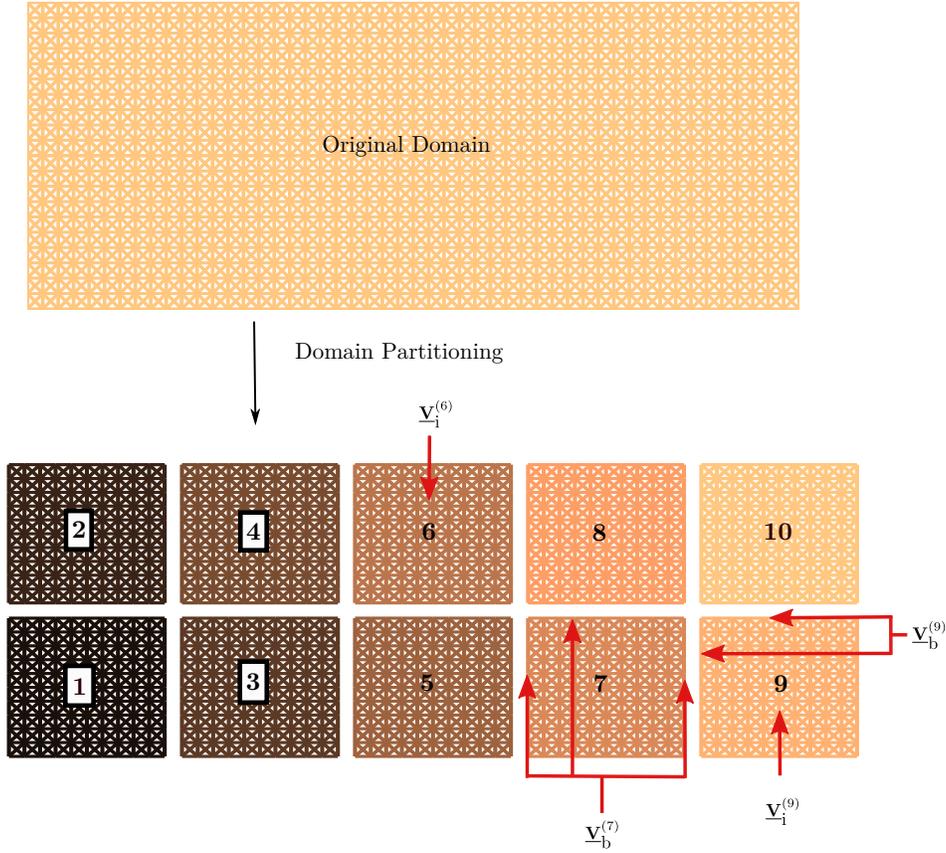}
\caption{Subdivision of the domain of interest into $10$ non-overlapping subdomains. $\vect{\boldsymbol \Delta  U}^{(e)}_\textrm{i}$ is the restriction of a vector $\vect{\boldsymbol \Delta  U}^{(e)}$ of nodal values of subdomain $e$ to the internal degrees of freedom of the subdomain, while $\vect{\boldsymbol \Delta  U}^{(e)}_\textrm{b}$ corresponds to the interface nodal values. The superscript between brackets indicates the number of the subdomain.}
\label{DDMfigure}
\end{figure}

Let us now give an overview of the domain decomposition method for the solution of the ``truth" problems corresponding to parameters $\mu \in \mathcal{P}^\text{s}$ (\textit{i.e.:} the snapshot). 
Domain $\Omega$ is split into non-overlapping subdomains $(\Omega^{(e)})_{e \in \llbracket 1, n_\textrm{e} \rrbracket}$ such that $\bigcup_{e \in \llbracket 1,  n_\textrm{e} \rrbracket} \Omega^{(e)} = \Omega$, as illustrated in Figure \ref{DDMfigure}. Each bar of the lattice structure belongs to one and only one subdomain. Nodes that are shared by two adjacent subdomains are interface nodes. We later refer to the set of subdomain indexes $\llbracket 1, n_\textrm{e} \rrbracket$ as $\mathcal{E}$. Let $\vect{U}^{(e)}(t;\vect{\boldsymbol \mu}) \in \mathbb{R}^{n_\text{u}^{(e)}}$ be the vector of nodal displacements of $\Omega^{(e)}$, which is looked for at an arbitrary point $(t,\vect{\boldsymbol \mu}) \in \mathcal{\tilde{P}}$ of the time-parameter domain. Each subdomain carries its own nodal unknowns for the interface nodes, which means that, for now, the corresponding kinematic is allowed to jump at the interface.

The local equilibrium of subdomain $\Omega^{(e)}$ is expressed in an algebraic form as follows:
\begin{equation}
\label{eq:localSubP}
\qquad \vect{F}_{\textbf{int}}^{(e)} \left( \vect{U}^{(e)}(t;\vect{\boldsymbol \mu}) , 
\left( \vect{U}^{(e)}(\tau;\vect{\boldsymbol \mu}) \right)_{\tau \in \mathcal{T}^h , \,  \tau < t} ; 
\vect{\boldsymbol \mu} \right) + \vect{F}^{(e)}_{\textbf{ext}} (t ; \vect{\boldsymbol \mu}) = {\mat{t}^{(e)}}^T \vect{ \boldsymbol \lambda}^{(e)}  \, , 
\end{equation}
with $\vect{ \boldsymbol \lambda}^{(e)} \in \mathbb{R}^{n_\text{b}^{(e)}}$ a vector of reaction forces from adjacent subdomains and $ \mat{t}^{(e)} \in \{0,1\}^{n_\text{b}^{(e)}} \times  \{0,1\}^{n_\text{u}^{(e)}} $ a trace operator which extracts the entries of vector of local nodal values corresponding to the interface nodes (\textit{i.e.} an output vector $\vect{X}_\mathbf{b}^{(e)} \in \mathbb{R}^{n_\text{b}^{(e)}}$ defined by $\vect{X}_\mathbf{b}^{(e)} = \mat{t}^{(e)} \vect{X}^{(e)}$, with $\vect{X}^{(e)} \in \mathbb{R}^{n_\text{u}^{(e)}}$ an arbitrary vector of local nodal values). The reaction forces must satisfy the following global interface equilibrium property:
\begin{equation}
\label{contactforceEq}
\sum_{e \in \mathcal{E}} \mat{A}^{(e)}  \vect{\boldsymbol \lambda}^{(e)} = \vect{0} \, 
\end{equation}
where $\{  \mat{A}^{(e)} \in \{0,1\}^{n_\text{b}}  \times \{0,1\}^{n_\text{b}^{(e)}}  \, | \, e \in \mathcal{E} \}$ is a set of assembly operators, with $n_\text{b}$ the number of interface equilibrium conditions, which is equal to the number of interface nodes. The set of subproblems is closed by the condition of equality of nodal displacements at an interface between two-subdomains (\textit{i.e.:} kinematic continuity in a continuous setting), which reads
\begin{equation}
\label{eq:ContiInterf}
\sum_{e \in \mathcal{E}} \mat{\bar{A}}^{(e)}  \mat{t}^{(e)} \vect{U}^{(e)} = \vect{0} \, 
\end{equation}
where $\{  \mat{\bar{A}}^{(e)} \in \{0,-1,1\}^{\bar{n}_\text{b}}  \times \{0,-1,1\}^{n_\text{b}^{(e)}} \, | \, e \in \mathcal{E} \}$, are signed boolean operators, with $\bar{n}_\text{b}$ the number of independent interface kinematic constraints. (see \cite{gosseletrey2006} for more details about the definition of properties of the assembly and trace operators).

In order to give expressions that are closer to the actual implementation of the method, we perform a linearisation of local problems \eqref{eq:localSubP} at iteration $i+1$ of the Newton algorithm. We look for iterates $\{  (\vect{U}^{(e),i+1},\vect{ \boldsymbol \lambda}^{(e),i+1}) \in \mathbb{R}^{n_\text{u}^{(e)}} \times \mathbb{R}^{n_\text{b}^{(e)}} \, | \, e \in \mathcal{E} \}$ of the local displacements and reaction forces satisfying both the local linearised systems
\begin{equation}
\label{eqOriginale0}
\quad \mat{K}^{(e),i} \vect{\boldsymbol \Delta  U}^{(e),i+1} = -\vect{R}^{(e)} + {\mat{t}^{(e)}}^T  \vect{ \boldsymbol \lambda}^{(e),i+1} \ \ , \forall \,  e \in \mathcal{E} \, ,
\end{equation}
and the global interface conditions \eqref{contactforceEq} and \eqref{eq:ContiInterf}. In the previous equation, the local tangent stiffness is $\mat{K}^{(e),i} \Def \left.  \frac{ \partial \vect{F}_{\textbf{int}}^{(e)}(\vect{U}^{(e)}) }{ \partial \vect{U}^{(e)}} \right|_{\vect{U}^{(e)} = \vect{U}^{(e),i}}$, the residual vector is $\vect{R}^{(e),i} \Def \vect{F}_{\textbf{int}}^{(e)} ( \vect{U}^{(e),i}) + \vect{F}^{(e)}_{\textbf{ext}}$, and the increment of displacement is defined by $\vect{\boldsymbol \Delta  U}^{(e),i+1} = \vect{U}^{(e),i+1}-\vect{U}^{(e),i}$. In the following, we will drop superscripts $i$ and $i+1$.

If we introduce the local operator $ \mat{E}^{(e)} \in \{ 0,1 \}^{n_\text{i}^{(e)}} \times\{ 0,1 \}^{n_\text{u}^{(e)}}$ ($n_\text{i}^{(e)} \Def n_\text{u}^{(e)} - n_\text{b}^{(e)}$ is the number of interior degrees of freedom of $e$) such that the output vector $\vect{X}_\mathbf{i}^{(e)} = \mat{E}^{(e)} \vect{X}^{(e)}$, with $\vect{X}^{(e)} \in \mathbb{R}^{n_\text{u}^{(e)}}$ arbitrary, is the restriction of $\vect{X}^{(e)}$ to the interior nodes of subdomain $\Omega^{(e)}$, for any $e \in \mathcal{E}$, we can recast the local systems \eqref{eqOriginale0} as follows:
\begin{equation}
\label{eqOriginale}
 \begin{bmatrix}
\mat{K}_\textbf{ii}^{(e)} & \mat{K}_\textbf{ib}^{(e)} \\
\mat{K}_\textbf{bi}^{(e)} & \mat{K}_\textbf{bb}^{(e)} 
\end{bmatrix}\begin{bmatrix}
\vect{\boldsymbol \Delta  U}_\textbf{i}^{(e)} \\
\vect{\boldsymbol \Delta  U}_\textbf{b}^{(e)}
\end{bmatrix} 
= 
\begin{bmatrix}
-\vect{R}_\textbf{i}^{(e)} \\
-\vect{R}_\textbf{b}^{(e)} + \vect{\boldsymbol \lambda}^{(e)}
\end{bmatrix} \ \ , \forall \,  e \in \mathcal{E} \, ,
\end{equation}
where $\vect{\boldsymbol \Delta  U}_\textbf{i}^{(e)} \Def \mat{E}^{(e)} \vect{\boldsymbol \Delta  U}^{(e)}$, $\vect{\boldsymbol \Delta  U}_\textbf{b}^{(e)} \Def \mat{t}^{(e)} \vect{\boldsymbol \Delta  U}^{(e)}$, $\vect{R}_\textbf{i}^{(e)} \Def \mat{E}^{(e)} \vect{R}^{(e)}$, 
$\vect{R}_\textbf{b}^{(e)} \Def \mat{t}^{(e)} \vect{R}^{(e)}$, 
$\mat{K}_\textbf{ii}^{(e)} \Def { \mat{E}^{(e)}} \mat{K}^{(e)} {\mat{E}^{(e)}}^T$, 
$\mat{K}_\textbf{ib}^{(e)} \Def { \mat{E}^{(e)}} \mat{K}^{(e)} {\mat{t}^{(e)}}^T$,  
$\mat{K}_\textbf{bi}^{(e)} \Def { \mat{t}^{(e)}} \mat{K}^{(e)} {\mat{E}^{(e)}}^T$ and 
$\mat{K}_\textbf{bb}^{(e)} \Def { \mat{t}^{(e)}} \mat{K}^{(e)} {\mat{t}^{(e)}}^T$.
The interior degrees of freedom $\vect{\boldsymbol \Delta  U}_\textbf{i}^{(e)}$ are eliminated from local systems \eqref{eqOriginale}  by static condensation, which is obtained by writing
\begin{equation}
\label{eq:TOTO}
\vect{\boldsymbol \Delta  U}_\textbf{i}^{(e)} = {\mat{K}_\textbf{ii}^{(e)}}^{-1}\left(-\vect{R}_\textbf{i}^{(e)} - \mat{K}_\textbf{ib}^{(e)}\vect{\boldsymbol \Delta  U}_\textbf{b}^{(e)}\right) \, ,
\end{equation}
where $\mat{K}_\textbf{ii}^{(e)}$ is assumed to be invertible. The condensed local problem is obtained by substitution of expression \eqref{eq:TOTO} in the second line of \eqref{eqOriginale}:
\begin{equation}
\label{schurEq}
 \mat{S}_\textbf{p}^{(e)} \vect{\boldsymbol \Delta  U}_\textbf{b}^{(e)} = \vect{F}_\textbf{c}^{(e)} + \vect{\boldsymbol \lambda}^{(e)} \, ,
\end{equation}
where the primal Schur complement $\mat{S}_\textbf{p}^{(e)}$ is defined by  $\mat{S}_\textbf{p}^{(e)}=\mat{K}_\textbf{bb}^{(e)}-\mat{K}_\textbf{bi}^{(e)}{\mat{K}_\textbf{ii}^{(e)}}^{-1}\mat{K}_\textbf{ib}^{(e)}$ , and the condensed forces $\vect{F}_\textbf{c}^{(e)}$ are defined by $\vect{F}_\textbf{c}^{(e)}=-\vect{R}_\textbf{b}^{(e)} - \mat{K}_\textbf{bi}^{(e)}\mat{K}_\textbf{ii}^{{(e)}^{-1}} (-\vect{R}_\textbf{i}^{(e)})$.


We now apply the primal domain decomposition methodology by enforcing the interface kinematic continuity \eqref{eq:ContiInterf} in a strong sense, which is done by writing that the local trace of the unknown displacement vectors $\{  \vect{U}^{(e)} \, | \, e \in \mathcal{E} \}$ are obtained by extraction from a global interface vector $\vect{U}_\textbf{b} \in \mathbb{R}^{n_\text{b}}$
\begin{equation}
\vect{\boldsymbol \Delta U}_\textbf{b}^{(e)}  \Def {\mat{t}^{(e)}} \vect{ \boldsymbol \Delta U}^{(e)}  = {\mat{A}^{(e)}}^T \vect{ \boldsymbol \Delta U}_\textbf{b} \ \ , \forall \,  e \in \mathcal{E} \, ,
\end{equation}
which implies the fulfilment of  \eqref{eq:ContiInterf} provided that the previous Newton iterate of the underlying displacement field is continuous.


A global assembled interface problem is obtained when left multiplying each of the local condensed systems (equation \eqref{schurEq}) by  assembly operators $\{  \mat{A}^{(e)}  \, | \, e \in \mathcal{E} \}$ and summing up over all subdomains, which reads
\begin{align}
 &\sum_{e \in \mathcal{E}} \mat{A}^{(e)} \mat{S}_\textbf{p}^{(e)} \vect{\boldsymbol \Delta  U}_\textbf{b}^{(e)} = \sum_{e \in \mathcal{E}} \mat{A}^{(e)}\vect{F}_\textbf{c}^{(e)} + \underbrace{\sum_{e \in \mathcal{E}} \mat{A}^{(e)} \vect{\boldsymbol \lambda}^{(e)}}_{ = \ \vect{0} \quad \textrm{from} \quad\eqref{contactforceEq}}\\
\Longleftrightarrow & \sum_{e \in \mathcal{E}} \left( \mat{A}^{(e)} \mat{S}_\textbf{p}^{(e)} {\mat{A}^{(e)}}^T \right) \vect{\boldsymbol \Delta  U}_\textbf{b} = \sum_{e \in \mathcal{E}} \mat{A}^{(e)}\vect{F}_\textbf{c}^{(e)} \label{schurAssembleEq} \, .
\end{align}
In a compact form, we look for an interface vector $\vect{\boldsymbol \Delta  U}_\textbf{b} \in \mathbb{R}^{n_\text{b}}$ satisfying
\begin{equation}
\label{globSchurEq}
   \mat{S}_\textbf{p} \vect{\boldsymbol \Delta  U}_\textbf{b} = \vect{F}_\textbf{c} \quad  \text{with} \quad \begin{cases} 
\mat{S}_\textbf{p} = \displaystyle \sum\limits_{e \in \mathcal{E}} \mat{A}^{(e)} \mat{S}_\textbf{p}^{(e)} {\mat{A}^{(e)}}^T \\
 \vect{F}_\textbf{c} = \displaystyle \sum \limits_{e \in \mathcal{E}} \mat{A}^{(e)} \vect{F}_\textbf{c}^{(e)} \, . 
 \end{cases} 
\end{equation}

Interface problem \eqref{globSchurEq} can be solved iteratively in parallel  using a Krylov-subspace method such as the conjugate gradient in a symmetric case or GMRes \cite{saadschultz1986}(or BiCGStab \cite{vandervorst1992}) in a non-symmetric case. 
In this framework, the global Schur complement need not be assembled. Instead, whenever it is needed in a matrix/vector multiplication, the multiplication is performed locally on each subdomain using the local Schur complements. The outcome of these local multiplications is then assembled:

\begin{equation}
\label{globSchurMult}
\forall \, \vect{X}_\textbf{b} \, \in \, \mathbb{R}^{n_\textrm{b}}, \qquad 
   \mat{S}_\textbf{p} \vect{X}_\textbf{b} = \displaystyle \sum\limits_{e \in \mathcal{E}} \mat{A}^{(e)} \mat{S}_\textbf{p}^{(e)} \underbrace{{\mat{A}^{(e)}}^T \vect{X}_\textbf{b}}_{=\vect{X}_\textbf{b}^{(e)}} \, .
\end{equation}
The local inversions involved in the computation of the local Schur complements are performed directly (using a Cholesky factorisation for instance). Using this method it is possible to perform the matrix/vector multiplications (computationally the most demanding part of a Krylov-subspace method) in parallel. In a similar way, the dot products involved in the iterative algorithm can be performed in parallel.
\begin{equation}
\label{ScalMult}
\forall \, \vect{X}_\textbf{b} \, \in \, \mathbb{R}^{n_\textrm{b}}, \qquad 
  {\vect{X}_\textbf{b}}^T \, \vect{X}_\textbf{b} = \displaystyle \sum\limits_{e \in \mathcal{E}}  {\vect{X}_\textbf{b}^{(e)}}^T \mat{D}^{(e)}{\vect{X}_\textbf{b}^{(e)}} \, ,
\end{equation}
where $\{ \mat{D}^{(e)} \, | \, e \in \mathcal{e} \}$ are diagonal matrices whose natural entries depend on the geometric multiplicity of the interface nodes.


\subsection{Formulation of reduced order modelling in the domain decomposition framework}
\label{subsec:PPOD}

\subsubsection{Local snapshot POD reduced spaces}

We propose to use POD-based model order reduction on the interior degrees of freedom of each subdomain. We assume that a snapshot $\{ \vect{U}(t;\vect{\boldsymbol \mu}) \, | \, (t,\vect{\boldsymbol \mu}) \in \mathcal{\tilde{P}}^\textrm{s} \}$ is available. This snapshot has been computed by making use the domain decomposition preconditioner described previously. Local POD spatial bases $\left( \vect{C}_{\textbf{i},i}^{(e)} \right)_{i \in \llbracket 1,n_\textrm{c}^{(e)}\rrbracket}$ of dimensions $n_\textrm{c}^{(e)}$ are now computed for the interior degrees of freedom of each subdomain $e \in \llbracket 1,n_\textrm{e} \rrbracket$ as described in section \ref{sec:MORAndPOD}. Accordingly, the normalised truncation error of the local snapshot POD transforms are defined as follows:
\begin{equation}
\label{errSnapshotLocal}
\left( {\nu_{\textrm{snap}}^{(e)}} \right)^2 = \frac{ \displaystyle 
\sum_{\vect{\boldsymbol \mu} \in \mathcal{P}^\textrm{s}}  \sum_{t \in \mathcal{T}^\textrm{h}}  \left\| \vect{U}^{(e)}_\textbf{i}(t;\vect{\boldsymbol \mu}) - \sum_{j=1}^{n^{(e)}_\textrm{c}} \left( {\vect{C}_{\textbf{i},j}^{(e)}}^T \vect{U}^{(e)}_\textbf{i}(t;\vect{\boldsymbol \mu}) \right) \vect{C}_{\textbf{i},j}^{(e)} \right\|_2^2
}{ \displaystyle
 \sum_{t \in \mathcal{T}^\textrm{h}} \sum_{\vect{\boldsymbol \mu} \in \mathcal{P}^\textrm{s}} \| \vect{U}^{(e)}_\textbf{i}(t;\vect{\boldsymbol \mu})  \|_2^2
}  \, , \ \ \forall e \in \mathcal{E} \, ,
\end{equation}
where $\vect{U}^{(e)}_\textbf{i} \Def \mat{E}^{(e)} \vect{U}^{(e)}$ for any $e \in \mathcal{E}$. Let us define the local operators $\{ \mat{C}_{\textbf{i}}^{(e)} \, | \, e \in \mathcal{E} \}$ whose columns are the local POD basis vectors of subdomain $e$.


\subsubsection{Local projection}

In the ``online" stage, we look for the interior degrees of freedom corresponding to an arbitrary point of the time-parameter domain $\mathcal{\tilde{P}}$ in the local reduced spaces. The reduction technique is here directly described for the linearised problem for the sake of concision, but one could equivalently start from the nonlinear partitioned problem \eqref{eq:localSubP}, introduce the a local reduced basis approximation and linearise the result. 

The kinematic interior approximation for the linearised problem reads:
\begin{equation} 
\begin{bmatrix}
\vect{\boldsymbol \Delta  U}_\textbf{i}^{(e)} \\
\vect{\boldsymbol \Delta  U}_\textbf{b}^{(e)} 
\end{bmatrix} = \begin{bmatrix}
\mat{C}_\textbf{i}^{(e)} \vect{\boldsymbol {\Delta \alpha}}_\textbf{i}^{(e)} \\
\vect{\boldsymbol \Delta  U}_\textbf{b}^{(e)} 
\end{bmatrix} \, , \ \ \forall e \in \mathcal{E} \, ,
\end{equation}
where $\vect{\boldsymbol {\Delta \alpha}}_\textbf{i}^{(e)}$ is a vector of local reduced state variables. Therefore, the local linearised system  of equation \eqref{eqOriginale} corresponding to an arbitrary subdomain $e \in \mathcal{E}$ now reads
\begin{equation}
\label{eq:tototo}
 \begin{bmatrix}
\mat{K}_\textbf{ii}^{(e)} & \mat{K}_\textbf{ib}^{(e)} \\
\mat{K}_\textbf{bi}^{(e)} & \mat{K}_\textbf{bb}^{(e)} 
\end{bmatrix}
\begin{bmatrix}
\mat{C}_\textbf{i}^{(e)} \vect{\boldsymbol {\Delta \alpha}}_\textbf{i}^{(e)} \\
\vect{\boldsymbol \Delta  U}_\textbf{b}^{(e)}
\end{bmatrix} 
=
\begin{bmatrix}
-\vect{R}_\textbf{i}^{(e)} \\
-\vect{R}_\textbf{b}^{(e)} + \vect{\boldsymbol \lambda}^{(e)}  
\end{bmatrix} \, .
\end{equation}
This is a set of $n_\textrm{i}^{(e)} + n_\textrm{b}^{(e)}$ equations in $n_\textrm{c}^{(e)} +  n_\textrm{b}^{(e)}$ unknowns. As we expect that $n_\textrm{i}^{(e)} + n_\textrm{b}^{(e)} \gg n_\textrm{c}^{(e)} +  n_\textrm{b}^{(e)}$, this system is overdetermined in general. Consistently with the developments proposed in section \ref{sec:Proj-based_MOR}, we perform a Galerkin projection: the residual of local system \eqref{eq:tototo} is required to be orthogonal to the local reduced space, which reads
\begin{equation}
\begin{bmatrix}
\mat{C}_\textbf{i}^{(e)} & \mat{0}
\\
\mat{0}                  & \mat{I}_{d,\mathbb{R}^{n_\text{i}^{(e)}}} 
 \end{bmatrix}^T 
\left( \begin{bmatrix}
-\vect{R}_\textbf{i}^{(e)} \\
-\vect{R}_\textbf{b}^{(e)} + \vect{\boldsymbol \lambda}^{(e)}  
\end{bmatrix}
- 
 \begin{bmatrix}
\mat{K}_\textbf{ii}^{(e)} & \mat{K}_\textbf{ib}^{(e)} \\
\mat{K}_\textbf{bi}^{(e)} & \mat{K}_\textbf{bb}^{(e)} 
\end{bmatrix}
\begin{bmatrix}
\mat{C}_\textbf{i}^{(e)} \vect{\boldsymbol {\Delta \alpha}}_\textbf{i}^{(e)} \\
\vect{\boldsymbol \Delta  U}_\textbf{b}^{(e)}
\end{bmatrix} 
\right) = \vect{0} \, .
\end{equation}
We end up with the following linear, square and symmetric system for the expression of the reduced local equilibria:
\begin{equation}
\label{RedEq}
\left( \vect{F}_\textbf{r}^{(e)} 
+ 
\begin{bmatrix}
\vect{0} \\
\vect{ \boldsymbol \lambda}^{(e)}  
\end{bmatrix} \right)
 -
\mat{K}_\textbf{r}^{(e)} 
\begin{bmatrix}
\vect{\boldsymbol {\Delta \alpha}}_\textbf{i}^{(e)} \\
\vect{\boldsymbol \Delta  U}_\textbf{b}^{(e)}
\end{bmatrix}  
= \vect{0}
\quad \mbox{where} \quad
\begin{cases} 
\mat{K}_\textbf{r}^{(e)} \Def 
\begin{bmatrix}
{\mat{C}_\textbf{i}^{(e)}}^T \mat{K}_\textbf{ii}^{(e)}\mat{C}_\textbf{i}^{(e)} & {\mat{C}_\textbf{i}^{(e)}}^T \mat{K}_\textbf{ib}^{(e)} \\
\mat{K}_\textbf{bi}^{(e)} \mat{C}_\textbf{i}^{(e)} & \mat{K}_\textbf{bb}^{(e)} 
\end{bmatrix}
\\
\vect{F}_\textbf{r}^{(e)} \Def \begin{bmatrix}
-{\mat{C}_\textbf{i}^{(e)}}^T \vect{R}_\textbf{i}^{(e)} \\
-\vect{R}_\textbf{b}^{(e)} 
\end{bmatrix} \, ,
\end{cases}
\end{equation}

\subsubsection{Condensed interface problem}
\label{sec:RedInterfProb}

Similarly as described in section \ref{sec:BDD}, local systems \eqref{RedEq} are condensed on the interface degrees of freedom, and are formally assembled. To do so, the reduced state variables $\vect{\boldsymbol {\Delta \alpha}}_\textbf{i}^{(e)}$ are eliminated using the identity 
\begin{equation}
\vect{\boldsymbol {\Delta \alpha}}_\textbf{i}^{(e)} = {\mat{K}_\textbf{ii,r}^{(e)}}^{-1}\left(-{\mat{C}_\textbf{i}^{(e)}}^T \vect{R}_\textbf{i}^{(e)} - \mat{K}_\textbf{ib,r}^{(e)} \, \vect{\boldsymbol \Delta  U}_\textbf{b}^{(e)}\right) \, ,
\end{equation}
where
 $\mat{K}_\textbf{ii,r} \Def {\mat{C}_\textbf{i}^{(e)}}^T \mat{K}_\textbf{ii}^{(e)}\mat{C}_\textbf{i}^{(e)}$ is assumed to be invertible and 
 $\mat{K}_\textbf{ib,r} \Def {\mat{C}_\textbf{i}^{(e)}}^T \mat{K}_\textbf{ib}^{(e)}$. By making use of interface kinematic and equilibrium conditions, which are not unchanged in our reduced order modelling approach, the assembled condensed reduced system reads:
\begin{equation}
\label{globSchurEqred}
\text{Find} \ \vect{\boldsymbol \Delta  U}_\textbf{b} \in \mathbb{R}^{n_\text{b}} \ \text{such that} \quad  
   \mat{S}_\textbf{p,r} \, \vect{\boldsymbol \Delta  U}_\textbf{b} = \vect{F}_\textbf{c,r} \ \ \text{ with } \ \ \begin{cases} 
\mat{S}_\textbf{p,r} = \displaystyle \sum\limits_{e \in \mathcal{E}} \mat{A}^{(e)} \mat{S}_\textbf{p,r}^{(e)} \, {\mat{A}^{(e)}}^T \\
 \vect{F}_\textbf{c,r} = \displaystyle \sum \limits_{e \in \mathcal{E}} \mat{A}^{(e)} \vect{F}_\textbf{c,r}^{(e)} \, , 
 \end{cases} 
\end{equation}
with the expression of the local condensed operators   $\mat{S}_\textbf{p,r}^{(e)} \Def\mat{K}_\textbf{bb}^{(e)}-\mat{K}_\textbf{bi,r}^{(e)} \, {\mat{K}_\textbf{ii,r}^{(e)}}^{-1}\mat{K}_\textbf{ib,r}^{(e)}$, the local condensed forces $\vect{F}_\textbf{c,r}^{(e)} \Def -\vect{R}_\textbf{b}^{(e)}-\mat{K}_\textbf{bi,r}^{(e)} \, \mat{K}_\textbf{ii,r}^{{(e)}^{-1}} (-{\mat{C}_\textbf{i}^{(e)}}^T \vect{R}_\textbf{i}^{(e)})$
 and $\mat{K}_\textbf{bi,r} \Def \mat{K}_\textbf{bi}^{(e)} \mat{C}_\textbf{i}^{(e)}$, for any $e \in \mathcal{E}$. Problem \eqref{globSchurEqred} can be solved in parallel (if the snapshot data is distributed in memory) using a Krylov algorithm, as described in section \ref{sec:BDD}.

We can now go one step further and choose not to reduce the local problems corresponding to some of the subdomains. Indeed, if localised non-linearities arise (damage in our case), the local reduction based on the separation of variables might be inefficient: a prohibitively large number of spatial basis vectors might be required to obtained the desired accuracy over the whole parameter domain (recall the results of section \ref{subsubsec:example}). This particular issue will be addressed in section \ref{subsec:learning}. So far, we will assume that the subdomains are divided into two complementary sets $\mathcal{E}^\textrm{red} \cup \mathcal{E}^\textrm{nred} = \mathcal{E}$, where $\mathcal{E}^\textrm{red}$ is a set of subdomains for which reduction is numerically efficient, while $\mathcal{E}^\textrm{nred}$ is the complementary set of subdomains, for which a direct solution to the corresponding local problem is preferred. The resulting hybrid condensed reduced problem consists in finding $ \vect{\boldsymbol \Delta  U}_\textbf{b} \in \mathbb{R}^{n_\text{b}}$ satisfying
\begin{equation}
\label{globSchurEqred2}
\mathit{
   \mat{S}_\textbf{p,hr} \, \vect{\boldsymbol \Delta  U}_\textbf{b} = \vect{F}_\textbf{c,hr} \mbox{ with } \begin{cases} 
\mat{S}_\textbf{p,hr} = \displaystyle \sum\limits_{e \in \mathcal{E}^\textrm{red}} \mat{A}^{(e)} \mat{S}_\textbf{p,r}^{(e)} \, {\mat{A}^{(e)}}^T + \sum\limits_{e \in \mathcal{E}^\textrm{nred}} \mat{A}^{(e)} \mat{S}_\textbf{p}^{(e)} \, {\mat{A}^{(e)}}^T \\
 \vect{F}_\textbf{c,hr} = \displaystyle \sum \limits_{e \in \mathcal{E}^\textrm{red}} \mat{A}^{(e)} \vect{F}_\textbf{c,r}^{(e)} + \sum\limits_{e \in \mathcal{E}^\textrm{nred}} \mat{A}^{(e)} \vect{F}_\textbf{c}^{(e)}
 \end{cases} \, .}
\end{equation}

\subsection{Local error estimation by Cross-Validation}
\label{subsec:learning}

\subsubsection{Principle}


The partitioned projection approach described in section \ref{subsec:PPOD} allows us to construct reduced spaces that are independent for each subdomain. 
We propose here a simple scheme in order to (i) determine independently  the dimension of the local reduced space that is necessary to achieve a predefined accuracy for the solution of each of the subproblems (ii) evaluate whether a subproblem is reducible or not in the sense of the usual separation of variables assumed by the POD. 

These two points are addressed while considering that a relevant snapshot is \textit{a priori} available. This relevant snapshot should explore the parameter domain sufficiently. At the same time, one does not want to compute too many snapshot solutions, in order for the ``offline/online'' strategy to remain affordable. Ultimately, a third point has to be added for the design of a  substructured learning strategy: (iii) assess whether the snapshot contains a sufficient quantity of information, and generate additional, well-chosen data if required. This last issue is extremely complicated to address. Some recent propositions have been made in \cite{kunischvolkwein2011,braconnierferrier2001,ryckelynckbenziane2011}, but most of the studies on the POD, or the Principal Component Analysis in the statistics community (a recent review is provided in \cite{abdiwilliams2010}) consider that a sufficiently rich snapshot is available, and perform the spectral analysis without considering the need, or the possibility, to regenerate data \textit{a posteriori}.

We will here address points (i) and (ii), while point (iii) will be left to the perspectives of this work. The particular technique used in this paper relies heavily on cross-validation (CV, see \cite{krzanowski1987} in the context of the PCA), and more precisely the Leave-One-Out (LOOCV) technique. In order to validate the predictivity of statistical models, one usually divide the available data into a training set and a validation set. In our application, the training set is the snapshot: the set of solutions to the parametric problem of evolution that corresponds to parameter values in $\mathcal{P}^{\text{s}}$. The relevancy of the reduced spaces generated by the snapshot-POD can then be evaluated on a set of additional fine-scale solutions: the training set. Using independent training and validation sets permits to avoid the overfitting behaviour (or ``Type-III error'' in statistics) that is classically observed in any regression-type model. In our context, the Snapshot POD  only minimises the mean square error of projection of the snapshot solutions in the reduced space \eqref{eq:SnapPODMetric}. Therefore, the associated error estimate \eqref{errSnapshot1} is expected to underestimate the error of projection associated to a hierarchically enriched snapshot, and in the limit, to underestimate the integral form \eqref{eq:DistancePOD} of the error of projection. Using a different set of solutions to identify the reduced space and to compute the error of projection permits to avoid this effect, but at the cost of additional data, which means further cost-intensive fine-scale solutions in our case.

The cross-validation error estimate avoids these additional computations by emulating the independence of training and validation sets using the same dataset. In order to do so, the summand in equation \eqref{errSnapshotLocal} is calculated using the local reduced basis obtained by a snapshot POD transform of all the available snapshot solutions but the one corresponding to the value of the summation variable. This is the usual LOOCV strategy applied to the POD. This can be written formally, for any subdomain $e \in \mathcal{E}$:

\begin{equation}
\label{errSnapshotLocal2}
\left( { {\tilde{\nu}}^{(e)}_{\textrm{snap}} } \right)^2= \frac{ \displaystyle 
\sum_{\vect{\boldsymbol \mu} \in {\mathcal{P}}^\text{s}  }  \sum_{t \in \mathcal{T}^\textrm{h}} \left\| \vect{U}_\textbf{i}(t;\vect{\boldsymbol \mu}) - \sum_{j=1}^{n^{(e)}_\textrm{c}} \left( \left. {\vect{\tilde{C}}_{\textbf{i},j}^{(e),(\vect{\boldsymbol \mu})}} \right.^T \vect{U}_\textbf{i}(t;\vect{\boldsymbol \mu}) \right)  {\vect{\tilde{C}}_{\textbf{i},j}^{(e),(\vect{\boldsymbol \mu})}} \right\|_2^2
}{ \displaystyle
 \sum_{t \in \mathcal{T}^\textrm{h}} \sum_{\vect{\boldsymbol \mu} \in \mathcal{P}^{\text{s} }} \| \vect{U}_\textbf{i}(t;\vect{\boldsymbol \mu})  \|_2^2
} \, ,
\end{equation}
 the modified reduced basis vectors $\left( {\vect{\tilde{C}}_{\textbf{i},j}^{(e),(\vect{\boldsymbol \mu})}} \right)_{j \in \llbracket 1,  n^{(e)}_\textrm{c} \rrbracket}$, which are parametrised by the summation variable $\vect{\boldsymbol \mu} \in {\mathcal{P}}^\text{s}$, are the $n^{(e)}_\textrm{c}$ first eigenvectors of the following modified POD operator:
\begin{equation}
\mat{\tilde{H}}^{(\vect{\boldsymbol \mu})} = \sum_{\vect{\boldsymbol \mu}^\star \in \left( \mathcal{P}^\textrm{s} \backslash {\vect{\boldsymbol \mu}} \right) } \sum_{t \in \mathcal{T}^\textrm{h}}  \vect{U}(t;\vect{\boldsymbol \mu}^\star) \, \vect{U}(t;\vect{\boldsymbol \mu}^\star)^T \, .
\end{equation}

Technically speaking, the computation of this estimate requires to perform an SVD for each of the snapshot solutions (and for each subdomain). 


Let us remark that statistical error estimates are commonly used in the context of deterministic parametric problem. For instance, classical Kriging interpolations are based on a randomisation of the field to interpolate. We refer to \cite{xiaobreitkopf2010,braconnierferrier2001} for recent combinations of Kriging and POD. The later contribution uses the LOOCV both as  an error estimate and as a criterion to refine the snapshot space in a hierarchical manner.


\subsubsection{Application}

\begin{figure}[htb]
 \centering
 \includegraphics[width= 0.6 \textwidth]{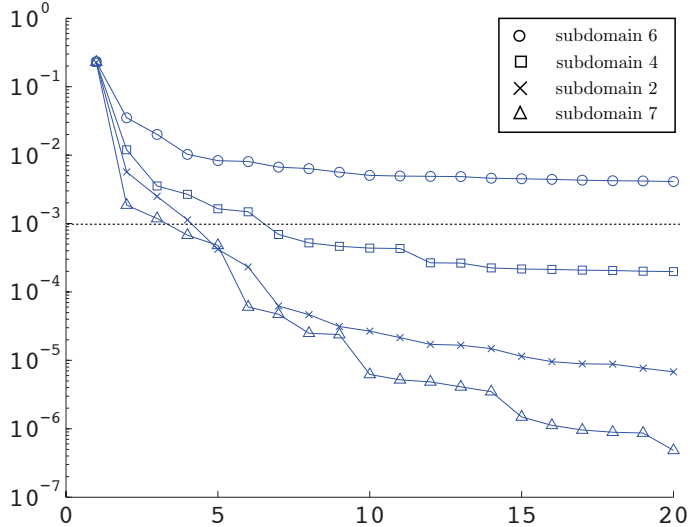}
 \caption{Cross-validation error estimate as a function of the order of the POD transforms for 4 of the 10 subdomains. The snapshot comprises 5 instances of the solution to the parametric problem of evolution. Subdomains are numbered as in Figure \ref{DDMfigure}.}
 \label{fig:5CV5}
 \end{figure}

The LOOCV error estimate is now applied to the problem of fracture. The parameter domain is sampled using a regular grid of 5 parameter values including the extremities of $\mathcal{P} = [15^\circ \ 45^\circ]$, which is, for now, assumed to be sufficiently fine for our purpose. In figure \ref{fig:5CV5}, the corresponding LOOCV estimate is plotted as a function of the dimension of the local reduced spaces for 4 different subdomains: subdomain 6, which is the most affected by the damage propagation, subdomain 4, which contains the ``tip of the crack'' for a range of parameter angles, and subdomains 2 and 7, which are further away from the source of nonlinearity (or lack of correlation, depending on the point of view). Again, we emphasise that we treat all subdomains in the same manner. We do not make use of an \textit{a priori} knowledge of the spatial distribution of damage. The lack of reducibility of certain parametric subproblems must be an output of the method.

The effect of the localised damage on the error estimates of each subdomain is relatively clear. For subdomains that are far away from the crack, we observe a fast convergence of the LOOCV error estimate with the dimension of the local POD reduced spaces. A satisfyingly level of predictivity, set here to the threshold $ { {\tilde{\nu}}^{(e)}_{\textrm{snap}} } \leq 10^{-3}$, is obtained with 4 to 5 reduced basis vectors. It is interesting to notice that we do not obtain a clear ``elbow'' in the convergence curve, which is often used to define the ``dimensionality'' of the underlying parametric problem. This is, to our best knowledge, due to the far effect of the crack. The lack of correlation due to the local damage tends to pollute the remote area. Further evidence of this fact can be found in our recent investigations about this particular effect  \cite{kerfridenschmidt2012}. For the subdomains that contain most of the damage, the observed convergence curves are much flatter. The required accuracy for subdomain 4 is obtained with 7 local POD basis vectors. In the case of subdomain 6, the LOOCV error estimate does not reach the predefined threshold. This indicates that the corresponding subproblem should not be reduced. 

We have now achieved our objective of choosing the dimension of the local reduced spaces based on a CV error estimate, and identifying non-reducible subproblems, based on an assumed sufficiently fine sampling of the parameter domain. The local reduced spaces obtained in this section will be the one used in the following to demonstrate the numerical efficiency of the partitioned model order reduction approach.

\section{System approximation  in the partitioned model order reduction approach}
\label{sec:SA-PPOD}




\subsection{Local ''gappy'' approximations}

We propose here to extend the concept of ``system approximation'' to the partitioned model order reduction introduced in section \ref{sec:DDM-MOR}. As mentioned previously, we choose to apply a tailored version of  the ``gappy'' reconstruction technique presented in different contexts in \cite{nguyenpatera2008,chaturantabutsorensen2010,carlbergbou-mosleh2011}. It is important to realise that the gappy technique approximates the Galerkin projection framework described in section \ref{sec:DDM-MOR}. Therefore, the system approximation will systematically be compared, or optimised, with respect to this framework and not with respect to the ``truth" modelling. This approach to system approximations is characterised as ``consistent'' in \cite{carlbergbou-mosleh2011}. 

The starting point of the gappy technique is to compute local ``static" reduced bases \\$\left\{ \mat{D}_\textbf{i}^{(e)} \in \mathbb{R}^{n_\text{u}^{(e)}} \times \mathbb{R}^{n_\text{d}^{(e)}}  \, | \, e \in \mathcal{E}^\text{red}  \right\}$ to approximate the vectors of internal forces $\left\{ \vect{F}^{(e)}_\bold{int,i} \, | \, e \in \mathcal{E}^\text{red}  \right\} \Def \left\{ \mat{E}^{(e)}\vect{F}^{(e)}_\bold{int} \, | \, e \in \mathcal{E}^\text{red}  \right\}$, as detailed previously in the non-partitioned case (see section \ref{sec:MORAndPOD}). Once the local bases are computed, the approximation reads
\begin{equation}
\label{eq:SA-DDM}
\begin{array}{l}
\displaystyle
\forall \, e \in \mathcal{E}^\text{red} \, , \forall \, t \in \mathcal{T}^\text{h} \, ,  \forall \, {\vect{\boldsymbol{\alpha}}^{(e)}}^\star \in \mathbb{R}^{n_\text{c}^{(e)}}  , \, \forall \, {\vect{U}_\textbf{b}^{(e)}}^\star \in \mathbb{R}^{n_\text{b}^{(e)}}  ,
\\ \displaystyle \quad
\vect{F}^{(e)}_\bold{int,i} \left( 
\begin{pmatrix}
\mat{C}_\textbf{i}^{(e)} {\vect{\boldsymbol{\alpha}}^{(e)}}^\star  \\
{\vect{U}_\textbf{b}^{(e)}}^\star 
\end{pmatrix}
, \left( \vect{U}^{(e)} (\tau ; \vect{\boldsymbol \mu}) \right)_{\tau \in \mathcal{T}^\text{h}, \tau < t } ; \vect{\boldsymbol \mu} \right) \approx 
\mat{D}_\textbf{i}^{(e)} \, \vect{\boldsymbol \beta}_\textbf{i}^{(e)} \left(
\begin{pmatrix}
{\vect{\boldsymbol{\alpha}}^{(e)}}^\star  \\
{\vect{U}_\textbf{b}^{(e)}}^\star 
\end{pmatrix}
, \left( \vect{U}^{(e)} (\tau ; \vect{\boldsymbol \mu}) \right)_{\tau \in \mathcal{T}^\text{h}, \tau < t }  , \vect{\boldsymbol \mu} \right) \, ,
\end{array}
\end{equation}
where $n_\text{b}^{(e)}$ is the number of interface degrees of freedom of subdomain $e$. We assume that the ``static" reduced bases are available. In the ``online" stage, the ``static" interpolation coefficients $\left\{ \vect{\boldsymbol \beta}_\textbf{i}^{(e)} \in \mathbb{R}^{n_\text{d}^{(e)}} \, | \, e \in \mathcal{E}^\text{red} \right\}$ are obtained at an arbitrary point along the reduced kinematic trajectory by minimisation of a distance between the previous approximation and the exact local vector of internal forces evaluated. This distance is measured at a set of sample spatial points, which yields the partitioned gappy approximation
\begin{equation}
\label{eq:Localgappy}
\begin{array}{l}
\displaystyle
 \forall \, e \in \mathcal{E}^\text{red} \, ,  \forall \, {\vect{\boldsymbol{\alpha}}^{(e)}}^\star \in \mathbb{R}^{n_\text{c}^{(e)}}  , \, \forall \, {\vect{U}_\textbf{b}^{(e)}}^\star \in \mathbb{R}^{n_\text{b}^{(e)}}  , 
\\ \displaystyle \quad 
\vect{F}^{(e)}_\bold{int,i}
\left( \begin{pmatrix}
\mat{C}_\textbf{i}^{(e)} {\vect{\boldsymbol{\alpha}}^{(e)}}^\star  \\
{\vect{U}_\textbf{b}^{(e)}}^\star 
\end{pmatrix} \right)
 \approx 
\mat{D}_\textbf{i}^{(e)} \left( {\mat{D}_\textbf{i}^{(e)}}^T  \mat{P}_\textbf{i}^{(e)} \,  {\mat{D}_\textbf{i}^{(e)}}  \right)^{-1} {\mat{D}_\textbf{i}^{(e)}}^T \mat{P}_\textbf{i}^{(e)} \, \vect{F}^{(e)}_\bold{int,i} \left(
\begin{pmatrix}
\mat{C}_\textbf{i}^{(e)} {\vect{\boldsymbol{\alpha}}^{(e)}}^\star  \\
{\vect{U}_\textbf{b}^{(e)}}^\star 
\end{pmatrix} \right)
\, ,
\end{array}
\end{equation}
The local boolean operator $\mat{P}_\textbf{i}^{(e)}$ operating on the subdomain $e \in \mathcal{E}^\text{red}$ is such that only the diagonal entries that correspond to all the degrees of freedom of a small set of internal nodes of subdomain $e$ are set to one. These nodes are called ``control points'' or ``control nodes''. We define the local ``gappy" operator of subdomain $e$ by $\mat{G}_\textbf{i}^{(e)} = \mat{D}_\textbf{i}^{(e)}  \left( {\mat{D}_\textbf{i}^{(e)}}^T \mat{P}_\textbf{i}^{(e)} \mat{D}_\textbf{i}^{(e)} \right)^{-1}{\mat{D}_\textbf{i}^{(e)}}^T \mat{P}_\textbf{i}^{(e)}$.

Let us explain how this approximation is employed to reduce the ``online" numerical complexity of the partitioned Galerkin-POD technique. Upon linearisation of the local nonlinear subproblems (i.e.: derivation of the vector of internal forces with respect to the reduced state variables and interface degrees of freedom), and taking into account the gappy approximation \eqref{eq:Localgappy}, one gets a modified expression of the local tangent systems (compare equation \eqref{eq:tototo}) at Newton iteration $i+1$ of an arbitrary time-parameter point of $\mathcal{\tilde{P}}$, for any subdomain $e \in \mathcal{E}$:
\begin{equation}
\label{eq:tototo2}
 \begin{bmatrix}
\mat{G}_\textbf{i}^{(e)} \, \mat{K}_\textbf{ii}^{(e)} & \mat{G}_\textbf{i}^{(e)} \, \mat{K}_\textbf{ib}^{(e)} \\
\mat{K}_\textbf{bi}^{(e)} & \mat{K}_\textbf{bb}^{(e)} 
\end{bmatrix}
\begin{bmatrix}
\mat{C}_\textbf{i}^{(e)} \vect{\boldsymbol {\Delta \alpha}}_\textbf{i}^{(e)} \\
\vect{\boldsymbol \Delta  U}_\textbf{b}^{(e)}
\end{bmatrix} 
=
\begin{bmatrix}
-\mat{G}_\textbf{i}^{(e)} \vect{F}_\textbf{int,i}^{(e)} \left( \vect{U}^{(e),i} \right) -\vect{F}_\textbf{ext,i}^{(e)} \\
-\vect{R}_\textbf{b}^{(e)}  + \vect{\boldsymbol \lambda}^{(e)}  
\end{bmatrix} \, ,
\end{equation}
with $\vect{F}_\textbf{ext,i}^{(e)} \Def \mat{E}^{(e)}\vect{F}_\textbf{ext}^{(e)}$

As mentioned in section \ref{sec:MORAndPOD}, this system is overdetermined but solutions can be obtained by making use of optimum arguments. We use a  Galerkin projection, which, together with the gappy approximation, yields the following matrix  formulation of the tangent subproblem corresponding to subdomain $e \in \mathcal{E}^\text{red}$:
\begin{equation}
\label{RedSAEq}
 \left(\vect{F}_\textbf{r,sa}^{(e)} +
\begin{bmatrix}
\vect{0} \\
\vect{ \boldsymbol \lambda}^{(e)}  
\end{bmatrix} \right) - \mat{K}_\textbf{r,sa}^{(e)} \begin{bmatrix}
\vect{\boldsymbol {\Delta \alpha}}^{(e)}_{\textbf{i}}\\
\vect{\boldsymbol \Delta  U}_\textbf{b}^{(e)}
\end{bmatrix} = \vect{0} 
\quad \text{with} 
\begin{cases}
\mat{K}_\textbf{r,sa}^{(e)} = 
\begin{bmatrix}
{\mat{C}_\textbf{i}^{(e)}}^T \mat{G}_{\textbf{i}}^{(e)}  \, \mat{K}_\textbf{ii}^{(e)} \, \mat{C}_\textbf{i}^{(e)}  & {\mat{C}_\textbf{i}^{(e)}}^T  \mat{G}_{\textbf{i}}^{(e)} \, \mat{K}_\textbf{ib}^{(e)} \\
\mat{K}_\textbf{bi}^{(e)}\mat{C}_\textbf{i}^{(e)} & \mat{K}_\textbf{bb}^{(e)} 
\end{bmatrix}
\\
\vect{F}_\textbf{r,sa}^{(e)} =
\begin{bmatrix}
-{\mat{C}_\textbf{i}^{(e)}}^T  \left( \mat{G}_\textbf{i}^{(e)} \vect{F}_\textbf{int,i}^{(e)} (\vect{U}^{(e),i} ) + \vect{F}_\textbf{ext,i}^{(e)} \right)
\\
-\vect{R}_\textbf{b}^{(e)} 
\end{bmatrix} \, .
\end{cases} 
\end{equation}
A condensed linearised interface problem is finally obtained as follows. We look for $ \vect{\boldsymbol \Delta  U}_\textbf{b} \in \mathbb{R}^{n_\text{b}}$ satisfying
\begin{equation}
\label{globSchurEqredSA}
   \mat{S}_\textbf{p,r,sa} \, \vect{\boldsymbol \Delta  U}_\textbf{b} = \vect{F}_\textbf{c,r,sa} 
   \quad \text{with} 
   \begin{cases}
\mat{S}_\textbf{p,r,sa} = \displaystyle \sum\limits_{e \in \mathcal{E}^\textrm{red}} \mat{A}^{(e)} \mat{S}_\textbf{p,r,sa}^{(e)} \, {\mat{A}^{(e)}}^T + \sum\limits_{e \in \mathcal{E}^\textrm{nred}} \mat{A}^{(e)} \mat{S}_\textbf{p}^{(e)} \, {\mat{A}^{(e)}}^T \\
\vect{F}_\textbf{c,r,sa}  = \displaystyle \sum\limits_{e \in \mathcal{E}^\textrm{red}} \mat{A}^{(e)} \vect{F}_\textbf{c,r,sa}^{(e)} + \sum\limits_{e \in \mathcal{E}^\textrm{nred}} \mat{A}^{(e)} \vect{F}_\textbf{c}^{(e)} \, .
   \end{cases} 
\end{equation}
The method to obtain the expression of the modified primal Schur complement $\mat{S}_\textbf{p,r,sa}$ and the corresponding condensed right-hand side is not detailed for the sake of concision. It follows exactly the method deployed  to get  their counterparts  whereby no system approximation was used (see equation \eqref{globSchurEqred}).

Notice that the symmetry of the condensed interface problem is lost when using the gappy technique. This issue can be alleviated by using a GMRes algorithm.

\begin{figure}[htb]
\centering
\includegraphics[width=0.8 \textwidth]{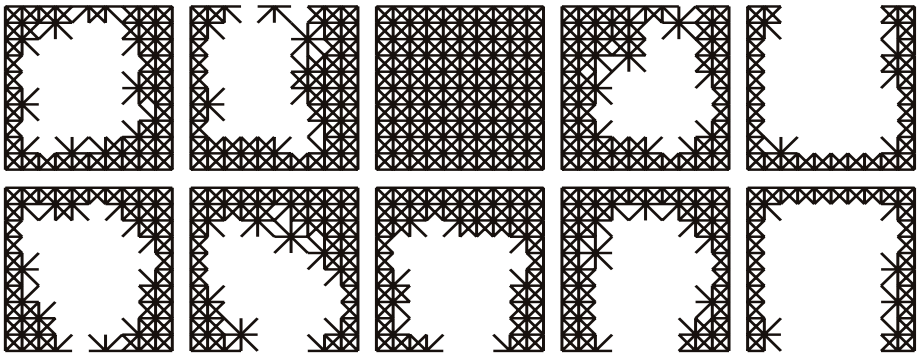}
\caption{Example of a reduced integration domain. Subdomain $6$ is not reduced. Therefore, all the associated elements belong to the integration domain. Since the interface between substructures is not reduced in the proposed primal version of the Schur-based partitioned model order reduction method, all the elements that are connected to the interface also belong to the reduced integration domain. The remaining controlled nodes are obtained by a Partitioned Discrete Empirical Interpolation Method.}
\label{redIntDom}
\end{figure}

The key benefit in using the gappy technique is that only the components of the local tangents and local residuals that are not filtered out by operators $\left\{ \mat{P}_\textbf{i}^{(e)} \, | \, e \in \mathcal{E}^\text{red}  \right\}$ need to be computed, the remainder being reconstructed by interpolation in the ``static" reduced spaces. In terms of implementation, the assembly of the tangents and residuals is performed via loops over all elements. With the system approximation, only contributions from elements that are connected to one of the ``control nodes''  are computed, which results in an online complexity that does not depend on the ``truth" number of unknowns.
The set of elements over which an integration of the internal forces is required is called the reduced integration domain. An example of such a domain is shown in Figure \ref{redIntDom}. The way this reduced integration domain was obtained is detailed in the following.

\subsection{Construction of the system approximation}

\subsubsection{Static POD bases}

To generate the local bases $\left\{ \mat{D}_\textbf{i}^{(e)} \, | \, e \in \mathcal{E}^\text{red}  \right\}$, we develop a technique that is strongly inspired by the one proposed in \cite{carlbergbou-mosleh2011}. Equation \eqref{eq:SA-DDM} indicates that we would like the system approximation to be optimal for any set of local reduced state variables. However, we can reasonably restrict ourselves to the state variables that are observed on a set of particular solutions to the Galerkin projection of the parametric problem in the kinematic reduced space. In order to do so, we first solve all time evolution problems corresponding to snapshot space $\mathcal{P}^\text{s}$ using the Galerkin framework described in section \ref{sec:DDM-MOR}, without system approximation. Such computations are expensive, but they are performed ``offline". The local solutions that are obtained in this fashion belong to the local POD reduced spaces and are considered as reference for the system approximation. We now want to approximate the spaces spanned by the local vectors of internal forces corresponding to the successive iterations of the Newton algorithm used to compute these reduced solutions. Let us call these spaces the ``static'' snapshot spaces. They can be represented mathematically, for any subdomain $e \in \mathcal{E}^\text{red}$, by the following set:
\begin{equation}
\label{eq:SetStatic}
\mathcal{F}^{s,(e)} = \left\{ 
\vect{F}^{(e)}_\bold{int,i}  \left( 
\begin{pmatrix}
\mat{C}_\textbf{i}^{(e)} {\vect{\boldsymbol{\alpha}}^{(e),i}}(t,\vect{\boldsymbol \mu})  \\
{\vect{U}_\textbf{b}^{(e),i}}(t; \vect{\boldsymbol \mu})  
\end{pmatrix}
, \left( \vect{U}^{(e)} (\tau ; \vect{\boldsymbol \mu}) \right)_{\tau \in \mathcal{T}^\text{h}, \tau < t } ; \vect{\boldsymbol \mu} \right)
 \, \left| \, \vect{\boldsymbol \mu} \in \mathcal{P}^\text{s} \, , t \in \mathcal{T}^\text{h}, \, i \in \llbracket 1, n_\text{new}^{(t),(\vect{\boldsymbol \mu})} \rrbracket \right\} \right. \, .
\end{equation}

In the previous expression, $n_\text{new}^{(t),(\vect{\boldsymbol \mu})}$ denotes the number of iterations of the Newton algorithm used to solve the problem of evolution at time $t \in \mathcal{T}^\text{h}$ and for parameter $\vect{\boldsymbol \mu} \in \mathcal{P}^\text{s}$. A singular value decomposition can now be used to compress and  hierarchically order the information contained in this set, which is similar to the technique used to obtain the reduced bases for the displacements and constitutes a keystone for the greedy selection of the reduced integration domain proposed in \cite{nguyenpatera2008,chaturantabutsorensen2010}. Technically, for each subdomain $e \in \mathcal{E}^\text{red}$, a matrix whose columns are the vectors of set \eqref{eq:SetStatic} is constructed. This matrix is decomposed by singular value decomposition. The left-singular vectors associated to singular values that are larger than a certain tolerance define the columns of operator $\mat{D}_\textbf{i}^{(e)}$.


\subsubsection{Selection of the control points}

For each subdomain $e \in \mathcal{E}^\text{red}$, given the ``static" reduced basis $\mat{D}_\textbf{i}^{(e)}$, we can now choose which subset of interior nodes will be defined as control nodes. This choice completely defines boolean operator $\mat{P}_\textbf{i}^{(e)}$ and, together with $\mat{D}_\textbf{i}^{(e)}$ obtained in the previous subsection, the required gappy reconstruction operator $\mat{G}_\textbf{i}^{(e)}$.

In the context of the DEIM \cite{chaturantabutsorensen2010}, the selection is performed in a greedy manner, for increasing rank of operator $\mat{D}_\textbf{i}^{(e)}$, where we recall that the columns of this operator are hierarchically ordered by SVD. More precisely, at  iteration $j>0$ of the greedy algorithm, the degree of freedom for which the gappy interpolation error 
\begin{equation}
\vect{\boldsymbol \epsilon}_\textbf{i,gap}^{(e),j}= \mat{D}_{\textbf{i},[1,j]}^{(e)} \, \vect{ \boldsymbol \beta }^{j} - \vect{D}_{\textbf{i},j+1}^{(e)} \, ,
\end{equation}
is maximum is defined as a ``control degree of freedom''. Operator $\mat{D}_{\textbf{i},[1,j]}^{(e)}$ is composed of the $j$ first columns of $\mat{D}_\textbf{i}^{(e)}$, while $\vect{D}_{\textbf{i},j+1}^{(e)}$ is the $j+1^\text{th}$ column of $\mat{D}_\textbf{i}^{(e)}$.
Interpolation coefficient $\vect{ \boldsymbol \beta }^j$ is obtained by solving the following optimisation problem:
\begin{equation}
\vect{ \boldsymbol \beta }^j = \underset{\vect{ \boldsymbol \beta }^\star \in \mathbb{R}^{j}}{\text{argmin}} \left( \left\|    \mat{D}_{\textbf{i},[1,j]}^{(e)} \, \vect{ \boldsymbol \beta }^\star - \vect{D}_{\textbf{i},j+1}^{(e)}
\right\|_{\mat{P}_\textbf{i}^{(e),j}} \right) \, , 
\end{equation}
The rank of the $j^\text{th}$ greedy iterate ${\mat{P}_\textbf{i}^{(e),j}}$ is $j$-times the number of scalar unknowns per interior node of subdomain $e$. In our implementation of the method, the node carrying the new ``control degree of freedom'' is added as a new ``control point'', and all its associated degrees of freedom are controlled, which means that the corresponding entries in ${\mat{P}_\textbf{i}^{(e),j+1}}$ are set to one. For an arbitrary subdomain $e$, the application of this method provides a number of ``control nodes'' equal to the rank of $\mat{D}_\textbf{i}^{(e)}$. We refer to reference \cite{chaturantabutsorensen2010} for more details about this technique, and in particular for a discussion about its optimality (in a greedy sense) and stability.

\subsubsection{Dimension of the local POD ``static spaces''}

One question that now arises is how to choose the order of truncation of the local SVD performed to approximate span($\mathcal{F}^{s,(e)}$), for any subdomain $e \in \mathcal{E}$. In other words, we need to choose the rank of the matrix of left singular vectors $\mat{D}_\textbf{i}^{(e)}$ for each subdomain $e \in \mathcal{E}^\text{red}$.
The simplest method is to truncate the local SVDs such that the truncation error becomes smaller than a predefined tolerance, or to use a cross-validation estimate, as proposed in section \ref{subsec:learning} when defining the dimension of the local reduced spaces for the displacements. 
However, we prefer here to link  the error generated by the gappy reconstruction technique to an error measured in terms of displacements, such that it can be compared to the error introduced by the truncation of the local snapshot POD performed to generate the local ``kinematic" reduced spaces.


In order to implement this idea, we proceed in an iterative manner. For a given truncation of the local ``static" SVDs, we evaluate the error introduced by the system approximation directly. This is done by solving the reduced problem when using the system approximation, and comparing the solution obtained in this fashion to the solution obtained when solving the reduced system of equations without system approximation. The error is of course only evaluated for parameter values belonging to the sampled parameter domain $\mathcal{P}^\text{s}$. If this error estimate is too large (in a sense to be defined later on), the dimensions of the ``static" reduced spaces is increased and the error estimation procedure is repeated.

More specifically, we initiate the iterative process with $n_\text{d}^{(e)} = n_\text{c}^{(e)}$ for all subdomains $e \in \mathcal{E}$. Local indicators for the total error introduced by the reduced order modelling technique are defined as follows:

\begin{equation}
\forall \, e \in \mathcal{E}, \quad \nu_\text{tot}^{(e)} = 
\sum_{\vect{\boldsymbol \mu} \in \mathcal{P}^\text{s}} \sum_{t \in \mathcal{T}^\text{h}} \left\| 
  \vect{U}_\textbf{ex}^{(e)}(t;\vect{\boldsymbol \mu})
-
  \vect{U}_\textbf{r,sa}^{(e)}(t;\vect{\boldsymbol \mu})
 \right\|_2 \, ,
\end{equation}
where $ \vect{U}_\textbf{ex}^{(e)}$ is the ``truth" solution to the parametric time-dependant problem, which has been computed to build the POD projection space for the displacement, and $ \vect{U}_\textbf{r,sa}^{(e)}$ denotes the solution obtained when using the reduced order model, with the current iterate of the system approximation, which needs to be computed. Performing simple algebraic manipulations, we can recast the expression of these estimates in the following manner:
\begin{equation}
\forall \, e \in \mathcal{E}, \quad \nu_\text{tot}^{(e)} = 
\sum_{\vect{\boldsymbol \mu} \in \mathcal{P}^\text{s}} \sum_{t \in \mathcal{T}^\text{h}} \left\| 
  \vect{U}_\textbf{ex}^{(e)}(t;\vect{\boldsymbol \mu})
 -   \vect{U}_\textbf{r}^{(e)}(t;\vect{\boldsymbol \mu})
 +   \vect{U}_\textbf{r}^{(e)}(t;\vect{\boldsymbol \mu})
-  \vect{U}_\textbf{r,sa}^{(e)}(t;\vect{\boldsymbol \mu})
 \right\|_2 \, ,
\end{equation}
with $\vect{U}_\textbf{r}^{(e)}$ the solution to the parametrised problem obtained when using the reduced order model without system approximation, which has been computed to generate the ``static" snapshot.
We can now use the triangle inequality, which yields the following relationship:
\begin{equation}
\forall \, e \in \mathcal{E}, \quad \nu_\text{tot}^{(e)} \leq 
\nu_\text{r}^{(e)} + \nu_\text{r,sa}^{(e)} 
\quad \text{with} \quad 
\left\{ \begin{array}{l} \displaystyle
\nu_\text{r}^{(e)} = \sum_{\vect{\boldsymbol \mu} \in \mathcal{P}^\text{s}} \sum_{t \in \mathcal{T}^\text{h}} \left\| 
  \vect{U}_\textbf{ex}^{(e)}(t;\vect{\boldsymbol \mu})
 -   \vect{U}_\textbf{r}^{(e)}(t;\vect{\boldsymbol \mu})
 \right\|_2
\\  \displaystyle
\nu_\text{r,sa}^{(e)} = \sum_{\vect{\boldsymbol \mu} \in \mathcal{P}^\text{s}} \sum_{t \in \mathcal{T}^\text{h}} \left\| 
  \vect{U}_\textbf{r}^{(e)}(t;\vect{\boldsymbol \mu})
-  \vect{U}_\textbf{r,sa}^{(e)}(t;\vect{\boldsymbol \mu})
 \right\|_2
\end{array} \right.
\end{equation}
Now, the term $\nu_\text{r,sa}^{(e)} $ measures the local error introduced by the system approximation, while $\nu_\text{r}^{(e)}$ measures the local error introduced by the kinematic approximation, which is monitored by the cross-validation estimate defined in section \ref{subsec:learning}, and can be decreased by enriching the ``kinematic" reduced space. The idea is then to compare these two estimates and to make sure that they are of the same order of magnitude, which can be formulated as follows:
\begin{equation}
\frac{ \nu_\text{r,sa}^{(e)} }{\nu_\text{r}^{(e)}} \leq 1
\end{equation}
If this condition is not satisfied with the current iterate of the system approximation, for any subdomain $e \in \mathcal{E}$, the rank $n_\text{d}^{(e)}$ of the corresponding ``static" operator $\mat{D}_\textbf{i}^{(e)}$ is increased (by one in our current implementation), and the error estimation procedure is repeated.

Notice that this simple strategy to control the accuracy of the gappy technique requires to compute a certain number of solutions to the evolution problem corresponding to parameters in $\mathcal{P}^\text{s}$. However, this is performed ``offline", and at reduced cost as we make use of the the gappy technique to compute the iterates of $\{ \nu_\text{r,sa}^{(e)} \, | \, e \in \mathcal{E}  \}$, while the set $\{ \nu_\text{r}^{(e)} \, | \, e \in \mathcal{E}  \}$ is computed once and for all and only requires information that is already available.

The reduced integration domain obtained by applying the methodology described in this section is represented in figure \ref{redIntDom} and will be the one used in the next section.

\section{Results}
\label{sec:results}

\subsection{Online numerical costs (``speed-up")}

We now solve the parametric, time-dependent lattice problem described in section \ref{sec:example} using the partitioned model order reduction approach, and report the speed-up in terms of run time. Speed-up is here to be understood as the ratio between the CPU time that is necessary to solve the ``truth" model, and the CPU time required to solve the reduced order model. The high numerical costs of the ``offline" phase are not considered in this definition. 

\begin{figure}[htb]
\centering
   \includegraphics[width= 0.8 \textwidth]{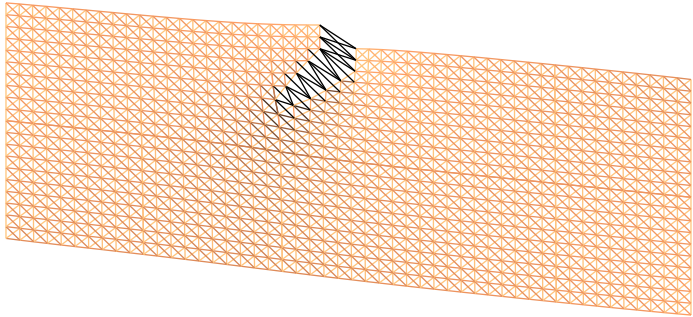}
\caption{Solution corresponding to the last time step of the fully discrete time-dependent problem for a load angle of $45^\circ$. The lattice structure represented here is composed of $121$ nodes per subdomain. The darkest bars correspond to a completely damaged state of the material, while the lightest bars are undamaged.}
\label{fig:sol10cells45deg}
\end{figure}

\begin{figure}[htb]
\centering
   \includegraphics[width= 0.8 \textwidth]{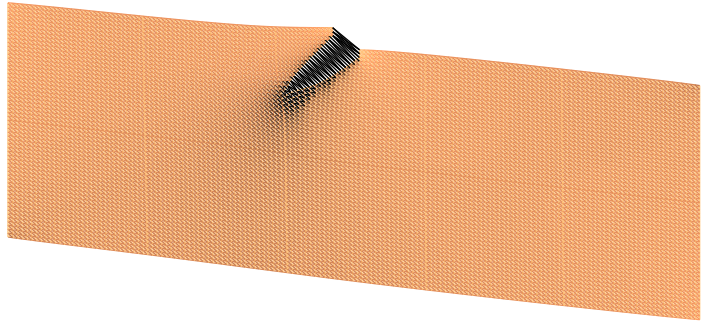}
\caption{Solution correspoding to the last time step of the fully discrete time-dependent problem for a load angle of $45^\circ$, using $961$ lattice nodes per subdomain.}
\label{fig:sol30cells45deg}
\end{figure}

We propose four different lattice structures, using $121$ (figure \ref{fig:sol10cells45deg}), $256$, $441$ and $961$ (figure \ref{fig:sol30cells45deg}) nodes for each of the $10$ subdomains. The snapshot that is used to compute the local reduced spaces is the one chosen in section \ref{subsec:learning}. Let us recall that the cross-validation procedure leads us to omit any reduction in subdomain 6, whose associated subproblem will be solved exactly. The remainder of the  subproblems are projected in the appropriate reduced spaces identified in section \ref{subsec:learning}, using the Petrov-Galerkin formulation (system approximation) developed in section \ref{sec:SA-PPOD}. We present speed-up results for the simulations corresponding to $\theta = 40^\circ$ and $\theta = 27^\circ$. These time solutions are not in the snapshot, and we can reasonably extrapolate that the observed speed-ups are representative of what can be expected for an arbitrary value of the parameter.


The proposed methodology is implemented in the commercial package Matlab, in a pseudo parallel fashion: the required operations that are local per subdomains are performed sequentially using a single processor. In this setting, we choose to solve the non-symmetric condensed interface problems using a direct LU factorisation. The reason for this is that no reduction of this problem has been developed so far. The number of interface degrees of freedom remains unchanged after the projection of the subproblems in the local reduced spaces. We therefore chose the implementation of the method that favored the observed speed-up, keeping in mind that it is pseudo-parallel. We will come back to this point in the conclusion of this work.

\begin{figure}[p]
\centering
\subfigure[Relative error for the different models using $121$ nodes per subdomain]{
   \includegraphics[width=0.48\textwidth]{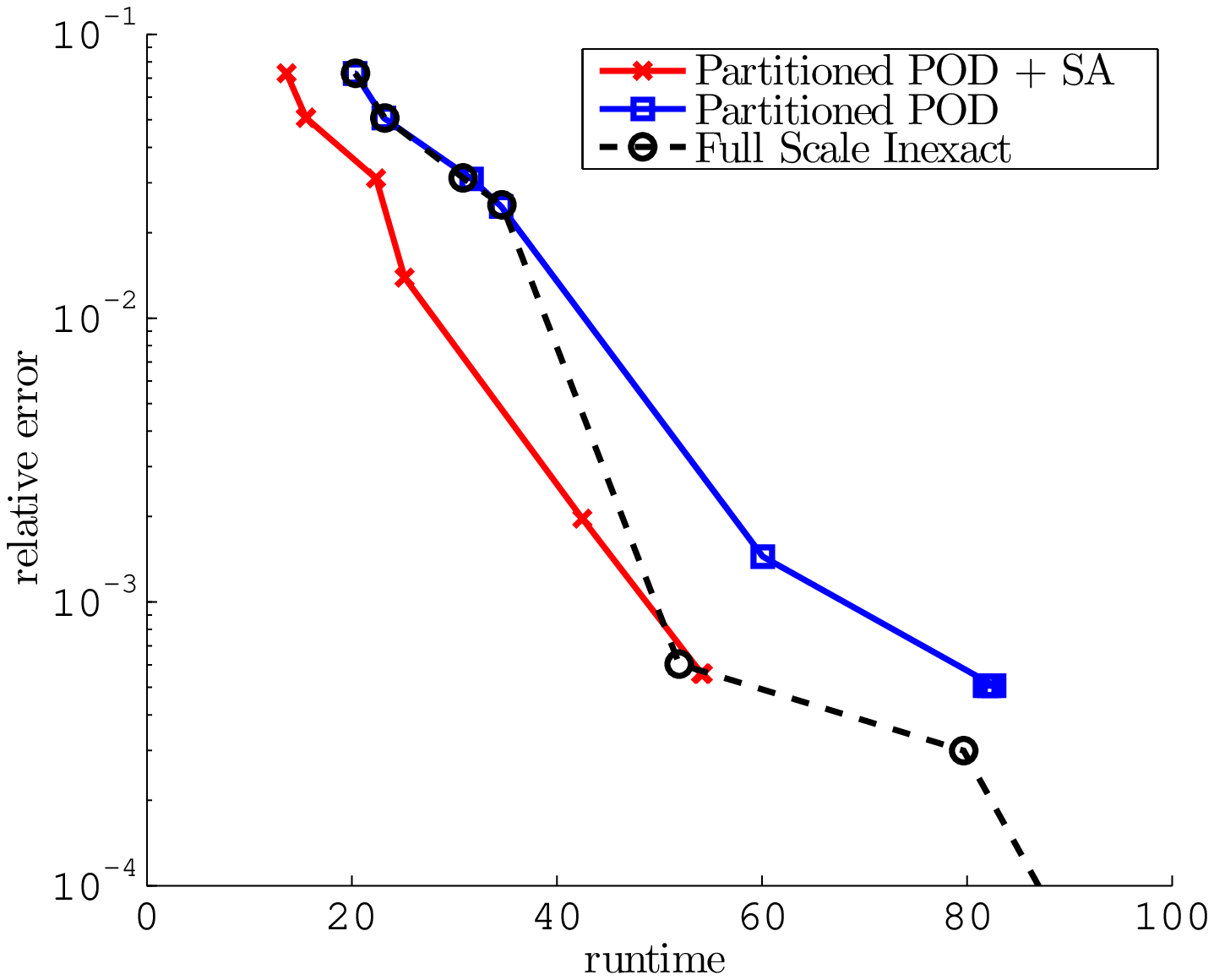}
 }
 \subfigure[Relative error for the different models using $256$ nodes per subdomain]{
   \includegraphics[width=0.48\textwidth]{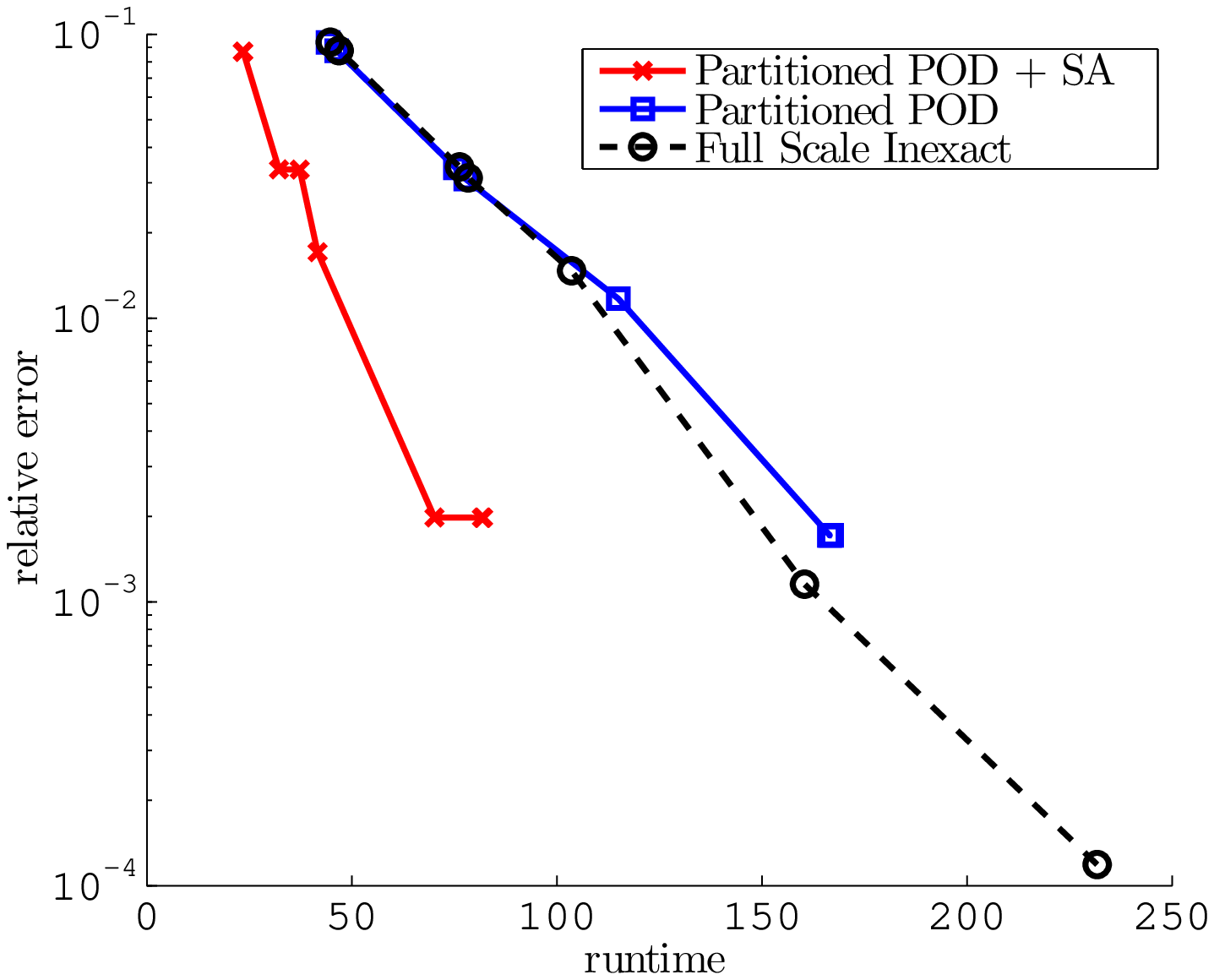}
 }
\subfigure[Relative error for the different models using $441$ nodes per subdomain]{
   \includegraphics[width=0.48\textwidth]{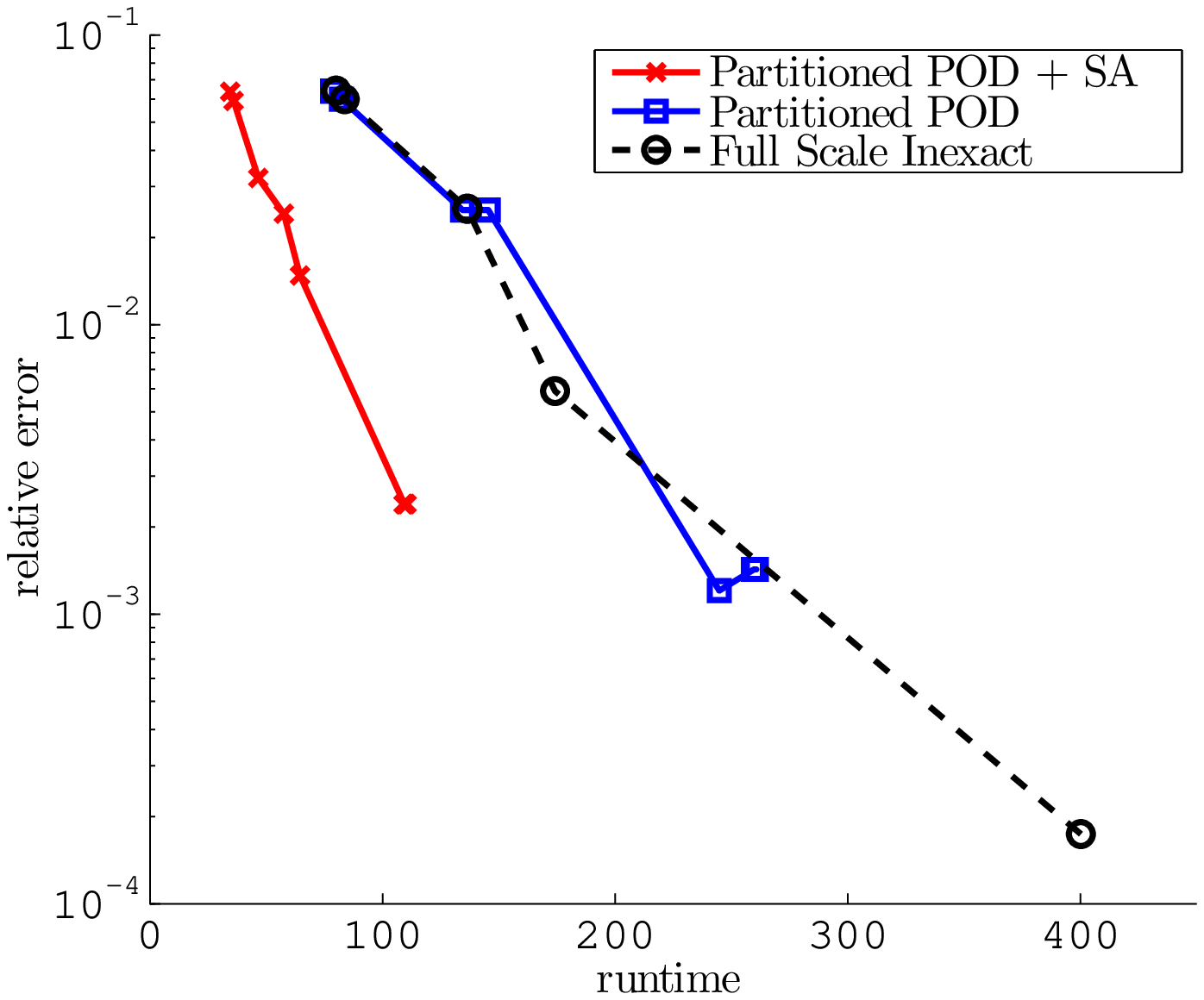}
 }
\subfigure[Relative error for the different models using $961$ nodes per subdomain]{
   \includegraphics[width=0.48\textwidth]{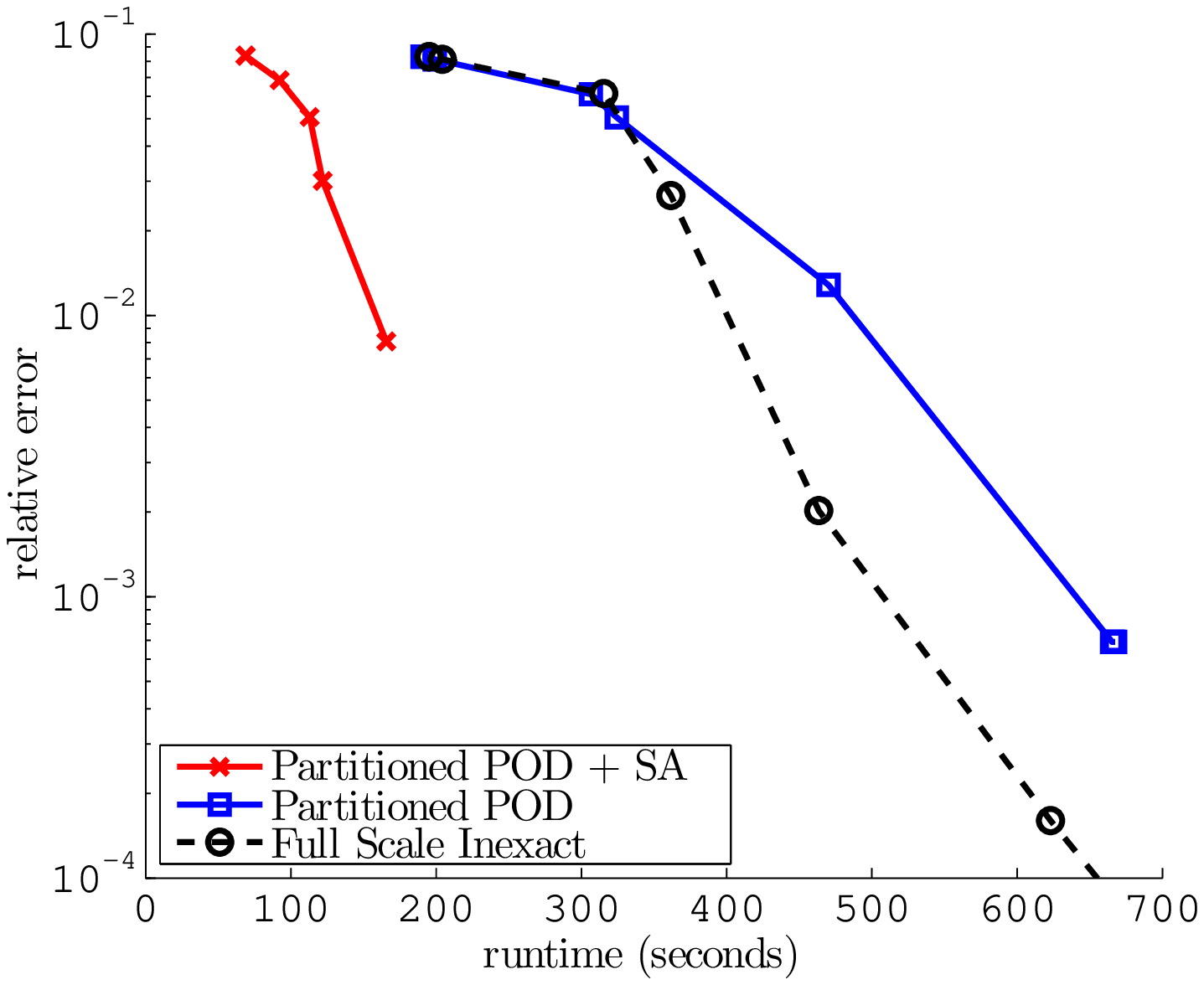}
 }
\caption{Relative error for the reference model and for the reduced order model as a function of  runtime for a load angle $\theta = 40^\circ$. The different points of the curves are generated by loosening the convergence of the Newton algorithms.
}
\label{errorGraph40}
\end{figure}

\begin{figure}[p]
\centering
\subfigure[Relative error for the different models using $121$ nodes per subdomain]{
   \includegraphics[width=0.48\textwidth]{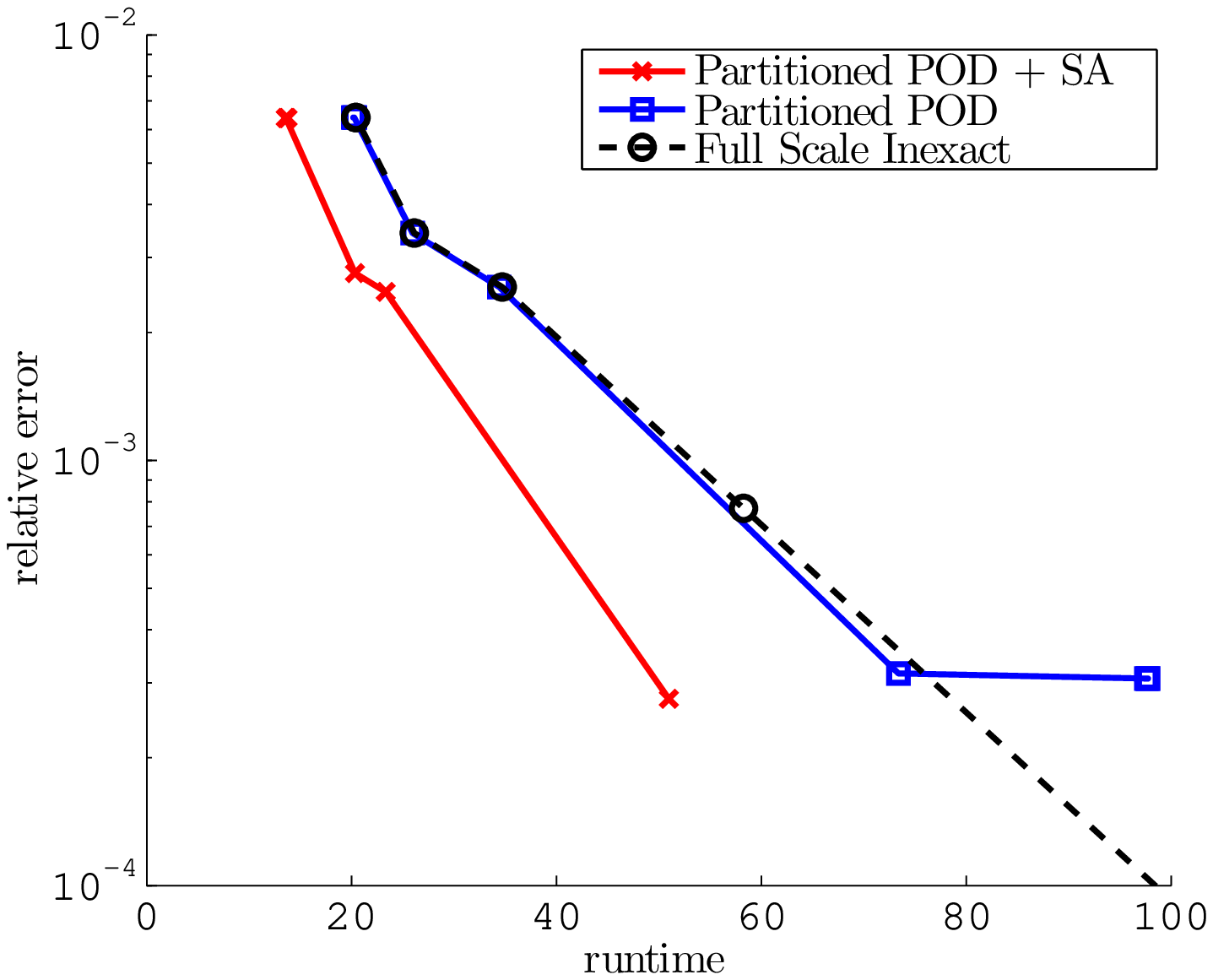}
 }
 \subfigure[Relative error for the different models using $256$ nodes per subdomain]{
   \includegraphics[width=0.48\textwidth]{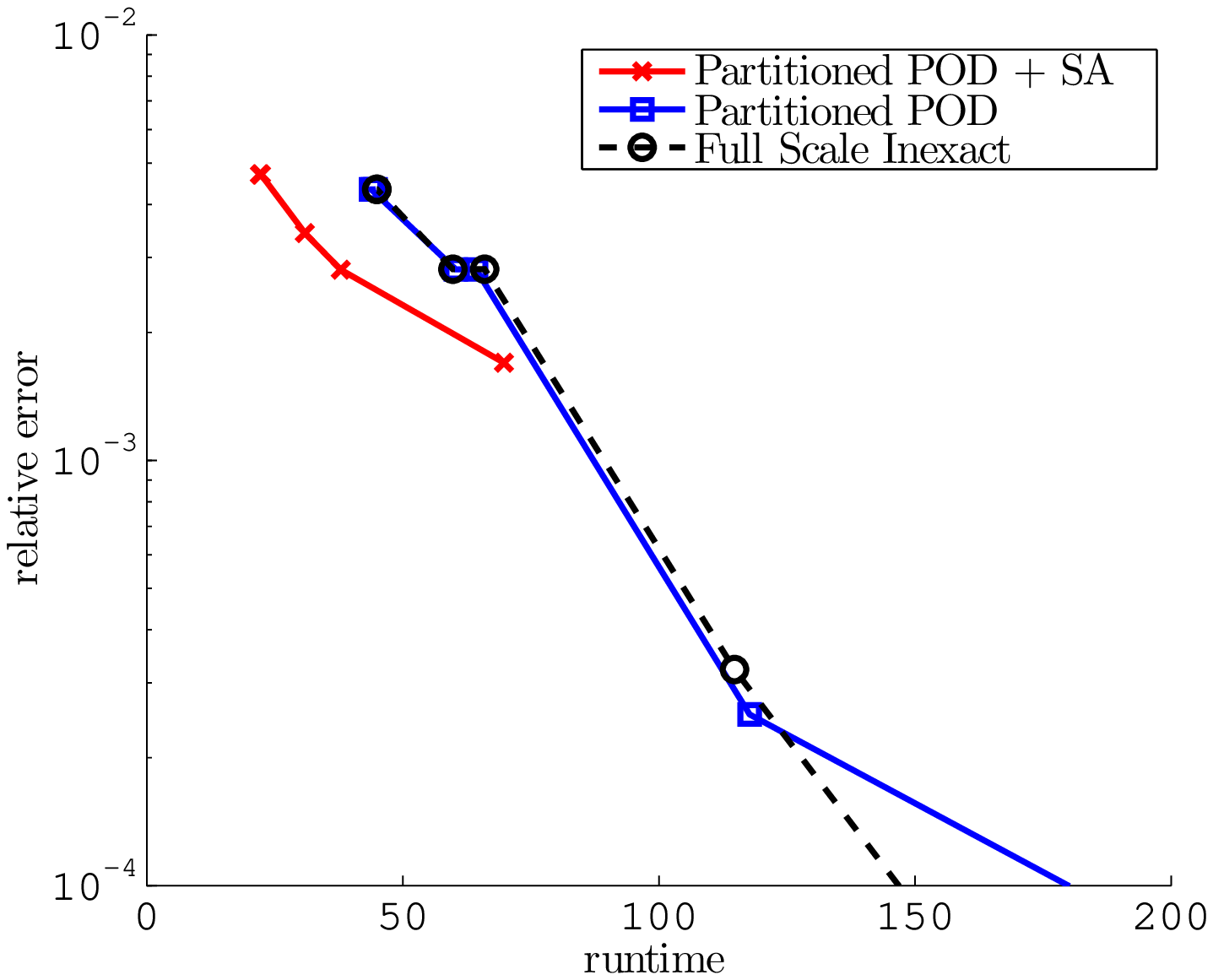}
 }
\subfigure[Relative error for the different models using $441$ nodes per subdomain]{
   \includegraphics[width=0.48\textwidth]{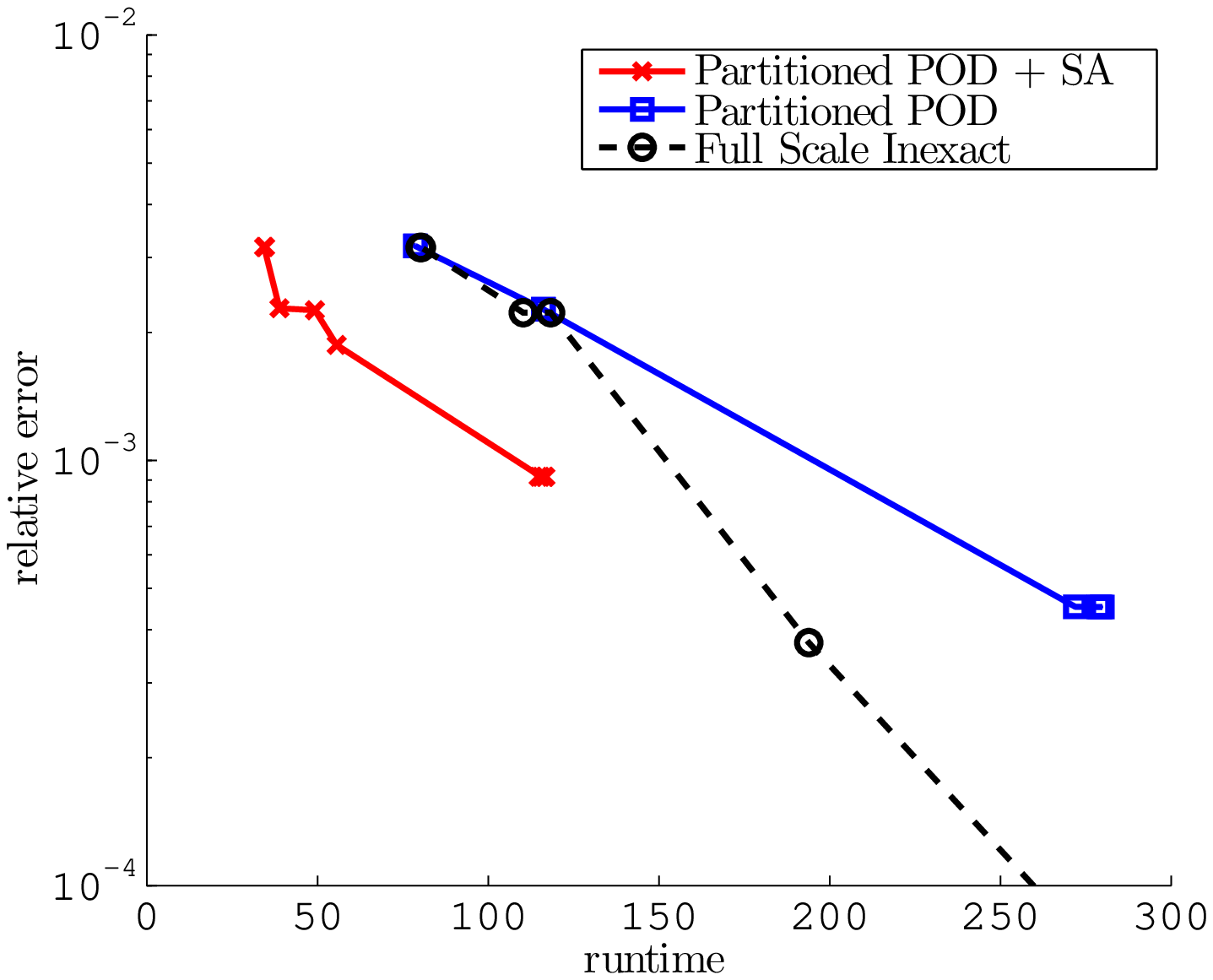}
 }
\subfigure[Relative error for the different models using $961$ nodes per subdomain]{
   \includegraphics[width=0.48\textwidth]{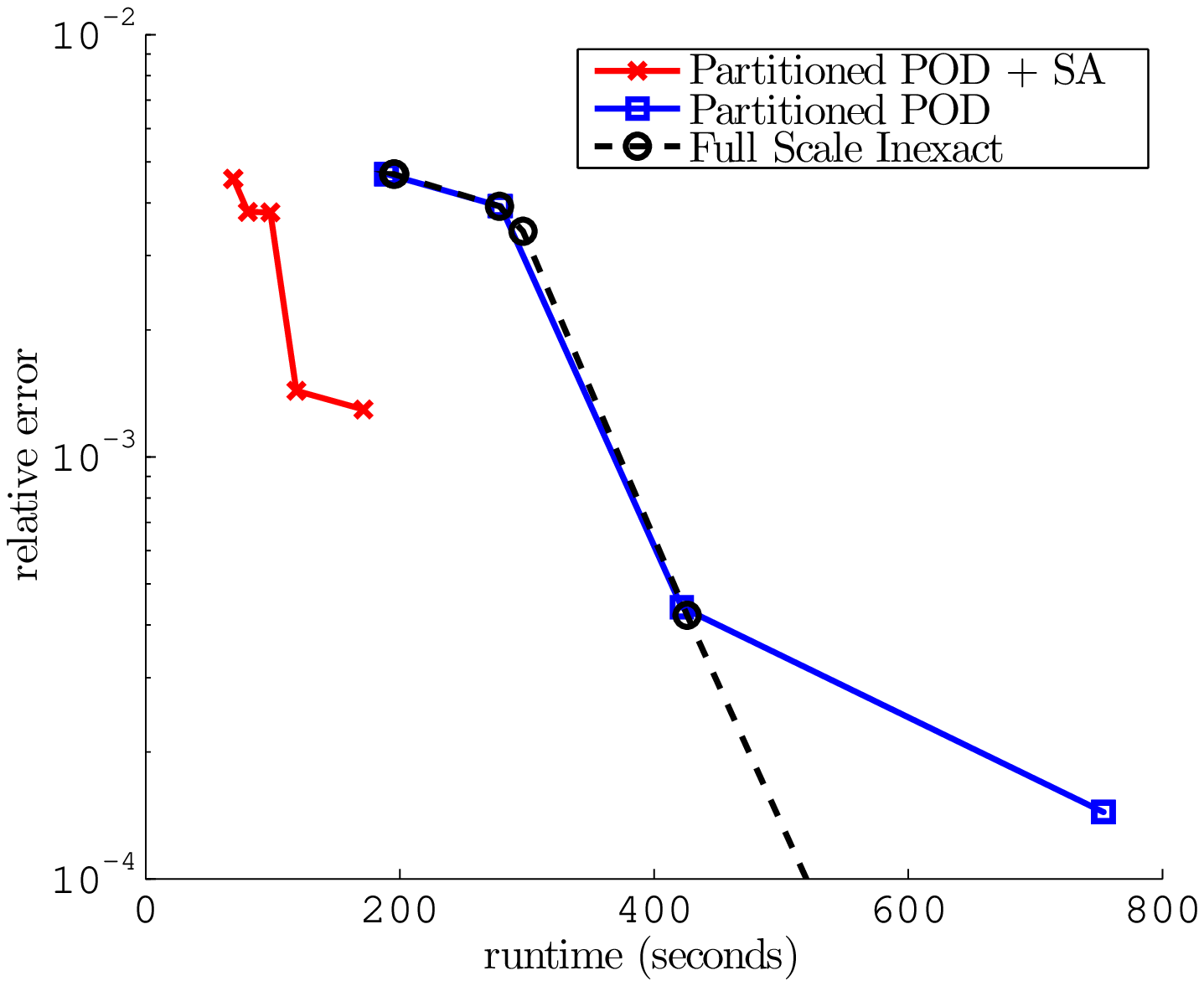}
 }
\caption{Relative error for the reference model and for the reduced order model as a function of  runtime for a load angle $\theta =27^\circ$}
\label{errorGraph27}
\end{figure}

In order to show the performance of the reduced order model, we first compute the ``truth" solution of the fully discrete problem that corresponds to the first of the two particular load angles mentioned previously. Note that this fine solution is computed using the partitioned model, but with no reduction. The convergence tolerance for the Newton algorithm used at each time step (euclidean norm of the residual divided by the norm of the vector of external forces) is set to $10^{-7}$. This is the reference solution $\vect{U}_\textbf{ex}$. The accuracy of any approximate solution $\vect{U}_\textbf{app}$ will be quantified using the following normalised error function:
\begin{equation}
\left. \nu_\text{app}^{(\vect{\boldsymbol \mu})}(\vect{U}_\textbf{app}) \right.^2 = 
\frac{ 
\displaystyle \sum_{t \in \mathcal{T}^h} \left\| \vect{U}_\textbf{app}(t;\vect{\boldsymbol \mu}) - \vect{U}_\textbf{ex}(t;\vect{\boldsymbol \mu}) \right\|^2_2
 }{
\displaystyle \sum_{t \in \mathcal{T}^h}  \left\|  \vect{U}_\textbf{ex}(t;\vect{\boldsymbol \mu})  \right\|^2_2
} \, .
\end{equation}
 
 Secondly, an approximate solution $\vect{U}_\textbf{inex}$ is obtained by a straightforward time-reduction technique: the Newton algorithms are solved to a loose tolerance, and the error $ \nu_\text{app}^{(\vect{\boldsymbol \mu})}(\vect{U}_\textbf{inex})$ is reported as a function of run time in figure \ref{errorGraph40}. This result is entitled ``Full Scale Inexact'' (notice that our use of the term ``inexact'' is not to be confused with the Inexact Newton Method, whereby one loosens the convergence tolerance of an iterative linear solver associated with the successive predictors of a Newton algorithm \cite{demboeisenstat1982}) .

Finally, we compare the speed-up obtained when using this straightforward approach to the one obtained with the projection-based partitioned reduction approaches. The error between the reference solution $\vect{U}_\textbf{ex}$ and the one obtained by the Galerkin projection-based partitioned model order reduction (without system approximation), denoted by $\vect{U}_\textbf{r}$, is the output $\nu_\text{app}^{(\vect{\boldsymbol \mu})}(\vect{U}_\textbf{r})$ of the previously defined error function. The corresponding result is labelled ``Partitioned POD'' in figure \ref{errorGraph40}. The error $\nu_\text{app}^{(\vect{\boldsymbol \mu})}(\vect{U}_\textbf{r,sa})$ of solution $\vect{U}_\textbf{r,sa}$ obtained with the partitioned reduction technique and the system approximation is reported next, under the label ``Partitioned POD + System Approximation''. All these curves are reproduced for the second test load angle in figure \ref{errorGraph27}.

The errors described previously are plotted for different levels of convergence of the Newton algorithms, in both the approximate full-scale case and the reduced cases, which provides a fair comparison ground for the various domain decomposition algorithms.



Observing the two figures of results, the following remarks can be made:
\begin{itemize}
 \item a significant speed-up is obtained when using the partitioned model order reduction approach together with the system approximation. This observation is only valid for certain range of accuracy. Indeed, the projection-based approach is limited, in terms of reachable accuracy, by the snapshot approximation of the POD, and by its subsequent truncation at a low order. For instance, in the top-right result of figure \ref{errorGraph40}, the error obtained with the reduction method cannot decrease under $2 \times 10^{-3}$. This is of course to be expected, and the remedy to this problem, if necessary, is to increase the size of the local reduced spaces.
 On the contrary, the error versus CPU time corresponding to the ``truth" problems can reach machine precision when decreasing the convergence tolerance of the Newton solvers.
 \item the Galerkin version of the partitioned POD approach produces insignificant speed-ups. This is a well-known fact. The number of degrees of freedom is reduced compared to the full-scale system, but the costly integrations of the reduced generalised forces over the spatial domain forbids any benefit in terms of computational gain over the reference model.
  \item the speed-up, observed in the region of reachable accuracy for the POD-based reduced order models, increases with the number of degrees of freedom of the reference problem. This can be easily explained. The cost of solving the reference problem increases with the number of fine-scale degrees of freedom. However, the dimensions of the local reduced spaces do not depend on this model refinement, but on the statistical properties of the parametric problem. Typically, one would expect that the numerical cost associated with the reduction technique does not increase with the number of degrees of freedom of the ``truth" models. In practice, this is not the case as some computational overhead penalises our implementation of the partitioned model order reduction approach, not the least of which is the fact that the condensed interface problem is not reduced. This overhead becomes more important when one increases the number of subdomains while keeping the same mesh size, since the number of degrees of freedom on the interface increases. This will be discussed in the conclusion of the paper.
\end{itemize}

Notice that in practice, the simulations using the reduced models with system approximation are only performed with the lowest tolerance threshold for the Newton algorithm. The intermediate run times have only been given for demonstration purposes.

\subsection{Remarks about the numerical efficiency of the system approximation}



\begin{figure}[H]
\centering
   \includegraphics[width=0.75\textwidth]{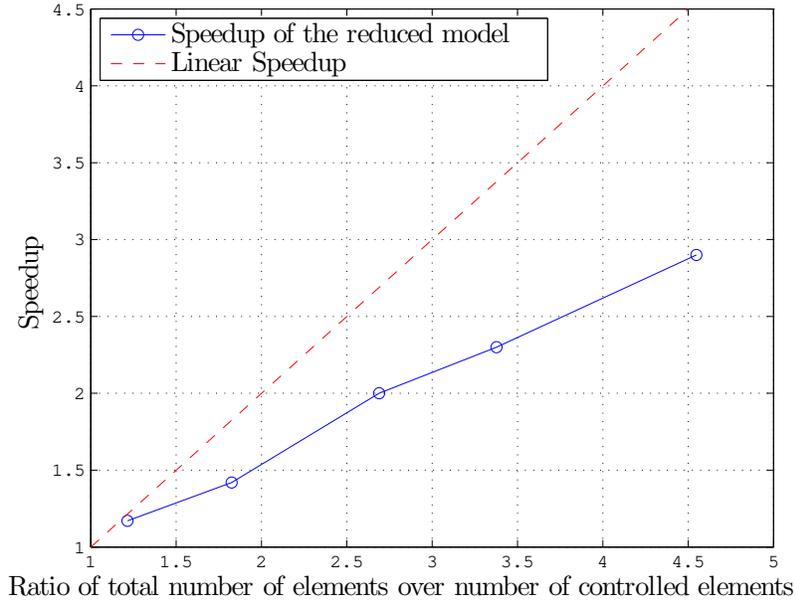}
\caption{Evolution of the speed-up with the ratio of the number of elements in the structure over the number of elements comprising the reduced integration domain.}
\label{fig:speedup}
\end{figure}

We now present the previous speed-up results in a different form. The aim is to show the trend in computational gain as a function of the number of degrees of freedom of the reference problem, when using the proposed reduction approach, in a unique graph. In order to so, the speed-up results reported previously are reported in figure \ref{fig:speedup} as a function of the ratio between the number of elements of the lattice and the number of elements that are connected to the control nodes of the system approximation. This ratio increases in a roughly linear manner with the number of degrees of freedom of the ``truth" problem. The different points of the curve are the one obtained with the lattice models comprising respectively $64$, $121$, $256$, $441$ and $961$ nodes per subdomain, with an appropriately low tolerance for the nonlinear solution algorithm.

The increase in the speed-up as function of the number of degrees of freedom of the full-scale problem appears clearly in this form. But more importantly, the graph shows that the observed speed-up is directly related to the size of the reduced integration domain. As mentioned previously, this is a clear indication that the main factor that prevents us from obtaining further speed-up with the proposed method is the fact that the interface problem is not reduced, which requires to perform integrations over a large number of elements. This is a path to explore in order to bring the idea of reduced order modelling in a partitioned framework to its full capability in the context of fracture.

%
%
%

\section{Conclusion and perspectives}

 In this paper, we have proposed a partitioned model order reduction strategy for parametrised problems of nonlinear fracture mechanics. The domain coupling has been performed using the tried and tested primal Schur-complement domain decomposition method. The local subproblems have been reduced by projection in low-dimensional subspaces obtained by the snapshot POD. We have shown that this approach permits to reduce,  in a flexible manner, the computational cost associated with highly nonlinear problems. In particular:
\begin{itemize}
\item the local reduced spaces are generated independently, and have independent dimensions, which allows us to focus the numerical effort where it is most needed. In fracture mechanics, subdomains that are close to highly damaged zones need a richer model to account for the effect of topological changes. The local POD transforms automatically generate local reduced spaces of relatively large dimensions in these zones.
\item the domain decomposition framework enables us to switch from reduced local solvers to ``truth" local solvers in a transparent manner. This is particularly useful for the subdomains that contain process zones, as a solution obtained by reduced order modelling would become more expensive than a direct solution for a desirable accuracy.
\item the transitition between ``offline'' and ``online'' computations becomes flexible. The reduced models can be used  in the zones where the local reduced spaces converge in a fast manner when enriching the snapshot space, while still computing snapshots and refining the reduced models via a direct local solver in the remaining subdomains.
\end{itemize}
We have shown that such a flexibility results in a significant speed-up in the case of parametric fracture mechanics problems. This speed-up naturally increases when the size of the highly damaged zone, in which the information is highly uncorrelated, is small compared to the scale of the structure.

This work is a step towards an optimal cost-reduction strategy for parametrised problems of fracture. Further work needs to be done to increase the understanding, robustness and performance of the method. Two main research avenues are particularly interesting from our point of view. Firstly, the interface problem itself was not reduced in our case, to guarantee the interface kinematic compatibility. This results in a suboptimal reduced order model and, in the case of parallel computing, would generate expensive communications through the network. A reduction of the interface problem using the POD can be performed, associated with a system approximation that is similar to the one we have used in this paper. Alternatively, using a dual Schur-complement domain decomposition method would allow the kinematic approximation of the subproblems to include the interface as well. One then needs to identify a relevant Lagrange multiplier space to ensure optimality and stability of  the Galerkin projection of the reference equations. This idea is our current direction of research.

An other difficulty is the load balancing mismatch that would occur when using such a strategy in parallel. CPUs which support domains that are not reduced, or domains for which the corresponding subproblem need to be projected in a space of relatively high dimension, would require to perform more operations. Hence, the domain partitioning itself should be performed jointly with the model reduction in order to distribute the load evenly.

Finally, we outlined throughout the paper some points that need further investigations but which are not directly related to the topic of reduced model partitioning addressed in this paper. The optimal choice of the snapshot samples used to construct \textit{a posteriori} reduced order models is currently a very active research area (see for instance the review \cite{quarteronirozza2011} concerning the reduced basis method, or the new developments proposed in \cite{kunischvolkwein2011} in the case of the snapshot POD). For arbitrary type of nonlinearity, a clear answer to this problem is, to date, not available. We have used a technique based on cross-validation, which, admittedly, requires a decently fine snapshot space in order to provide a relevant error estimate. In addition, our technique does not help find particular zones of non-smoothness in the parameter domain. It only provides a general trend for the projection error. Furthermore, an important point related to this issue is that the error criteria that have been used in this work are all based on global euclidean norms, without consideration for the physical phenomenon of interest. We believe that developing a ``goal-oriented'' domain-decomposition-based reduced order modelling would help alleviate a certain number of issues related to the certification of reduced order modelling for general nonlinearities.

\section*{Acknowledgements}

The authors thank the financial support of EPSRC High End Computing Studentship for Mr. Olivier Goury as well as the support of Cardiff and Glasgow University’s Schools of Engineering. Pierre Kerfriden and St\'ephane Bordas would like to also acknowledge the financial support of the Royal Academy of Engineering and of the Leverhulme Trust for S. Bordas' Senior Research Fellowship entitled ``Towards the next generation surgical simulator'' which funded Pierre Kerfriden's post in 2009-2010 and allowed him to lay the foundations for the work presented in this paper.

\bibliographystyle{unsrt}
\bibliography{MORbib}

\begin{thebibliography}{10}

\bibitem{antoulassorensen2001}
C.~Antoulas and D.C. Sorensen.
\newblock Approximation of large-scale dynamical systems: an overview.
\newblock {\em International Journal of Applied Mathematics and Computer
  Science}, 11(5):1093--1121, 2001.

\bibitem{sirovich1987}
L.~Sirovich.
\newblock Turbulence and the dynamics of coherent structures. part {I}:
  coherent structures.
\newblock {\em Quarterly of Applied Mathematics}, 45:561--571, 1987.

\bibitem{beattieborggaard2006}
C.A. Beattie, J.~Borggaard, S.~Gugercin, and T.~Iliescu.
\newblock A domain decomposition approach to pod.
\newblock {\em Proceedings of the 45th IEEE Conference on Decision and
  Control}, 2006.

\bibitem{ansallemfarhat2008}
D.~Amsallem and C.~Farhat.
\newblock {An Interpolation Method for Adapting Reduced-Order Models and
  Application to Aeroelasticity}.
\newblock {\em AIAA Journal}, 46(7):1803--1813, 2008.

\bibitem{nguyenpatera2008}
NC~Nguyen, AT~Patera, and J.~Peraire.
\newblock A `best points' interpolation method for efficient approximation of
  parametrized functions.
\newblock {\em International Journal for Numerical Methods in Engineering},
  73(4):521--543, 2008.

\bibitem{buffonitelib2009}
M.~Buffoni, H.~Telib, and A.~Iollo.
\newblock {Iterative methods for model reduction by domain decomposition}.
\newblock {\em Computers \& Fluids}, 38(6):1160--1167, June 2009.

\bibitem{craigbampton1968}
R.~Craig and M.~Bampton.
\newblock Coupling of substructures for dynamic analysis.
\newblock {\em American Institute of Aeronautics and Astronautics},
  6(7):1313--1319, 1968.

\bibitem{dickensnakagawa1997}
J~M Dickens, J~M Nakagawa, and M~J Wittbrodt.
\newblock {A critique of mode acceleration and modal truncation augmentation
  methods for modal response analysis}.
\newblock {\em Computers and Structures}, 62(6):985--998, 1997.

\bibitem{meyermatthies2003}
M.~Meyer and H.G. Matthies.
\newblock {Efficient model reduction in non-linear dynamics using the
  Karhunen-Loeve expansion and dual-weighted-residual methods}.
\newblock {\em Computational Mechanics}, 31(1):179--191, 2003.

\bibitem{barbonegivoli2003}
P.~E. Barbone, D.~Givoli, and I.~Patlashenko.
\newblock Optimal modal reduction of vibrating substructures.
\newblock {\em International Journal for Numerical Methods in Engineering},
  57:341--369, 2003.

\bibitem{rixen2004}
D.~Rixen.
\newblock A dual craig-bampton method for dynamic substructuring.
\newblock {\em Journal of Computational and Applied Mathematics}, 168:383--391,
  2004.

\bibitem{markovicpark2007}
D.~Markovic, K.C. Park, and A.~Ibrahimbegovic.
\newblock {Reduction of substructural interface degrees of freedom in
  flexibility-based component mode synthesis}.
\newblock {\em International journal for numerical methods in engineering},
  70(2):163--180, 2007.

\bibitem{hurty1960}
WC~Hurty.
\newblock {Vibrations of Structural Systems by Component Mode Synthesis}.
\newblock {\em Journal of the Engineering Mechanics Division}, 86(4):51--70,
  1960.

\bibitem{prudhommerovas2002}
C.~Prud'homme, D.~V. Rovas, K.~Veroy, L.~Machiels, Y.~Maday, A.~T. Patera, and
  G.~Turinici.
\newblock {Reliable Real-Time Solution of Parametrized Partial Differential
  Equations: Reduced-Basis Output Bound Methods}.
\newblock {\em Journal of Fluids Engineering}, 124(1):70--80, 2002.

\bibitem{barraultmaday2004}
M.~Barrault, Y.~Maday, N.C. Nguyen, and A.T. Patera.
\newblock An 'empirical interpolation' method: application to efficient
  reduced-basis discretization of partial differential equations.
\newblock {\em Comptes Rendus de Math\'ematiques}, 339(9):667--672, 2004.

\bibitem{constantinewang2012}
P.~G. Constantine and Q.~Wang.
\newblock Residual minimizing model interpolation for parameterized nonlinear
  dynamical systems.
\newblock {\em SIAM Journal on Scientific Computing}, 34:118--144, 2012.

\bibitem{pearson1901}
K.~Pearson.
\newblock On lines and planes of closest fit to systems of points in space.
\newblock {\em Philosophical Magazine}, 2(6):559--572, 1901.

\bibitem{hotelling1933}
H.~Hotelling.
\newblock Analysis of a complex of statistical variables into principal
  components.
\newblock {\em Journal of Educational Psychology}, 24:417--441, 1933.

\bibitem{ryckelynck2008}
D.~Ryckelynck.
\newblock Hyper-reduction of mechanical models involving internal variables.
\newblock {\em International Journal for Numerical Methods in Engineering},
  77(1):75 -- 89, 2008.

\bibitem{ryckelynckbenziane2010}
D~Ryckelynck and D~M Benziane.
\newblock Multi-level a priori hyper-reduction of mechanical models involving
  internal variables.
\newblock {\em Computer Methods in Applied Mechanics and Engineering},
  199(17-20):1134--1142, 2010.

\bibitem{ladevezepassieux2009}
P.~Ladev\`eze, J.C. Passieux, and D.~N\'eron.
\newblock The latin multiscale computational method and the proper generalized
  decomposition.
\newblock {\em Computer Methods in Applied Mechanics and Engineering},
  199(21):1287--1296, 2009.

\bibitem{chinestaammar2010}
F.~Chinesta, A.~Ammar, and E.~Cueto.
\newblock Recent advances and new challenges in the use of the proper
  generalized decomposition for solving multidimensional models.
\newblock {\em Archives of Computational Methods in Engineering - State of the
  Art Reviews}, 17(4):327--350, 2010.

\bibitem{nouy2010}
A.~Nouy.
\newblock {A priori model reduction through Proper Generalized Decomposition
  for solving time-dependent partial differential equations}.
\newblock {\em Computer Methods in Applied Mechanics and Engineering},
  199(23-24):1603--1626, 2010.

\bibitem{yvonnethe2007}
J~Yvonnet and Q.C. He.
\newblock {The reduced model multiscale method (R3M) for the non-linear
  homogenization of hyperelastic media at finite strains}.
\newblock {\em Journal of Computational Physics}, 223(1):341--368, 2007.

\bibitem{kerfridengosselet2010}
P.~Kerfriden, P.~Gosselet, S.~Adhikari, and S.~Bordas.
\newblock {Bridging proper orthogonal decomposition methods and augmented
  Newton-Krylov algorithms: an adaptive model order reduction for highly
  nonlinear mechanical problems}.
\newblock {\em Computer Methods in Applied Mechanics and Engineering},
  200(5-8):850--866, 2011.

\bibitem{gallandgravouil2010}
F.~Galland, A.~Gravouil, E.~Malvesin, and M.~Rochette.
\newblock {A global model reduction approach for 3D fatigue crack growth with
  confined plasticity}.
\newblock {\em Computer Methods in Applied Mechanics and Engineering},
  200(5-8):699--716, 2011.

\bibitem{parkpark2004}
KC~Park and YH~Park.
\newblock {Partitioned component mode synthesis via a flexibility approach}.
\newblock {\em AIAA journal}, 42(5), 2004.

\bibitem{rickeltreese2006}
Christian Rickelt and Stefanie Reese.
\newblock {A simulation strategy for life time calculations of large, partially
  damaged structures}.
\newblock In {\em III European Conference on Computational Mechanics}, 2006.

\bibitem{kerfridenpassieux2011}
P.~Kerfriden, J.C. Passieux, and S.~Bordas.
\newblock Local/global model order reduction strategy for the simulation of
  quasi-brittle fracture.
\newblock {\em International Journal for Numerical Methods in Engineering},
  89(2):154--179, 2011.

\bibitem{haryadikapania1998}
S.G. Haryadi, R.K. Kapania, and S.G. Haryadi.
\newblock {Global/local analysis of composite plates with cracks}.
\newblock {\em Composites Part B}, 29(B):271--276, 1998.

\bibitem{legresleyalonso2003}
P.A. LeGresley and J.J. Alonso.
\newblock Dynamic domain decomposition and error correction for reduced order
  models.
\newblock {\em AIAA 41st Aerospace Sciences Meeting}, 2003.

\bibitem{ammar2011}
A~Ammar, F~Chinesta, and E~Cueto.
\newblock {Coupling finite elements and proper generalized decompositions}.
\newblock {\em International Journal for Multiscale Computational Engineering},
  9(1):1--24, 2011.

\bibitem{kunishvolkwein2003}
K.~Kunisch and S.~Volkwein.
\newblock {Galerkin Proper Orthogonal Decomposition Methods for a General
  Equation in Fluid Dynamics}.
\newblock {\em SIAM Journal on Numerical analysis}, 40(2):492--515, 2003.

\bibitem{astridweiland2008}
P.~Astrid, S.~Weiland, K.~Willcox, and A.C.P.M. Backx.
\newblock Missing point estimation in models described by proper orthogonal
  decomposition.
\newblock {\em IEEE Transactions on Automatic Control}, 53(10):2237--2251,
  2008.

\bibitem{xiaobreitkopf2010}
Manyu Xiao, Piotr Breitkopf, Rajan Filomeno~Coelho, Catherine Knopf-Lenoir,
  Maryan Sidorkiewicz, and Pierre Villon.
\newblock {Model reduction by CPOD and Kriging}.
\newblock {\em Structural and Multidisciplinary Optimization}, 41:555--574,
  2010.

\bibitem{carlbergbou-mosleh2011}
K.~Carlberg, C.~Bou-Mosleh, and C.~Farhat.
\newblock Efficient non-linear model reduction via a least-squares
  petrov--galerkin projection and compressive tensor approximations.
\newblock {\em International Journal for Numerical Methods in Engineering},
  86(2):155--181, 2011.

\bibitem{farhatroux1991}
C.~Farhat and F.X. Roux.
\newblock A method of finite element tearing and interconnecting and its
  parallel solution algorithm.
\newblock {\em International Journal for Numerical Methods in Engineering},
  32:1205--1227, 1991.

\bibitem{mandel1993}
J.~Mandel.
\newblock Balancing domain decomposition.
\newblock {\em Communications in Numerical Methods in Engineering},
  9(3):233--241, 1993.

\bibitem{le-tallec1994a}
P.~{Le Tallec}.
\newblock Domain decomposition methods in computational mechanics.
\newblock In {\em Computational Mechanics Advances}, volume~1. Elsevier, 1994.

\bibitem{ladevezedureisseix2000}
P.~Ladev{\`e}ze and D.~Dureisseix.
\newblock A micro/macro approch for parallel computing of heterogeneous
  structures.
\newblock {\em International Journal for computational Civil and Structural
  Engineering}, 1:18--28, 2000.

\bibitem{germainbesson2007}
N.~Germain, J.~Besson, F.~Feyel, and P.~Gosselet.
\newblock {High-performance parallel simulation of structure degradation using
  non-local damage models}.
\newblock {\em International journal for numerical methods in engineering},
  71(3):253--276, 2007.

\bibitem{allixkerfriden2010}
O.~Allix, P.~Kerfriden, and P.~Gosselet.
\newblock On the control of the load increments for a proper description of
  multiple delamination in a domain decomposition framework.
\newblock {\em International Journal for Numerical Methods in Engineering},
  DOI:10.1002/nme.2884, 2010.

\bibitem{lloberas-vallsrixen2011}
O.~Lloberas-Valls, D.J. Rixen, Simone A., and L.J. Sluys.
\newblock Domain decomposition techniques for the efficient modeling of brittle
  heterogeneous materials.
\newblock {\em Computer Methods in Applied Mechanics and Engineering},
  200:1577--1590, 2011.

\bibitem{guidaultallix2008}
P.-A. Guidault, O.~Allix, L.~Champaney, and S.~Cornuault.
\newblock A multiscale extended finite element method for crack propagation.
\newblock {\em Computer Methods in Applied Mechanics and Engineering},
  197(5):381--399, 2008.

\bibitem{allixkerfriden2010b}
O.~Allix, P.~Kerfriden, and P.~Gosselet.
\newblock A relocalization technique for the multiscale computation of
  delamination in composite structures.
\newblock {\em Computer Modeling in Engineering and Sciences}, 55(3):271--291,
  2010.

\bibitem{braconnierferrier2001}
J-C.~Jouhaud T.~Braconnier, M.~Ferrier and P.~Sagaut.
\newblock Towards an adaptive pod/svd surrogate model for aeronautic design.
\newblock {\em Computers and Fluids}, 40(1):195--209, 2011.

\bibitem{quarteronirozza2011}
A~Quarteroni, G~Rozza, and A~Manzoni.
\newblock Certified reduced basis approximation for parametrized partial
  differential equations and applications.
\newblock {\em Journal of Mathematics in Industry}, 1(3):1--44, 2011.

\bibitem{chenwhite2000}
Y~Chen and J~White.
\newblock {A Quadratic Method for Nonlinear Model Order Reduction}.
\newblock In {\em Proceeding of the international Conference on Modeling and
  Simulation of Microsystems}, pages 470--480, 2000.

\bibitem{rewienskiwhite2003}
M~Rewienski and J~White.
\newblock {A trajectory piecewise-linear approach to model order reduction and
  fast simulation of nonlinear circuits and micromachined devices}.
\newblock {\em IEEE Transactions on Computer-Aided Design of Integrated
  Circuits and Systems}, 22(2):155--170, 2003.

\bibitem{niroomandialfaro2008}
S.~Niroomandi, I.~Alfaro, E.~Cueto, and F.~Chinesta.
\newblock Real-time deformable models of non-linear tissues by model reduction
  techniques.
\newblock {\em Computer Methods and Programs in Biomedicine}, 91(3):223--231,
  2008.

\bibitem{chaturantabutsorensen2010}
S.~Chaturantabut and D.C. Sorensen.
\newblock Nonlinear model reduction via discrete empirical interpolation.
\newblock {\em SIAM Journal on Scientific Computing}, 32:2737--2764, 2010.

\bibitem{legresleyalonso2001}
P.A. LeGresley and J.J. Alonso.
\newblock Investigation of non-linear projection for pod based reduced order
  models for aerodynamics.
\newblock {\em AIAA paper}, 926:2001, 2001.

\bibitem{ryckelynck2005}
D.~Ryckelynck.
\newblock A priori hyperreduction method: an adaptive approach.
\newblock {\em Journal of Computational Physics}, 202(1):346--366, 2005.

\bibitem{astrid2004}
P.~Astrid.
\newblock Reduction of process simulation models-a proper orthogonal
  decomposition approach.
\newblock {\em PhD report, Technical University of Eindhoven}, 2004.

\bibitem{lemaitrechaboche1990}
J.~Lema\^itre and J.-L. Chaboche.
\newblock {\em Mechanics of Solid Materials}.
\newblock Cambridge University Press, 1990.

\bibitem{ryckelynckbenziane2011}
D.~Ryckelynck, D.~Missoum~Benziane, S.~Cartel, and J.~Besson.
\newblock A robust adaptive model reduction method for damage simulations.
\newblock {\em Computational Materials Science}, 50(5):1597--1605, 2011.

\bibitem{gosseletrey2006}
P.~Gosselet and C.~Rey.
\newblock Non-overlapping domain decomposition methods in structural mechanics.
\newblock {\em Archives of Computational Methods in Engineering}, 13:515--572,
  2006.

\bibitem{saadschultz1986}
Y.~Saad and M.H. Schultz.
\newblock Gmres: A generalized minimal residual algorithm for solving
  nonsymmetric linear systems.
\newblock {\em SIAM J. Sci. Stat. Comput.}, 7(3):856--869, 1986.

\bibitem{vandervorst1992}
H.A. Van~der Vorst.
\newblock Bi-cgstab: A fast and smoothly converging variant of bi-cg for the
  solution of nonsymmetric linear systems.
\newblock {\em SIAM Journal on scientific and Statistical Computing}, 13:631,
  1992.

\bibitem{kunischvolkwein2011}
K.~Kunisch and S.~Volkwein.
\newblock Optimal snapshot location for computing pod basis functions.
\newblock {\em ESAIM: Mathematical Modelling and Numerical Analysis},
  44(3):509--529, 2010.

\bibitem{abdiwilliams2010}
Herv{\'e} Abdi and Lynne~J. Williams.
\newblock Principal component analysis.
\newblock {\em Wiley Interdisciplinary Reviews: Computational Statistics},
  2(4):433--459, 2010.

\bibitem{krzanowski1987}
W.~J. Krzanowski.
\newblock Cross-validation in principal component analysis.
\newblock {\em Biometrics}, 43(3):575--584, 1987.

\bibitem{kerfridenschmidt2012}
P.~Kerfriden, K.M. Schmidt, T.~Rabczuk, and Bordas S.P.A.
\newblock Statistical extraction of process zones and representative subspaces
  in fracture of random composites.
\newblock {\em Accepted for publication in International Journal for Multiscale
  Computational Engineering, arXiv:1203.2487v2}, 2012.

\bibitem{demboeisenstat1982}
R.S. Dembo, S.C. Eisenstat, and T.~Steihaug.
\newblock Inexact newton methods.
\newblock {\em SIAM Journal on Numerical Analyses}, 19(2):400--408, 1982.

\end{thebibliography}

\end{document}